\documentclass[aps,prd,amsfonts,onecolumn,preprintnumbers,superscriptaddress,%
showpacs,floatfix,nofootinbib]{revtex4}

\usepackage{epsfig}
\usepackage{graphicx}% Include figure files
\usepackage{dcolumn}% Align table columns on decimal point
\usepackage{bm}% bold math
\usepackage{color}

\newcommand{\lsim}{\mbox{\raisebox{-.9ex}{~$\stackrel{\mbox{$<$}}{\sim}$~}}}
\newcommand{\gsim}{\mbox{\raisebox{-.9ex}{~$\stackrel{\mbox{$>$}}{\sim}$~}}}
\def\calg{{\cal G}}

\def\thebiblio#1{
\begin{center}\bf \large References
\end{center}
\list
{[\arabic{enumi}]}{\settowidth\labelwidth{#1.}\leftmargin\labelwidth
 \advance\leftmargin\labelsep
 \usecounter{enumi}}
 \def\newblock{\hskip .11em plus .33em minus -.07em}
 \sloppy
 \sfcode`\.=1000\relax}

\begin{document}
\preprint{}
\title{%
Vector Curvaton with varying Kinetic Function}
% Force line breaks with \\

\author{Konstantinos Dimopoulos}%
\email{k.dimopoulos1@lancaster.ac.uk}
\affiliation{%
Physics Department, Lancaster University, Lancaster LA1 4YB, U.K.\\}%

\author{Mindaugas Kar\v{c}iauskas}%
\email{m.karciauskas@lancaster.ac.uk}
\affiliation{%
Physics Department, Lancaster University, Lancaster LA1 4YB, U.K.\\}%

\author{Jacques M. Wagstaff}%
\email{wagstafj@exchange.lancs.ac.uk}
\affiliation{%
Physics Department, Lancaster University, Lancaster LA1 4YB, U.K.\\}%

\date{\today}% It is always \today, today,
             % but any date may be explicitly specified

\begin{abstract}
A new model realisation of the vector curvaton paradigm is presented and
analysed. The model consists of a single massive Abelian vector field, with a
Maxwell type kinetic term. By assuming that the kinetic function and the mass
of the vector field are appropriately varying during inflation, it is shown 
that a scale invariant spectrum of superhorizon perturbations can be generated.
These perturbations can contribute to the curvature perturbation of the 
Universe. If the vector field remains light at the end of inflation it is 
found that it can generate substantial statistical anisotropy in the spectrum 
and bispectrum of the curvature perturbation. In this case the non-Gaussianity
in the curvature perturbation is predominantly anisotropic, which will be 
a testable prediction in the near future. If, on the other hand, the vector 
field is heavy at the end of inflation then it is demonstrated that particle
production is approximately isotropic and the vector field alone can give
rise to the curvature perturbation, without directly involving any fundamental 
scalar field. The parameter space for both possibilities is shown to be 
substantial. Finally, toy-models are presented which show that the desired 
variation of the mass and kinetic function of the vector field can be 
realistically obtained, without unnatural tunings, in the context of 
supergravity or superstrings.
\end{abstract}

\maketitle

\begin{widetext}

\section{Introduction}

Cosmic inflation is the most compelling solution to the horizon and flatness
problems of the standard, hot big bang cosmology. As an additional bonus, 
inflation can also generate the primordial curvature perturbation $\zeta$ of
the Universe, which is necessary for the formation of the observed structures
such as galaxies and galactic clusters. Fortunately, a snapshot of the 
primordial curvature perturbation is reflected onto the Cosmic Microwave 
Background (CMB) radiation, whose temperature perturbations can provide
detailed information about $\zeta$, such as the amplitude and scale dependence
of its spectrum. The importance of this information lies in its discriminatory 
power as it offers, so far, the only tool which allows the testing of 
individual inflation models. Detailed cosmological observations, not only of 
the CMB but also of the distribution of structures in the Universe (e.g.
galaxy surveys), are in 
agreement with the generic predictions of inflation for $\zeta$. Indeed, the
observational evidence supports the existence of a nearly scale-invariant
spectrum of predominantly adiabatic and Gaussian superhorizon perturbations. 
Inflation typically produces such perturbations through the amplification of 
the quantum fluctuations of suitable fields during a period of accelerated
expansion of space. The process is called particle production and it is a 
manifestation of the Hawking radiation process in de Sitter space. Now that 
observations of the products of this process are becoming ever more precise, 
model-builders are forced to construct inflationary models which are better and
more closely connected with the theory, more detailed and rigorous, and
more testable and falsifiable by comparison with the data. The question to be 
asked then is, which are those ``suitable fields'', whose quantum fluctuations
are the ultimate sources of the curvature perturbation and all the observed 
structures in the Universe?

A very useful tool in theoretical physics and, by extension, in cosmology
has been the fundamental scalar quantum field. Such scalar fields are 
ubiquitous in theories beyond the standard model such as supersymmetry
(the scalar partners of the observed fermion fields) or string theory
(string moduli fields). In cosmology, a scalar field is usually assumed to
drive the dynamics of inflation (inflaton field) or even give rise to the 
accelerated expansion at present (quintessence field). 
It is not surprising, therefore, that the particle production process,
responsible for the generation of $\zeta$, has until recently been considered
only in the context of scalar fields. So useful a tool is a scalar field that
one tends to forget that, as yet, it is no more than a theoretical 
construction, whose physical existence is only conjectured, since no 
fundamental scalar quantum field has ever been observed. One could wonder
therefore whether other kinds of fields contributed to the generation of the
curvature perturbation in the Universe. What if scalar fields did not exist?
Could the galaxies form without them?

Recently, the contribution of vector boson fields to the curvature perturbation
is attracting growing attention. In the pioneering work in Ref.~\cite{vecurv} a
mechanism was introduced, which could potentially allow a single vector field 
to generate the observed curvature perturbation. The mechanism was the first 
such suggestion in the literature and was based on the curvaton idea, which was
developed for scalar fields in Ref.~\cite{curv} (see also Ref.~\cite{early}) 
in order to 
alleviate the fine-tunning problems of inflation model-building \cite{liber}.
The vector 
curvaton mechanism assumes that the vector field has a negligible contribution
to the energy budget of the Universe during inflation, when on the one hand it 
becomes homogenised, but on the other hand can also obtain a scale-invariant 
superhorizon spectrum of perturbations due to some appropriate breaking of its 
conformality. After inflation
the density parameter of the vector field grows, especially after the mass of 
the vector field becomes important (larger than the Hubble scale). When this
happens, the zero-mode undergoes coherent oscillations during which the vector 
field acts as pressureless, {\em isotropic} matter \cite{vecurv}. Hence, it can
dominate (or nearly dominate) the Universe without generating excessive 
%large-scale anisotropy, 
anisotropic expansion, avoiding thus the primary obstacle for using vector 
fields to generate $\zeta$. When the field dominates the Universe it imposes
its curvature perturbation spectrum according to the curvaton scenario 
\cite{curv}. 

The vector curvaton mechanism is a paradigm in search of a model, much like 
inflation itself. The main ingredient necessary is a mechanism for
the generation of a flat spectrum of vector field perturbations during 
inflation. This is a problem which was originally addressed by efforts to 
generate a coherent primordial magnetic field during inflation, in order to 
source the magnetic fields of the galaxies (for a review see 
Ref.~\cite{pmfrev}). As realised in Ref.~\cite{TW}, the particle production 
process can operate with vector fields only when their conformal invariance is
appropriately broken. A number of proposals have been put forward for this to 
occur during inflation \cite{mine,pmfinf,gaugekin}. For example, 
as discussed in Ref.~\cite{vecurv}, a flat spectrum of vector field 
perturbations is generated if the effective mass-squared of the vector field 
during inflation is \mbox{$m_{\rm eff}^2=-2H_*^2$}, where $H_*$ is the
inflationary Hubble scale. In Ref.~\cite{TW} it has been shown that this could 
be achieved if the vector field is non-minimally coupled to gravity with a 
coupling of the form $\frac16RA^2$, where $R$ is the Ricci scalar. 
This idea was employed in the vector curvaton mechanism in Ref.~\cite{nonmin}.
%and \mbox{$A^2\equiv A_\mu A^\mu$}
Another possibility was to consider a varying kinetic function as 
discussed in Ref.~\cite{gaugekin}. The first attempt to implement this idea
to the vector curvaton mechanism is in Ref.~\cite{sugravec}.

In the above early works on the vector curvaton, attention was paid mainly on 
the transverse components of the vector field perturbations. The reason was 
that the vector curvaton mechanism employs the fractional perturbation of 
the density of the vector field, which is a scalar quantity, i.e. isotropic. 
Indeed, the mechanism does not cause any anisotropic expansion because,
when the zero-mode of the field undergoes rapid coherent oscillations, so 
do the field's perturbations. Thus, the transverse components were deemed 
enough to
generate $\zeta$ and the longitudinal component was ignored. This turned out
to miss a key effect introduced to the curvature perturbation from vector 
fields, namely statistical anisotropy.

Statistical anisotropy in the curvature perturbation is a new observable and
it amounts to
%is due to anisotropic particle production and would produce 
direction dependent patterns in the CMB temperature perturbations as well as in
the large scale structure (e.g. rows of galaxies) if it is intense enough. 
The observational limit on such a signal is rather weak as statistical 
anisotropy in the spectrum of $\zeta$ is allowed up to as much as 30\% 
\cite{GE}. This implies that it is likely to be observed by the forthcoming 
observations of the Planck satellite mission which will detect 
statistical anisotropy in the spectrum if it is larger than $2\%$
\cite{stanisplanck}, while there is no reason to think that this will be the 
ultimate limit. Another possible indication of statistical anisotropy in the 
curvature perturbation  is the alignment of the quadrupole and octupole moments
of the CMB (the so-called ``Axis of Evil'' \cite{AoE}) and of galaxy spins
\cite{spins,align}.
The alignment of the low CMB multipoles was shown to persist beyond
foreground removal \cite{foreground} despite being statistically extremely 
unlikely with isotropic perturbations (see also \cite{hansen}). Moreover,
in Ref.~\cite{align} it is claimed that the preferred direction of the 
alignment of galactic spins appears to coincide with the ``Axis of Evil'', as 
one would expect if they were both due to statistical anisotropy in $\zeta$.
It has to be noted that statistical anisotropy cannot be introduced by scalar 
fields alone, so if its observation is confirmed, 
%is indeed observed, 
this would be a powerful indication
that vector fields at least affect $\zeta$. 

The first study of the generation of statistical anisotropy by a vector field 
is in  Ref.~\cite{soda}, where a vector field is assumed to modulate the decay 
of the inflaton. In a separate development  \cite{stanis0}
it was pointed out   that even a homogeneous  vector field will generate
statistical anisotropy in $\zeta$, because it will break the statistical
isotropy of the scalar field perturbation generated from the vacuum
fluctuation. This is further explored in Ref.~\cite{stanis+}.
A comprehensive model-independent study of statistical anisotropy can be found
in Ref.~\cite{stanis}. The study employs the $\delta N$ formalism \cite{dN},
extending it to include the effects of vector fields. In Ref.~\cite{stanis} 
it is shown that a vector field can alone generate $\zeta$ through the vector 
curvaton mechanism only if {\em the particle production process is 
approximately isotropic}. Otherwise, the contribution of the vector curvaton 
field has to be subdominant and its significance amounts to the generation of 
statistical anisotropy only.
As an application, the $\frac16RA^2$ vector curvaton model was revisited and 
shown to result in statistical anisotropy of order unity, which would be 
excessive if the vector field were to provide the dominant contribution to 
$\zeta$. In such a case, the only way that excessive statistical anisotropy can
be avoided without introducing another dominant (and statistically isotropic) 
source of curvature perturbations (such as scalar fields) is by considering
either a triad of identical vector fields each perpendicular to the other two 
\cite{triad} or hundreds of identical vector fields, randomly oriented in 
space. In such a case anisotropy in the spectrum or in the expansion can be 
avoided or suppressed, which means that vector fields could also act as 
inflatons, without generating anisotropic expansion.

Using vector fields as inflatons was first considered in Ref.~\cite{VI}.
More recently, vector field inflation has been studied in Ref.~\cite{vinf}
with the $\frac16RA^2$ model, involving hundreds of randomly oriented vector 
fields. For more studies of similar vector inflation models see 
Refs.~\cite{vinf+,mota}.\footnote{Another way to avoid anisotropy is to 
consider a time-like vector field \cite{vinf0}.} In the same spirit, vector 
fields have been considered as candidates for dark energy \cite{mota,vde}.
Furthermore, Refs.~\cite{YMinf} and \cite{YMde} investigate inflation or dark 
energy respectively, using Yang-Mills fields. Finally, inflation with
$P$-forms is explored in Ref.~\cite{forms}.

Also obtained in Ref.~\cite{stanis} is the contribution of vector fields in 
the bispectrum of the curvature perturbation (for the trispectrum see 
Ref.~\cite{tri}). Using these results, in
Ref.~\cite{fnlanis} the non-Gaussianity in the spectrum is studied in a 
model-independent way, though the results are applied to the models of 
Refs.~\cite{nonmin} and \cite{soda}. A recent study in Ref.~\cite{nonAbel}, 
extends this treatment in the case of non-Abelian vector fields, applying the 
results in the $\frac16RA^2$ model. In all cases, it is shown that 
statistical anisotropy in the spectrum and bispectrum are correlated. This is a
smoking gun for the contribution of a vector field to the curvature 
perturbation. 
It is likely that the Planck mission will observe non-zero 
non-Gaussianity in the CMB temperature perturbations. If this is the case then 
simple single-scalar-field models of inflation will be ruled out. If an angular
modulation of the non-linearity parameter $f_{\rm NL}$, which quantifies 
non-Gaussianity, is also observed, then this is a strong indication of the 
contribution of vector fields to $\zeta$. 

An important aspect of the mechanisms which break the conformality of vector
fields and result in particle production of their perturbations during 
inflation is the appearance of instabilities. For example, in 
Ref.~\cite{peloso} the $\frac16RA^2$ model was criticised for giving rise to 
ghosts and other instabilities, e.g. when the modes of the perturbations exit 
the horizon, although in Ref.~\cite{stanis} the latter was demonstrated not to 
be a problem (also see Ref.~\cite{stanis} for a discussion on ghosts). For a 
comprehensive analysis on ghosts and tachyon instabilities see also 
Ref.~\cite{carroll}.

In this paper we introduce a new model of vector curvaton, which does not
suffer from such instabilities (see also Ref.~\cite{inst}). 
The model consists of a massive Abelian
vector field with a Maxwell type kinetic term, whose kinetic function is 
modulated during inflation similarly to the models in Ref.~\cite{gaugekin}. 
Our model is a follow up of Ref.~\cite{sugravec}, which did not consider 
statistical anisotropy. We attempt a detailed analytical investigation and 
confirm our findings numerically. As we demonstrate, our vector curvaton
model gives rise to a rich phenomenology. In particular, when the mass of 
the vector field is smaller than the inflationary Hubble scale the vector 
curvaton can give rise to statistical anisotropy in both the spectrum and 
bispectrum, with the angular modulation dominating $f_{\rm NL}$, rendering 
thus the model, in this case,
falsifiable by observations in the near future. When the mass of the vector 
field is larger than the inflationary Hubble scale though, particle production 
is isotropic and the vector field can alone generate the observed curvature 
perturbation. The parameter space for both regimes is shown to be substantial.

The structure of our paper is as follows. In Sec.~II we present our model and
obtain the equations of motion of the perturbations of the vector field
components in momentum space. In Sec.~III we study particle production in order
to obtain the desired scaling of the mass and the kinetic function of the 
vector field, which results in the production of a scale invariant superhorizon
spectrum of perturbations. In Secs.~IV and V we calculate in detail the power 
spectra of the perturbations in the two possible alternatives for the scaling 
of the kinetic function. To this end we obtain the evolution of the mode 
functions of the perturbations, through analytic approximations of high 
precision. In Sec.~VI we calculate the statistical anisotropy in the spectrum 
and bispectrum of the produced perturbations in all cases. In Sec.~VII we 
study the evolution of the zero mode of the vector boson condensate, during
and after inflation. The evolution of the zero mode is necessary in 
order to determine the cosmology after inflation as well as calculate the 
curvature perturbation. In Sec.~VIII we apply the vector curvaton mechanism
in order to obtain constraints on the model parameters such as the mass of
the vector field and the inflationary Hubble scale. In Sec.~IX we present
two toy-models which demonstrate how the desired scaling of the mass and the 
kinetic function of the vector field can be obtained in a natural way in the
context of supergravity and superstrings. In Sec.~X we summarise and discuss 
our findings. Finally, in Sec.~XI we present our conclusions.
Throughout the paper we use natural units such that \mbox{$c=\hbar=k_B=1$} and
Newton's gravitational constant is \mbox{$8\pi G=m_P^{-2}$}, where 
\mbox{$m_P=2.44\times 10^{18}\,$GeV} is the reduced Planck mass.

\section{Equations of motion}

Consider the Lagrangian density for a massive, Abelian vector field
\begin{equation}
{\cal L}=-\frac{1}{4}fF_{\rm \mu\nu}F^{\mu\nu}+\frac{1}{2}m^2A_\mu A^\mu\;,
\label{L}
\end{equation}
where $f$ is the kinetic function, $m$ is the mass and 
the field strength tensor is
\mbox{$F_{\mu\nu}=\partial_\mu A_\nu - \partial_\nu A_\mu$}. The above can be 
the Lagrangian density of a massive Abelian gauge field, in which case $f$ is 
the gauge kinetic function. However, we need not restrict ourselves to gauge 
fields only. If no gauge symmetry is considered the argument in support of the 
above Maxwell type kinetic term is that it is one of the few (three) choices 
\cite{carroll} which avoids introducing instabilities, such as ghosts 
\cite{peloso}. Also, we note here that a massive vector field which is
not a gauge field is renormalizable only if it is Abelian \cite{tikto}.

We focus, at first, on a period of cosmic inflation, during which we assume 
that the
contribution of the vector field to the energy budget of the Universe is minute
and can be ignored. Thus, we take the inflationary expansion to be isotropic. 
We also assume that inflation is of
quasi-de Sitter type, i.e. the Hubble parameter is $H\approx\,$constant.
We will consider that, during inflation, \mbox{$f=f(t)$} and \mbox{$m=m(t)$} 
can be functions of cosmic time $t$.
%\footnote{A similar setup is employed in 
%so-called dilaton electromagnetism \cite{bambayoko}, where 
%\mbox{$f=e^{-\lambda\Phi/m_P}$} with $\Phi$ being the dilaton. This setup has 
%been used to break the conformality of electromagnetism and generate a 
%primordial magnetic field during inflation \cite{bambasasa} (see 
%also~\cite{ratra}).}

Inflation is expected to homogenise the vector field. Following 
Ref.~\cite{sugravec}, we find that the temporal component of the homogeneous
vector field has to be zero, while the spatial components satisfy the equation 
of motion:
\begin{equation}
\mbox{\boldmath $\ddot A$}+\left(H+\frac{\dot f}{f}\right)
\mbox{\boldmath $\dot A$}+\frac{m^2}{f}\mbox{\boldmath $A$}=0\,,
\label{EoMhom}
\end{equation}
where the dot denotes derivative with respect to $t$. From the above it is 
evident that the effective mass of the vector field is
\begin{equation}
M\equiv\frac{m}{\sqrt f}\,,
\label{M}
\end{equation}
where we assumed that \mbox{$m,f>0$}. 

In order to study 
particle production we need to perturb the vector field around the homogeneous
zero mode $A_\mu(t)$ as:
\begin{equation}
A_\mu(t,\mbox{\boldmath $x$})=A_\mu(t)+
\delta A_\mu(t,\mbox{\boldmath $x$})\,.
\label{pert}
\end{equation}
Now, let us switch to momentum space by Fourier expanding the perturbations:
\begin{equation}
\delta A_\mu(t, \mbox{\boldmath $x$})=
\int\frac{d^3k}{(2\pi)^{3/2}}\;\delta{\cal A}_\mu (t, \mbox{\boldmath $k$})\,
\exp(i\mbox{\boldmath $k\cdot x$})\,.
\label{fourier}
\end{equation}
Then, according to Ref.~\cite{sugravec} the equations of motion for
the spatial components of the vector field perturbations in momentum space are:
\begin{equation}
\left[\partial_t^2+\left(H+\frac{\dot f}{f}\right)\partial_t+
\frac{m^2}{f}+\left(\frac{k}{a}\right)^2\right]
\delta\mbox{\boldmath $\cal A$}^\perp 
=0
\label{EoMperp}
\end{equation}
and
%\begin{widetext}
\begin{equation}
\left\{\partial_t^2+
\left[H+\frac{\dot f}{f}+
\left(2H+2\frac{\dot m}{m}-\frac{\dot f}{f}\right)
\frac{\left(\frac{k}{a}\right)^2%k^2
}{\left(\frac{k}{a}\right)^2+%k^2
\frac{m^2}{f}}\right]
\partial_t\,+
\frac{m^2}{f}+\left(\frac{k}{a}\right)^2\right\}
\delta\mbox{\boldmath $\cal A$}^\parallel=0\,,
\label{EoMparal}
\end{equation}
%\end{widetext}
where \mbox{$a=a(t)$} is the scale factor of the Universe,
\mbox{$k\equiv|\mbox{\boldmath $k$}|$} and
the longitudinal and transverse components 
%(i.e. parallel and perpendicular to \mbox{\boldmath $k$}) 
are defined as:
\begin{equation}
\delta\mbox{\boldmath $\cal A$}^\parallel\equiv
\frac{\mbox{\boldmath $k$}(\mbox{\boldmath $k\cdot$}%
\delta\mbox{\boldmath $\cal A$})}{k^2}
\quad{\rm and}\quad
\delta\mbox{\boldmath $\cal A$}^\perp\equiv
\delta\mbox{\boldmath $\cal A$}-\delta\mbox{\boldmath $\cal A$}^\parallel.
\label{translong}
\end{equation}

To continue we need to employ the canonically normalised vector field $B_\mu$
and the physical (in contrast to comoving) vector field $W_\mu$, whose
spatial components are:
\begin{equation}
\mbox{\boldmath $B$}\equiv\sqrt f\mbox{\boldmath $A$}\quad{\rm and}\quad
\mbox{\boldmath $W$}\equiv\mbox{\boldmath $B$}/a=
\sqrt f\,\mbox{\boldmath $A$}/a\,.
\label{BW}
\end{equation}
Expressing Eqs.~(\ref{EoMperp}) and (\ref{EoMparal}) in terms of the 
physical vector field we obtain the following equations for the spatial 
components of the physical vector field perturbations:
%
%\begin{widetext}
\begin{equation}
%\ddot{\delta W}_k^\perp
\left\{
\partial_t^2+3H\partial_t+
%+3H\dot{\delta W}_k^\perp+
%\left\{
\frac12\left[\frac12\left(\frac{\dot f}{f}\right)^2-
\frac{\ddot f}{f}-\frac{\dot f}{f}H+4H^2\right]+
\frac{m^2}{f}+\left(\frac{k}{a}\right)^2\right\}
\delta \mbox{\boldmath $\cal W$}%_k\mbox{\boldmath $k$}
^\perp=0
\label{EoMtrans}
\end{equation}
and
\begin{eqnarray}
%\ddot{\delta W}_k^\parallel+
\left\{\partial_t^2+
\left[3H+\left(2H+2\frac{\dot m}{m}-\frac{\dot f}{f}\right)
\frac{\left(\frac{k}{a}\right)^2}{\left(\frac{k}{a}\right)^2+\frac{m^2}{f}}
\right]
%\dot{\delta W}_k^\parallel+
\partial_t+ 
\frac12\left[\frac12\left(\frac{\dot f}{f}\right)^2-
\frac{\ddot f}{f}-\frac{\dot f}{f}H+4H^2\right]\;+
\right.& &\nonumber\\
%\left\{
\left.
\left(H-\frac12\frac{\dot f}{f}\right)
\left(2H+2\frac{\dot m}{m}-\frac{\dot f}{f}\right)
\frac{\left(\frac{k}{a}\right)^2}{\left(\frac{k}{a}\right)^2+\frac{m^2}{f}}+
\frac{m^2}{f}+\left(\frac{k}{a}\right)^2\right\}
\delta \mbox{\boldmath $\cal W$}%_k
^\parallel & = & 0\,,
\label{EoMlong}
\end{eqnarray}
%\end{widetext}
where
\begin{equation}
\delta\mbox{\boldmath $W$}(t, \mbox{\boldmath $x$})=
\int\frac{d^3k}{(2\pi)^{3/2}}\;\delta\mbox{\boldmath $\cal W$}%_k 
(t, \mbox{\boldmath $k$})\,\exp(i\mbox{\boldmath $k\cdot x$})\,.
\label{Fourier}
\end{equation}

\section{Particle Production}

In this section we attempt a preliminary study of the particle production
process during inflation in order to ascertain what kind of time dependence
do $f(t)$ and $m(t)$ need to have in order to result in scale-invariant power 
spectra for all the components of the superhorizon vector field perturbations.

To study particle production of the vector field during inflation, we first
need to promote $\delta\mbox{\boldmath $W$}$ to a quantum %-mechanical 
operator. To do that we expand in creation and annihilation operators as
%
%\begin{widetext}
\begin{equation}
\hat{\delta\mbox{\boldmath $W$}}=
\int\frac{d^3k}{(2\pi)^%{3/2}
3}\;\sum_\lambda\left[
\mbox{\boldmath $e$}_\lambda(\hat{\mbox{\boldmath $k$}})
\hat a_\lambda(\mbox{\boldmath $k$})
w_\lambda(t,k)e^{i\mbox{\scriptsize\boldmath $k\cdot x$}}
+\mbox{\boldmath $e$}^*_\lambda(\hat{\mbox{\boldmath $k$}})
\hat a^\dag_\lambda(\mbox{\boldmath $k$})
w^*_\lambda(t,k)e^{-i\mbox{\scriptsize\boldmath $k\cdot x$}}\right],
\label{expand}
\end{equation}
%\end{widetext}
where $\hat{\mbox{\boldmath $k$}}\equiv\mbox{\boldmath $k$}/k$, with 
$k\equiv|\mbox{\boldmath $k$}|$ and $\lambda=L,R,\parallel$ with $L,R$ denoting
the Left and Right transverse polarisations respectively such that
\begin{equation}
(\delta\mbox{\boldmath $\cal W$}^\perp)^2=
(\delta\mbox{\boldmath $\cal W$}^L)^2+
(\delta\mbox{\boldmath $\cal W$}^R)^2.
\label{LR}
\end{equation}
The polarisation vectors 
$\mbox{\boldmath $e$}_\lambda$ can be chosen as
\begin{equation}
e_L\equiv\frac{1}{\sqrt2}(1,i,0),\;\;%/\sqrt2,\;
e_R=\frac{1}{\sqrt2}(1,-i,0),\;\;%/\sqrt2,\; 
e_\parallel=(0,0,1)\,,
\label{polar}
\end{equation}
while we have canonical quantisation with
\begin{equation}
\left[\hat a_\lambda(\mbox{\boldmath $k$}),
\hat a^\dag_{\lambda'}(\mbox{\boldmath $k$}')\right]=(2\pi)^3
\delta(\mbox{\boldmath $k$}-\mbox{\boldmath $k$}')\delta_{\lambda\lambda'}\;.
\label{quant}
\end{equation}

Since Eqs.~(\ref{EoMtrans}) and (\ref{EoMlong}) are linear they will be
satisfied also by the corresponding mode functions $w_\lambda(t,k)$. To study 
particle production in this theory we need to solve these equations with the
appropriate initial conditions. After obtaining the solutions we will find
the appropriate constraints on $f$ and $m$, which can provide us with a scale 
invariant spectrum of superhorizon perturbations.
The strategy is to begin with the transverse equation
(\ref{EoMtrans}), since it is the simplest. Conditions obtained by requiring
a flat spectrum for the transverse components of the vector field perturbations
will hopefully simplify the longitudinal equation (\ref{EoMlong}) as well.

\subsection{The transverse components}

Let us assume that the time dependence of the kinetic function can be
parametrised as
\begin{equation}
f\propto a^\alpha,
\label{fa}
\end{equation}
where $\alpha$ is a real constant. 
We will also assume that 
\mbox{$f\rightarrow 1$} at (least by) the end of inflation so that, after 
inflation, the vector field is canonically normalised.
%\footnote{This means that 
%\mbox{$f=e^{-\alpha N}$}, where $N$ is the remaining e-folds of inflation.}
Then, according to Eq.~(\ref{EoMtrans}) the
equation of motion for the transverse mode functions is
%
%\begin{widetext}
\begin{equation}
\ddot w_{L,R}+3H\dot w_{L,R}+\left[-\frac14(\alpha+4)(\alpha-2)H^2+
%\frac{m^2}{f}
M^2+\left(\frac{k}{a}\right)^2\right]w_{L,R}=0\,,
\label{wLR}
\end{equation}
%\end{widetext}
where we used Eq.~(\ref{M}) and considered that the theory is parity invariant
so that 
\mbox{$\delta\mbox{\boldmath $\cal W$}_L=\delta\mbox{\boldmath $\cal W$}_R$},
i.e. \mbox{$w_L=w_R\equiv w_{L,R}$}.

In analogy to the equation of motion for the mode function of a scalar field
during quasi-de Sitter inflation, it can be readily deduced that
a scale invariant spectrum of perturbations is attained if
\begin{equation}
\alpha=-1\pm 3
\label{alpha}
\end{equation}
(i.e. either \mbox{$f\propto a^2$} or \mbox{$f\propto a^{-4}$}) and
\begin{equation}
M_*%=\left.\frac{m}{\sqrt f}\right|_*
\ll H\,,
\label{mbound}
\end{equation}
where the star denotes the time when the cosmological scales exit the horizon.
The latter condition simply requires that the physical vector field $W_\mu$ is 
effectively massless at that time.\footnote{Note that this is not the same as
having $A_\mu$ being effectively massless. In the latter case the vector field
is approximately conformally invariant and does not undergo particle 
production. However, the conformality of the massless physical vector field 
$W_\mu$ is broken.} The above condition can be obtained as follows.

Let us define %$r$ such that
\begin{equation}
r\equiv%\frac{am}{k\sqrt f}\,.
\frac{aM}{k}\,.
\label{r}
\end{equation}
If we assume that the field is effectively massless
\mbox{$M%=m/\sqrt{f}
\ll k/a$} (i.e. \mbox{$r\ll 1$})
until the end of inflation we
can straightforwardly calculate the power spectrum of the perturbations. 
First, we solve Eq.~(\ref{wLR}) using the Bunch-Davis vacuum boundary 
condition
\begin{equation}
\lim_{_{\hspace{.5cm}
\frac{k}{aH}\rightarrow+\infty}}\hspace{-.5cm}
w_{L,R}=\frac{a^{-1}}{\sqrt{2k}}\,e^{ik/aH},
\label{LRvac}
\end{equation}
which is valid well within the horizon, where the inflationary expansion 
is not felt and one can use flat spacetime quantum field theory. The solution 
is then
\begin{equation}
w_{L,R}=\frac{a^{-3/2}}{2}\sqrt{\frac{\pi}{H}}\,e^{i\frac{\pi}{2}(\nu+\frac12)}
H_\nu^{(1)}(k/aH)\,,
\label{wLRsmallm}
\end{equation}
where $H_\nu^{(1)}$ is the Hankel function of the first kind and $\nu$ is 
given by
\begin{equation}
\nu=\frac12|\alpha+1|\,.
\label{nu}
\end{equation}
At late times (superhorizon scales) the dominant term in 
the above solution approaches
\begin{equation}
\lim_{_{\hspace{.5cm}
\frac{k}{aH}\rightarrow 0^+}}\hspace{-.5cm}
w_{L,R}=-\frac{ia^{-3/2}}{2\Gamma(1-\nu)}\sqrt{\frac{\pi}{H}}
e^{i\frac{\pi}{2}(\nu+\frac12)}\left(\frac{k}{2aH}\right)^{-\nu}.
\label{wLRsupH}
\end{equation}
Hence, the power spectrum of the perturbations is
\begin{eqnarray}
{\cal P}_{L,R} & \equiv & \frac{k^3}{2\pi^2}
\left|\hspace{-.5cm}\lim_{_{\hspace{.5cm}
\frac{k}{aH}\rightarrow 0^+}}\hspace{-.5cm}
w_{L,R}\right|^2
%\;=\nonumber\\
%& = & 
=\frac{4\pi}{[\Gamma(1-\nu)]^2}\left(\frac{H}{2\pi}\right)^2
\left(\frac{k}{2aH}\right)^{3-2\nu}.
\end{eqnarray}
As evident, scale invariance is attained when \mbox{$\nu=3/2$}, which, in view 
of Eq.~(\ref{nu}), results in the values for $\alpha$ shown in 
Eq.~(\ref{alpha}). If \mbox{$\alpha=-1\pm 3$} then the above gives
\begin{equation}
{\cal P}_{L,R}=\left(\frac{H}{2\pi}\right)^2,
\label{PLR}
\end{equation}
i.e. the spectrum is scale invariant with amplitude given by the Hawking 
temperature for de Sitter space. 

As is shown later on, even if the
$r\ll 1$ condition is violated before the end of inflation (but after the 
cosmological scales exit the horizon), the scale dependence of the spectrum 
at the vicinity of the cosmological scales is not altered (only its amplitude 
is). The reason for this is that, when \mbox{$r\gg 1$}, the mass term in
Eq.~(\ref{wLR}) dominates the $k$-dependent term. Consequently, the equation 
loses its sensitivity on scale dependence, which means that the evolution of 
the perturbations in the \mbox{$r\gg 1$} regime will not affect their 
dependence on scale. Hence, a scale-invariant spectrum will remain so.
We demonstrate explicitly this in Sec.~\ref{fmpref}.
Thus, Eqs.~(\ref{alpha}) and (\ref{mbound}) are sufficient for the generation 
of a scale invariant spectrum of perturbations for the transverse component of 
our vector field.

\subsection{The longitudinal component}

Let us assume now that the time dependence of $m$ can be
parametrised as
\begin{equation}
m\propto a^\beta,
\label{ma}
\end{equation}
where $\beta$ is a real constant. Then, in view of our notation and 
Eq.~(\ref{EoMlong}), the equation of motion for the longitudinal mode function 
is
%
%\begin{widetext}
\begin{equation}
\ddot w_\parallel+
\left(3+\frac{2-\alpha+2\beta}{1+r^2}\right)H\dot w_\parallel+
\left[-\frac12(\alpha-2)\left(\alpha+4+\frac{2-\alpha+2\beta}{1+r^2}\right)H^2
+\left(\frac{k}{a}\right)^2(1+r^2)\right]
w_\parallel = 0\,.
\label{wlong}
\end{equation}
%\end{widetext}
Now, let us make use of Eq.~(\ref{alpha}), which is necessary
to obtain a scale-invariant spectrum for the transverse components. If we 
assume also that \mbox{$r\ll 1$}, the above equation simplifies to
%
%\begin{widetext}
\begin{equation}
\ddot w_\parallel+(5-\alpha+2\beta)H\dot w_\parallel+
\left[-\frac12(\alpha-2)(2-\alpha+2\beta)H^2+\left(\frac{k}{a}\right)^2\right]
w_\parallel = 0\,.
\label{wlongsmall}
\end{equation}
%\end{widetext}

Similarly to the transverse components, we can 
%evaluate the integration 
%constants $c_1$ and $c_2$ 
solve the above equation
by using the vacuum boundary condition, which
reads
\begin{equation}
\lim_{_{\hspace{.5cm}
\frac{k}{aH}\rightarrow+\infty}}\hspace{-.5cm}
w_\parallel=\gamma\frac{a^{-1}}{\sqrt{2k}}\,e^{ik/aH}.
\label{longvac}
\end{equation}
Note that, for the longitudinal component of the vector field perturbations,
the vacuum condition is multiplied by the Lorentz boost factor $\gamma$, which
takes us from the frame with \mbox{{\boldmath $k$}$\,=0$} (where there is no 
distinction between longitudinal and transverse components) to that of momentum
\mbox{{\boldmath $k$}$\,\neq 0$}. The Lorentz boost factor is
\begin{equation}
\gamma=\frac{E}{M}=
\frac{\sqrt{\left(\frac{k}{a}\right)^2+M^2%\frac{m^2}{f}
}}{M%\frac{m}{\sqrt f}
}=
\sqrt{1+\frac{1}{r^2}},
\label{gamma}
\end{equation}
where we considered that the effective mass of the physical vector field is 
%\mbox{$M=m/\sqrt f$} [c.f. Eq.~(\ref{M})].
given by Eq.~(\ref{M}).

Now, assuming that $r\ll 1$ at early times, Eq.~(\ref{wlongsmall}) remains 
valid within the horizon, where it can be 
matched to the vacuum expression in Eq.~(\ref{longvac}) with 
\mbox{$\gamma\simeq 1/r$}. By doing so we obtain
%
%\begin{widetext}
\begin{equation}
w_\parallel=
\frac{a^{-3/2}}{r}\sqrt{\frac{\pi}{4H}}
\frac{e^{-i\frac{\pi}{2}(\hat\nu-\frac32)}}{\sin(\pi\hat\nu)}
%\left[J_{\hat\nu}(k/aH)-e^{i\pi\hat\nu}J_{-\hat\nu}(k/aH)\right],
\left[J_{\hat\nu}\left(\frac{k}{aH}\right)-e^{i\pi\hat\nu}
J_{-\hat\nu}\left(\frac{k}{aH}\right)\right],
\label{wlongsmallm}
\end{equation}
%\end{widetext}
where $J_{\hat\nu}$ denotes Bessel function of the first kind, and
\begin{equation}
\!\,
\hat\nu=\frac12\sqrt{9+2(\alpha+1)(2-\alpha+2\beta)+(2-\alpha+2\beta)^2}.
\hspace{-1cm}
\label{hatnu}
\end{equation}

At late times (superhorizon scales) the dominant term in the above solution 
approaches
%
%\begin{widetext}
\begin{equation}
%\!\,\hspace{-.5cm}
\lim_{_{\hspace{.5cm}
\frac{k}{aH}\rightarrow 0^+}}\hspace{-.5cm}
w_\parallel=-\frac{a^{-3/2}}{\Gamma(1-\hat\nu)}\sqrt{\frac{\pi}{H}}
\frac{e^{i\frac{\pi}{2}(\hat\nu+\frac32)}}{\sin(\pi\hat\nu)}
%\left(\frac{H\sqrt f}{m}\right)
\left(\frac{H}{M}\right)
\left(\frac{k}{2aH}\right)^{1-\hat\nu},
%\hspace{-2cm}
\label{wlongsupH}
\end{equation}
where we used Eq.~(\ref{r}). Hence, the power spectrum of the perturbations is
%
%\begin{eqnarray}
\begin{equation}
{\cal P}_\parallel 
%& \equiv & 
\equiv\frac{k^3}{2\pi^2}
\left|\hspace{-.5cm}\lim_{_{\hspace{.5cm}
\frac{k}{aH}\rightarrow 0^+}}\hspace{-.5cm}
w_\parallel\right|^2
%\;=\nonumber\\
% & = & 
=\frac{16\pi}{\sin^2(\pi\hat\nu)[\Gamma(1-\hat\nu)]^2}
%\left(\frac{H\sqrt f}{m}\right)^2
\left(\frac{H}{M}\right)^2
\left(\frac{H}{2\pi}\right)^2
\left(\frac{k}{2aH}\right)^{5-2\hat\nu}.
\end{equation}
%\end{eqnarray}
%\end{widetext}
It is clear that scale invariance is attained when \mbox{$\hat\nu=5/2$}. If 
this is so then the above becomes
\begin{equation}
{\cal P}_\parallel=9%\left(\frac{H\sqrt f}{m}\right)^2
\left(\frac{H}{M}\right)^2
\left(\frac{H}{2\pi}\right)^2,
\label{Plong}
\end{equation}
i.e. the spectrum is scale invariant with amplitude given by the Hawking 
temperature for de Sitter space, but also determined by the mass fraction
$M/H$ of the physical vector field. As with the case of the transverse 
components, scale invariance is retained even in the regime when 
\mbox{$r\gg 1$} (see Sec.~\ref{fmpref}).

Now, using Eqs.~(\ref{alpha}) and (\ref{hatnu}), one finds that scale 
invariance (i.e. $\hat\nu=5/2$) requires
\begin{equation}
\beta=-\frac12(3\pm 5)\,.
\label{beta}
\end{equation}
We can discard one of the above values as follows.\footnote{The `$\pm$' sign 
in Eq.~(\ref{alpha}) is uncorrelated with the one in Eq.~(\ref{beta}), which 
means that discarding one of the values of $\beta$ does not imply doing so 
also for one of the values of $\alpha$.} 
According to Eqs.~(\ref{fa}), (\ref{r}) and (\ref{ma}) 
\begin{equation}
r\propto a^{1+\beta-\alpha/2}.
\label{rab}
\end{equation}
This means that the value \mbox{$\beta=-4$} is unacceptable, because it 
implies that $r$ is a decreasing function of $a$ (for \mbox{$\alpha=-1\pm 3$}),
which means that our assumption \mbox{$r\ll 1$} cannot hold true in the
subhorizon limit. Hence, we conclude that a scale invariant spectrum for the
longitudinal component of the vector field perturbations can be attained only 
if \mbox{$\beta=1$}, i.e.
\begin{equation}
m\propto a\,.
\label{m}
\end{equation}

In view of the above and Eqs.~(\ref{PLR}) and (\ref{Plong}), 
we see that, if \mbox{$\alpha=2$}, then (when \mbox{$r\ll 1$}) we have
\begin{equation}
{\cal P}_\parallel={\rm constant}\gg{\cal P}_{L,R}
\quad({\rm for}\;\alpha=2)\;,
\end{equation}
where we considered Eq.~(\ref{mbound}). On the other hand, if 
\mbox{$\alpha=-4$},
then \mbox{${\cal P}_\parallel\propto a^{-6}$}. Thus, even though the spectrum 
is scale invariant, its amplitude decreases in time. Scale invariance is not
spoiled by the piling up of more and more perturbations, while they exit the 
horizon, because their amplitude is reduced accordingly in time, as shown in 
Fig.~\ref{flatP}. The gradual decrease of ${\cal P}_\parallel$ implies that, 
even though
\mbox{$({\cal P}_\parallel/{\cal P}_{L,R})_*\gg 1$}, this ratio decreases 
towards the end of inflation and may allow for the longitudinal and transverse 
spectra to be eventually comparable. 

%This is why we put more emphasis in the 
%case when \mbox{$\alpha=-4$} in the following.

%\vspace{-2cm}

\begin{figure}[t]
\includegraphics[width=100mm,angle=0]{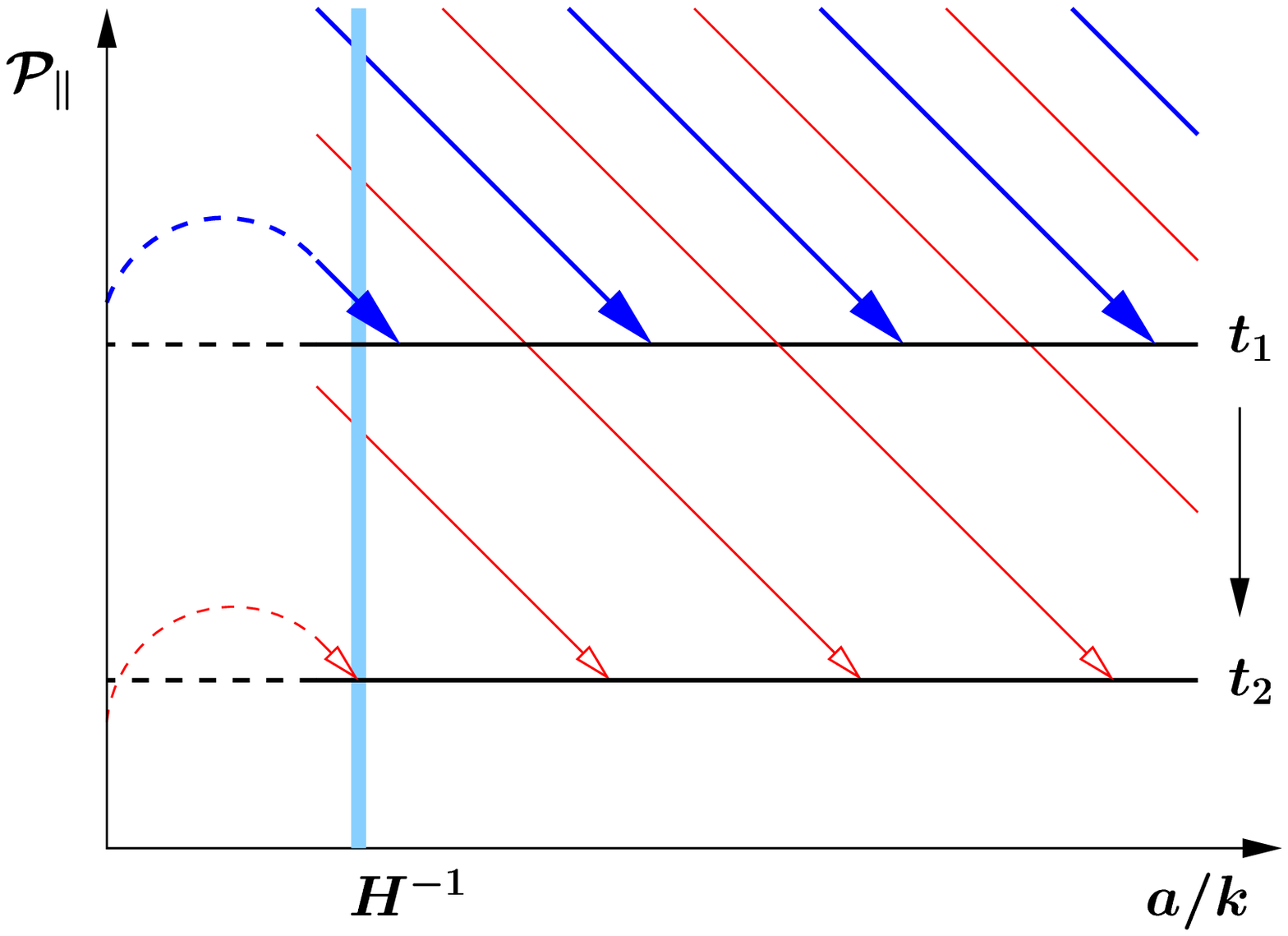}
\vspace{-4cm}
\caption{
Log-log plot of the superhorizon power spectrum 
\mbox{${\cal P}_\|\propto a^{-6}$} (case \mbox{$\alpha=-4$}) in terms of
the physical lengthscale \mbox{$\ell\sim a/k$} 
\underline{\em at a given fixed time $t$}. The spectrum is flat and it is
shown by the solid horizontal lines, which depict its value at two different 
times: $t_1$ and \mbox{$t_2>t_1$} for superhorizon scales \mbox{$\ell>H^{-1}$}.
The slanted arrows show the evolution of superhorizon modes of given, fixed
$k$. The figure attempts to show that, as time passes and more perturbation
modes exit the horizon, their amplitude at horizon crossing reduces in such 
a way that they end up on top of the flat spectrum at the time of exit.
%, while the latter gradually reduces in amplitude.
}\label{flatP}
\end{figure}

\section{\boldmath Case: $f\propto a^{-4}$ \& $m\propto a$}\label{fmpref}

In this section we assume the values 
\begin{equation}
\alpha=-4\quad\&\quad\beta=1\,,
\label{abpref}
\end{equation}
i.e. \mbox{$f\propto a^{-4}$} and \mbox{$M\propto a^3$} [cf. Eq.~(\ref{M})].
Using these we perform an analytic calculation of the power spectra for
all the components of the superhorizon vector field perturbations generated by 
the particle production process.

\subsection{The transverse components}

For the choice in Eq,~(\ref{abpref}), the equation of motion for the transverse
mode functions in Eq.~(\ref{wLR}) becomes
\begin{equation}
\ddot w_{L,R}+3H\dot w_{L,R}+
\left(\frac{k}{a}\right)^2(1+r^2)w_{L,R}=0\,.
\label{wLR+}
\end{equation}
The generic solution to the above can be obtained when \mbox{$r\not\approx 1$}.
Indeed, one finds
%
%\begin{widetext}
\begin{eqnarray}
w_{L,R}=a^{-3/2}\left[
c_1J_{3/2}\left(\frac{k}{aH}\right)+
c_2J_{-3/2}\left(\frac{k}{aH}\right)
\right] & {\rm for} & r\ll 1\qquad{\rm and}
\label{w+smallr}\\
w_{L,R}=a^{-3/2}\left[
\hat c_1J_{1/2}\left(\frac{M}{3H}\right)+%\left(\frac{m}{3H\sqrt f}\right)+
\hat c_2J_{-1/2}\left(\frac{M}{3H}\right)
%\left(\frac{m}{3H\sqrt f}\right)
\right] & {\rm for} & r\gg 1\,,
\label{w+bigr}
\end{eqnarray}
%\end{widetext}
where $c_i$ and $\hat c_i$ are integration constants.
Because of Eq.~(\ref{mbound}), we find that the switch-over between the two 
regimes occurs after horizon exit since
\begin{equation}
%\left.\frac{m}{H\sqrt f}\right|_*
\frac{M_*}{H}
\ll 1\equiv\frac{k}{a_*H}
\quad\Rightarrow\quad r_*\ll 1\,,
\label{r*}
\end{equation}
and, in view of Eqs.~(\ref{rab}) and (\ref{abpref}),
\begin{equation}
r\propto a^4.
\label{ra}
\end{equation}
Also, note that at the switch-over moment, denoted here with subscript $k$
(because it is $k$-dependent), we have
\begin{equation}
%\left.\frac{m}{H\sqrt f}\right|_k
\frac{M_k}{H}
\equiv\frac{k}{a_kH}\ll 1\,,
\label{ak}
\end{equation}
exactly because the switch-over moment for a given mode occurs when the 
mode in question is already of superhorizon size.

Using the above we can approximate the solutions in 
Eqs.~(\ref{w+smallr}) and (\ref{w+bigr}) in the vicinity of the switch-over as
%
%\begin{widetext}
\begin{eqnarray}
w_{L,R}=a^{-3/2}\sqrt{\frac{2}{\pi}}\left[
\frac13 c_1\left(\frac{k}{aH}\right)^{3/2}-
c_2\left(\frac{aH}{k}\right)^{3/2}
\right] & {\rm for} & r\lsim 1\qquad{\rm and}
\label{w+sr}\\
w_{L,R}=a^{-3/2}\sqrt{\frac{2}{\pi}}\left[
\hat c_1\left(\frac{M}{3H}\right)^{1/2}+
%\left(\frac{m}{3H\sqrt f}\right)^{1/2}+
\hat c_2\left(\frac{3H}{M}\right)^{1/2}
%\left(\frac{3H\sqrt f}{m}\right)^{1/2}
\right] & {\rm for} & r\gsim 1\,,
\label{w+br}
\end{eqnarray}
%\end{widetext}
where we used that \mbox{$k/a_kH\ll 1$}, %\mbox{$(m/H\sqrt f)_k\ll 1$} 
\mbox{$M_k/H\ll 1$} 
and \mbox{$\lim_{_{_{\hspace{-.5cm}x\rightarrow 0}}}
J_\nu(x)=\frac{x^\nu}{2^\nu\Gamma(1+\nu)}$}. 

To approximate the complete 
solution we can perform a matching at the switch-over of both the mode function
$w_{L,R}$ and its time-derivative $\dot w_{L,R}$. In effect, this will enable 
us to connect the $\hat c_i$ integration constants with the $c_i$. To perform 
the matching it is useful to consider that
\begin{equation}
%\frac{m}{3H\sqrt f}\propto a^3
\frac{M}{3H}\propto a^3
\quad\Rightarrow\quad\frac{a_k}{a}=r^{-1/4}.
\label{akcond}
\end{equation}
After matching we obtain
%
%\begin{eqnarray}
%\hat c_1 & = & -\frac{\sqrt 3}{4}\left(\frac{2a_kH}{k}\right)^2c_2
%\label{C1}\\
%\hat c_2 & = & \frac{4\sqrt 3}{9}\left(\frac{k}{2a_kH}\right)^2c_1
%\label{C2}
%\end{eqnarray}
%
%\begin{widetext}
\begin{equation}
\hat c_1=-\frac{\sqrt 3}{4}\left(\frac{2a_kH}{k}\right)^2c_2
\quad{\rm and}\quad
\hat c_2=\frac{4\sqrt 3}{9}\left(\frac{k}{2a_kH}\right)^2c_1\;.
\label{C1}
\end{equation}
%\end{widetext}
Now, to evaluate the $c_i$ we need to match the solution in 
Eq.~(\ref{w+smallr}) with the vacuum condition in Eq.~(\ref{LRvac}). 
Thus, we obtain
\begin{equation}
c_1=\frac12\sqrt{\frac{\pi}{H}}\quad{\rm and}\quad
c_1=-\frac{i}{2}\sqrt{\frac{\pi}{H}}\,.
\label{c1}
\end{equation}
Hence, the analytic expression for the transverse mode functions can be
written as
%
%\begin{widetext}
\begin{eqnarray}
w_{L,R}=a^{-3/2}\sqrt{\frac{\pi}{4H}}\left[
J_{3/2}
\left(\frac{k}{aH}\right)-i
J_{-3/2}
\left(\frac{k}{aH}\right)
\right] & {\rm for} & %r\ll 1
\frac{k}{aH}\gsim 1\,,
\label{wLRsr}
\\
w_{L,R}=\frac{i}{\sqrt{2k}}\left(\frac{H}{k}\right)
\left[1+\frac{i}{3}\left(\frac{k}{aH}\right)^3\right]
\simeq\frac{i}{\sqrt{2k}}\left(\frac{H}{k}\right)
 & {\rm for} & %r\sim 1
\frac{k}{aH}\ll 1\ll \frac{3H}{M}\,,
\label{wLR=r}
\\
w_{L,R}=a^{-3/2}\sqrt{\frac{\pi}{4H}}\left[
i\left(\frac{aH}{k}\right)^{3/2}
%\!\left(\frac{3H\sqrt f}{m}\right)^\frac12%{1/2}
\left(\frac{3H}{M}\right)^{1/2}
J_{1/2}
%\!\!\left(\frac{m}{3H\sqrt f}\right)+
\left(\frac{M}{3H}\right)+
\right.\;\;\;\;
 & & \nonumber\\
\left.+
\frac13\left(\frac{k}{aH}\right)^{3/2}
%\!\left(\frac{m}{3H\sqrt f}\right)^\frac12%{1/2}
\left(\frac{M}{3H}\right)^{1/2}
J_{-1/2}
%\!\!\left(\frac{m}{3H\sqrt f}\right)
\left(\frac{M}{3H}\right)
\right] & {\rm for} & %r\gg 1\,,
\frac{3H}{M}\lsim 1
\,,
\label{wLRbr}
\end{eqnarray}
%\end{widetext}
where we have used that
\begin{equation}
%\!\,
\left(\frac{2a_kH}{k}\right)^4=\frac{16}{3}
\left(\frac{aH}{k}\right)^3
%\left(\frac{3H\sqrt f}{m}\right)
\left(\frac{3H}{M}\right)
={\rm constant},%\hspace{-1cm}
\label{akcond1}
\end{equation}
as obtained by Eqs.~(\ref{r}) and (\ref{akcond}). Eq.~(\ref{wLR=r}) corresponds
to both the Eqs.~(\ref{w+sr}) and (\ref{w+br}) after the matching, as they 
are both of the form \mbox{$w_{L,R}=C_1+C_2a^{-3}$}, with $C_i$ being 
constants, which are identified in the two regimes by the matching. Calculating
the power spectrum using Eq.~(\ref{wLR=r}), one finds the result shown in 
Eq.~(\ref{PLR}). This demonstrates that the validity of this result extends 
beyond the \mbox{$r\ll 1$} regime of the previous section, to the region
with \mbox{$r\gg 1$} as long as \mbox{$M%=m/\sqrt f
\ll H$}.

The above analytic approximation is remarkably accurate as shown by comparison 
with the solution of Eq.~(\ref{wLR+}) obtained numerically 
(see Fig.~\ref{fig1}). The 
reason is that the physical scale $k/a$ evolves over an exponentially large 
range of values, whereas the approximation might be challenged only over a 
couple of orders of magnitude around the switch-over value 
\mbox{$k/a_k=M_k%=m/\sqrt f
$} (i.e. $r=1$).

\subsection{The longitudinal component}

Let us concentrate now on the longitudinal mode function.
For the choice in Eq,~(\ref{abpref}), the equation of motion for the 
longitudinal mode function in Eq.~(\ref{wlong}) becomes
%
%\begin{widetext}
\begin{equation}
\ddot w_\parallel+
\left(3+\frac{8}{1+r^2}\right)H\dot w_\parallel+
\left[\frac{24}{1+r^2}H^2+\left(\frac{k}{a}\right)^2(1+r^2)\right]
w_\parallel = 0\,.
\label{wlong+}
\end{equation}
The generic solution to the above can be obtained when \mbox{$r\not\approx 1$}
and \mbox{$r\not\approx r_c$} (cf. Eq.~(\ref{rc})).
Indeed, one finds
\begin{eqnarray}
w_\parallel=a^{-11/2}\left[
c_3J_{5/2}\left(\frac{k}{aH}\right)+
c_4J_{-5/2}\left(\frac{k}{aH}\right)
\right] & {\rm for} & r\ll 1\,,
\label{w0smallr}\\
w_\parallel=a^{-3/2}\left[
c_3'J_{3/8}\left(\sqrt\frac32\,\frac{1}{r}\right)+
c_4'J_{-3/8}\left(\sqrt\frac32\,\frac{1}{r}\right)
\right] & {\rm for} & 1\ll r\ll r_c\,,
\label{w0inter}\\
w_\parallel=a^{-3/2}\left[
\hat c_3J_{1/2}\left(\frac{M}{3H}\right)+
%\left(\frac{m}{3H\sqrt f}\right)+
\hat c_4J_{-1/2}\left(\frac{M}{3H}\right)
%\left(\frac{m}{3H\sqrt f}\right)
\right] & {\rm for} & r\gg r_c\,,
\label{w0bigr}
\end{eqnarray}
%\end{widetext}
where $c_i$, $c_i'$ and $\hat c_i$ are integration constants and $r_c$ is 
determined by the condition
\begin{equation}
r_c\equiv 2\sqrt 6\,
%\left.\frac{H\sqrt f}{m}\right|_c\;,
\frac{H}{M_c}\;,
\label{rc}
\end{equation}
which shows that \mbox{$r_c\gg 1$} (considering Eq.~(\ref{mbound})).
The above condition corresponds to the
moment when the first and last terms in the square brackets in 
Eq.~(\ref{wlong+}) become comparable.\footnote{%
To visualise the physical interpretation of $r_c$ consider the range 
\mbox{$[k/a,H]$}, whose lower limit reduces in time. The effective mass of the
physical vector field grows in time as \mbox{$M%=m/\sqrt f
\propto a^3$}.
When it becomes comparable to $k/a$ we have \mbox{$r=1$} and \mbox{$a=a_k$}.
After this moment, $M$ enters into the region in question and continues to grow
approaching $H$. The moment where \mbox{$r=r_c$} corresponds to when $M$ 
crosses the geometric mean of the region, i.e. \mbox{$M=\sqrt{H(k/a)}$}.} 
Note that 
Eq.~(\ref{w0bigr}) looks the same as Eq.~(\ref{w+bigr}). This is because
Eqs.~(\ref{wLR+}) and (\ref{wlong+}) become identical for \mbox{$r\gg r_c$}.

In this case we need to do two matchings of $w_\parallel$ and 
$\dot w_\parallel$; at \mbox{$r=1$} and \mbox{$r=r_c$}.
Let us perform the matching at $r_c$ first. In the vicinity of $r_c$ 
Eqs.~(\ref{w0inter}) and (\ref{w0bigr}) can be written respectively as
%
%\begin{widetext}
\begin{eqnarray}
w_\parallel=a^{-3/2}\left[
\frac{c_3'}{\Gamma(11/8)} 
\left(\sqrt\frac38\,\frac{1}{r}\right)^{3/8}+
\frac{c_4'}{\Gamma(5/8)} 
\left(\sqrt\frac83\,r\right)^{3/8}
\right] & {\rm for} & r\lsim r_c\qquad{\rm and}
\label{w0int}\\
w_\parallel=a^{-3/2}\sqrt{\frac{2}{\pi}}\left[
\hat c_3\left(\frac{M}{3H}\right)^{1/2}+
%\left(\frac{m}{3H\sqrt f}\right)^{1/2}+
\hat c_4\left(\frac{3H}{M}\right)^{1/2}
%\left(\frac{3H\sqrt f}{m}\right)^{1/2}
\right] & {\rm for} & r\gsim r_c\,.
\label{w0big}
\end{eqnarray}
%\end{widetext}
Both the above are of the form \mbox{$w_\parallel=C_1+C_2a^{-3}$}, which
means that they can be readily matched. Indeed, after matching one finds
%
%\begin{widetext}
\begin{equation}
c_3'=4\sqrt{\frac{3}{\pi}}\left(\frac83\right)^{3/16}\Gamma(11/8)
\left(\frac{2a_kH}{k}\right)^{1/2}\hat c_4
\quad{\rm and}\quad
c_4'=\frac13\sqrt{\frac{3}{\pi}}\left(\frac38\right)^{3/16}\Gamma(5/8)
\left(\frac{k}{2a_kH}\right)^{1/2}\hat c_3\,.
\label{C2'}
\end{equation}
%\end{widetext}

Now we can proceed similarly to the transverse case and write
the solutions in the vicinity of the \mbox{$r=1$} switch-over as
%
%\begin{widetext}
\begin{eqnarray}
w_\parallel=a^{-11/2}\sqrt{\frac{2}{\pi}}\left[
\frac{1}{15} c_3\left(\frac{k}{aH}\right)^{5/2}+
3c_4\left(\frac{aH}{k}\right)^{5/2}
\right] & {\rm for} & r\lsim 1\qquad{\rm and}
\label{w0sr}\\
w_\parallel=a^{-3/2}\sqrt{\frac{2}{\pi}}\left[
\hat c_3\left(\frac{M}{3H}\right)^{1/2}+
%\left(\frac{m}{3H\sqrt f}\right)^{1/2}+
\hat c_4\left(\frac{3H}{M}\right)^{1/2}
%\left(\frac{3H\sqrt f}{m}\right)^{1/2}
\right] & {\rm for} & r>1\,.
\label{w0br}
\end{eqnarray}
%\end{widetext}
Note that Eq.~(\ref{w0br}) incorporates both Eqs.~(\ref{w0int}) and 
(\ref{w0big}), by virtue of the matching in Eq.~(\ref{C2'}).

As in the transverse case, to approximate the complete solution, 
we perform a matching at the \mbox{$r=1$} switch-over of both the mode function
$w_\parallel$ and its time-derivative $\dot w_\parallel$, connecting thereby
the $\hat c_i$ integration constants with the $c_i$. After matching we obtain
%
%\begin{widetext}
\begin{equation}
\!\,
\hat c_3=-\frac{4\sqrt 3}{9}a_k^{-4}
\left(\frac{k}{2a_kH}\right)^2c_3
\quad{\rm and}\quad
\hat c_4=\sqrt 3\, a_k^{-4}
\left(\frac{2a_kH}{k}\right)^2
\left[\frac14 c_4+\frac{64}{135}
\left(\frac{k}{2a_kH}\right)^5c_3\right].
%\simeq
%\frac{\sqrt 3}{4}\, a_k^{-4}
%\left(\frac{2a_kH}{k}\right)^2c_4\;.
%\hspace{-1cm}
\label{C2}
\end{equation}
%\end{widetext}
To evaluate the $c_i$ we need to match the solution in 
Eq.~(\ref{w0smallr}) with the vacuum condition in Eq.~(\ref{longvac}). 
Thus, we obtain
\begin{equation}
c_3=-\frac{i}{2}a_k^4\sqrt{\frac{\pi}{H}}\quad{\rm and}\quad
c_4=-\frac12 a_k^4\sqrt{\frac{\pi}{H}}\,.
\label{c2}
\end{equation}
Hence, the analytic expression for the longitudinal mode function can be
written as
%
%\begin{widetext}
\begin{eqnarray}
w_\parallel=-ia^{-9/2}\sqrt{\frac{\pi}{4H}}
\left(\frac{aH}{k}\right)^2
%\left(\frac{H\sqrt f}{m}\right)
\left(\frac{H}{M}\right)
\left[
J_{5/2}
\!\left(\frac{k}{aH}\right)-i
J_{-5/2}
\!\left(\frac{k}{aH}\right)
\right] & {\rm for} & %r\ll 1
\frac{k}{aH}\gsim 1\,,
\label{w+0sr}
\\
w_\parallel\simeq-\frac{3a_k^4}{\sqrt{2H}}\left(\frac{H}{k}\right)^{5/2}a^{-3}
=-\frac{1}{\sqrt{2H}}\left(\frac{H}{k}\right)^{3/2}
%\left(\frac{3H\sqrt f}{m}\right)
\left(\frac{3H}{M}\right)
 & {\rm for} & %r\sim [1,r_c],
\label{w+0=r}
\frac{k}{aH}\ll 1\ll \frac{3H}{M}\,,
\\
%
%\hspace{-.7cm}
%w_\parallel=\frac{a^{-3/2}}{\sqrt H}
%\left(\frac{2aH}{k}\right)^{\frac{15}{8}}
%\!\left(\frac{2H\sqrt f}{m}\right)^\frac58
%\left\{-\frac32\left(\frac83\right)^{\frac{3}{16}}
%\Gamma\left(\frac{11}{8}\right)
%%\left[1+\frac{256i}{135}\left(\frac{k}{2a_kH}\right)^5\right]
%J_\frac38\!\left(\sqrt\frac32\,\frac{1}{r}\right)\right.+
% & & \nonumber\\
%\left.
%+\frac{2i}{9}\left(\frac38\right)^{\frac{3}{16}}\Gamma\left(\frac58\right)
%\left(\frac{k}{2a_kH}\right)^5
%J_{-\frac38}\!\left(\sqrt\frac32\,\frac{1}{r}\right)
%\right] & {\rm for} & 1\ll r\ll r_c
%\label{w+0in}\\
%
%w_\parallel=a^{-3/2}\sqrt{\frac{\pi}{4H}}\left\{
%\frac{i}{3}\left(\frac{k}{aH}\right)^\frac32%{3/2}
%\left(\frac{m}{3H\sqrt f}\right)^\frac12%{1/2}
%J_\frac12%{1/2}
%\!\left(\frac{m}{3H\sqrt f}\right)
%\right.- \hspace{3.8cm} & & \nonumber\\
%\left.
%-\left(\frac{aH}{k}\right)^\frac32
%\left(\frac{3H\sqrt f}{m}\right)^\frac12
%%\left[1+\frac{256i}{135}\left(\frac{k}{2a_kH}\right)^5\right]
%J_{-\frac12}%1/2}
%\!\left(\frac{m}{3H\sqrt f}\right)
%\right\} & {\rm for} & r\gg r_c
%\label{w+0br}\\
%
\hspace{-.5cm}
w_\parallel=a^{-3/2}\sqrt{\frac{\pi}{4H}}\left[
\frac{i}{3}\left(\frac{k}{aH}\right)^{3/2}
%\left(\frac{m}{3H\sqrt f}\right)^\frac12%{1/2}
\left(\frac{M}{3H}\right)^{1/2}
J_{1/2}
%\!\left(\frac{m}{3H\sqrt f}\right)
\left(\frac{M}{3H}\right)-
\right.\;\;\;\;
 & & \nonumber\\
\left.-
%\right.- \hspace{3.8cm} & & \nonumber\\
%\left.
\left(\frac{aH}{k}\right)^{3/2}
%\left(\frac{3H\sqrt f}{m}\right)^\frac12
\left(\frac{3H}{M}\right)^{1/2}
%\left[1+\frac{256i}{135}\left(\frac{k}{2a_kH}\right)^5\right]
J_{-1/2}
%\!\left(\frac{m}{3H\sqrt f}\right)
\left(\frac{M}{3H}\right)
\right] & {\rm for} & %r\gg r_c
\frac{3H}{M}\lsim 1
\,,
\label{w+0br}
\end{eqnarray}
%\end{widetext}
In the above, Eq.~(\ref{w+0=r}) corresponds to both Eqs.~(\ref{w+0sr}) and 
(\ref{w+0br}). This is because both equations, in the vicinity of the 
\mbox{$r=1$} switch-over, are of the form \mbox{$w_\parallel=Ca^{-3}$}, with 
$C$ a constant (see Eqs.~(\ref{w0sr}) and (\ref{w0br}), the latter also
incorporating Eq.~(\ref{w0inter}) in the limit \mbox{$r>1$}). 
This corresponds to the growing mode of Eq.~(\ref{w+0sr}) 
and the decaying mode of Eq.~(\ref{w+0br}), which is dominant near 
\mbox{$a\sim a_k$}. It can also be shown that this mode remains dominant
until \mbox{$M%=m/\sqrt f
\sim H$}, i.e. beyond the \mbox{$r\sim r_c$} 
switch-over. Calculating the power spectrum using Eq.~(\ref{w+0=r}), 
one finds the result shown in Eq.~(\ref{Plong}). As with the transverse case, 
this demonstrates that the validity of this result extends 
beyond the \mbox{$r\ll 1$} regime of the previous section, to the region
with \mbox{$r\gg r_c$} as long as \mbox{$M%=m/\sqrt f
\ll H$}.
As was the case of the transverse modes, the above analytic approximation is 
remarkably accurate which is demonstrated by comparison  with the solution of 
Eq.~(\ref{wlong+}) obtained numerically (see Fig.~\ref{fig1}). 

\begin{figure}[t]
\includegraphics[width=170mm,angle=0]{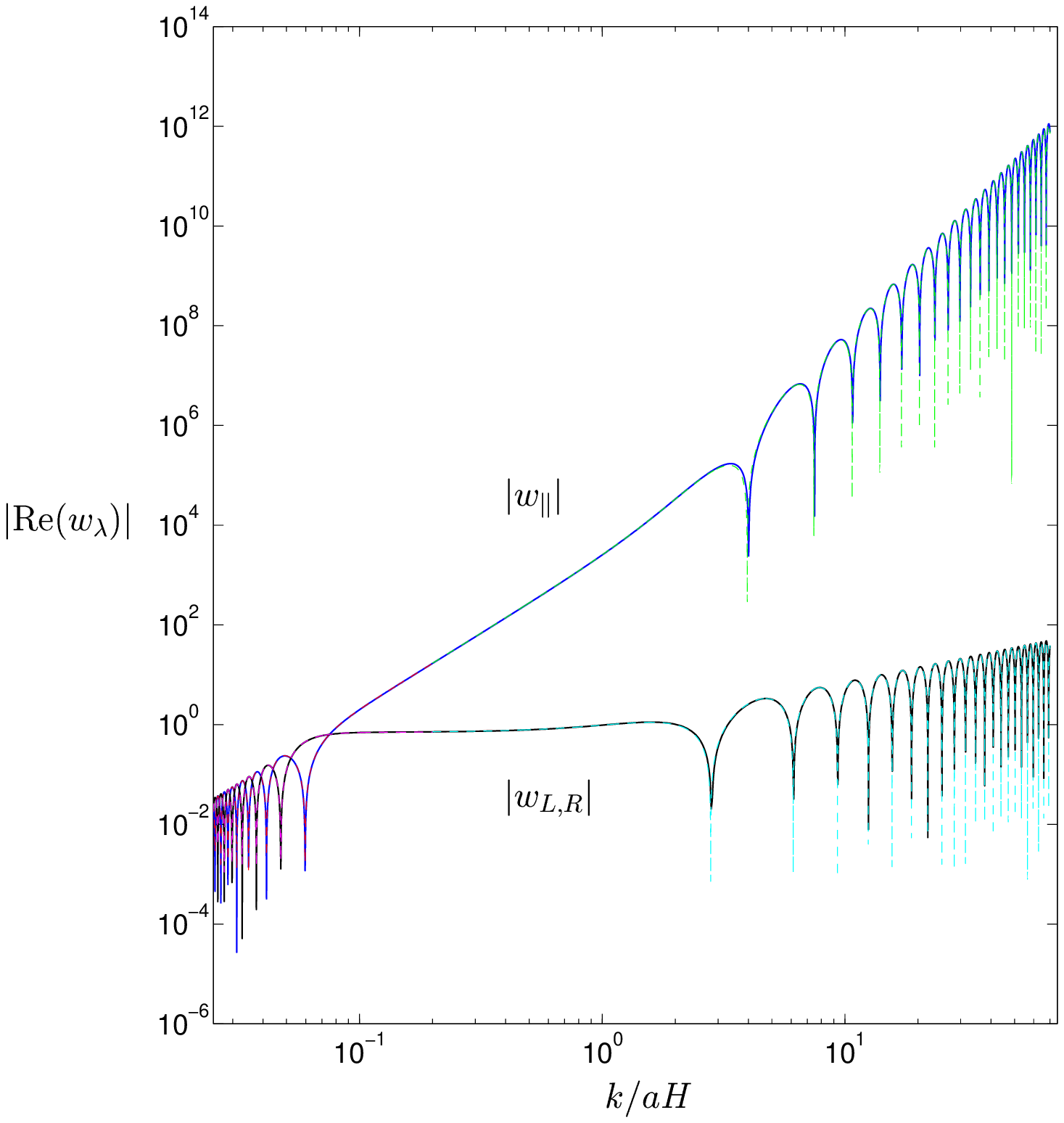}
\vspace{-7cm}
\caption{
Log-log plot of the evolution of the real part of the longitudinal (upper) and 
transverse (lower) mode functions $w_\lambda$ with \mbox{$\lambda=\|,L,R$} of 
the vector field perturbations \underline{\em at a given, fixed $k$}, in terms 
of the physical momentum scale $k/aH$ (weighted by the Hubble scale $H$) in the
case when \mbox{$\alpha=-4$} (\mbox{$f\propto a^{-4}$}). The plots have been 
normalised with respect to the amplitude of the transverse mode functions at 
horizon crossing (\mbox{$k/aH=1$}). The solid lines depict the numerical 
solutions of the equations of motion in Eqs.~(\ref{wLR+}) and (\ref{wlong+}),
while the dashed lines depict that corresponding analytic approximations
in Eqs.~(\ref{wLRsr}) - (\ref{wLRbr}) and (\ref{w+0sr}) - (\ref{w+0br})
respectively. The precision of the approximation is remarkable as the 
difference can hardly be seen. Notice that there are clearly three regimes for
the evolution of the mode functions: First, when subhorizon, they undergo 
oscillations until horizon crossing (\mbox{$k/aH\sim 1$}), when they enter a
power-law regime, which eventually is terminated by another phase of 
oscillations, when the vector field becomes heavy (\mbox{$M\gtrsim 3H$}).
The longitudinal mode function, when superhorizon, scales as 
\mbox{$w_\|\propto a^{-3}$} before oscillations, in agreement with 
Eq.~(\ref{w+0=r}). In contrast, the transverse mode function remains constant
with respect to $a$, as suggested by Eq.~(\ref{wLR=r}). Both mode functions
oscillate with amplitude \mbox{$||w_\lambda||\propto a^{-3}$} when the vector 
field becomes heavy, cf. Eq.~(\ref{dWequal}).
}\label{fig1}
\end{figure}

\subsection{The power spectra}\label{dWosc}

In the previous section we found that the transverse and the longitudinal
mode functions evolve in a similar manner. Indeed, the functions cease 
oscillating after horizon exit and until \mbox{$M%=m/\sqrt f
\sim H$}. During
this period they satisfy an equation of the form (cf. Eqs.~(\ref{w+br}) and
(\ref{w0br}))
\begin{equation}
w_\lambda=a^{-3/2}\sqrt{\frac{2}{\pi}}\left[
\hat c_A\left(\frac{M}{3H}\right)^{1/2}+
%\left(\frac{m}{3H\sqrt f}\right)^{1/2}+
\hat c_B\left(\frac{3H}{M}\right)^{1/2}
%\left(\frac{3H\sqrt f}{m}\right)^{1/2}
\right] \quad{\rm for}\quad M%=\frac{m}{\sqrt f}\,
,\,\frac{k}{a}\ll H\,,
\label{wlinear}
\end{equation}
where the $\hat c_i$ are appropriate constants. The above expression is of the 
form \mbox{$w_\lambda=C_A+C_Ba^{-3}$}, with $C_i$ constants. 

However, there is a crucial difference between the transverse and longitudinal 
modes. In the case of $w_{L,R}$, $\hat c_A$ and $\hat c_B$ in 
Eq.~(\ref{wlinear}) are identified with $\hat c_1$ and $\hat c_2$ respectively,
which are shown in Eq.~(\ref{C1}). Using these values it can be readily 
confirmed that the dominant term in Eq.~(\ref{wlinear}) in the regime
\mbox{$M,k/a\ll H$} 
is the ``growing'' mode: 
\mbox{$w_{L,R}=\,$constant}. In contrast, in the case of $w_\parallel$, 
$\hat c_A$ and $\hat c_B$ in Eq.~(\ref{wlinear}) are identified with 
$\hat c_3$ and $\hat c_4$ respectively, which are shown in Eq.~(\ref{C2}).
Using these values it can be readily confirmed that the dominant term in 
Eq.~(\ref{wlinear}) in the regime \mbox{$M,k/a\ll H$} is the 
decaying mode: \mbox{$w_\parallel\propto a^{-3}$} (by virtue of the condition
\mbox{$M/H\ll 1\ll aH/k$}).

This difference in scaling is reflected, of course, on the power spectra, as
already shown in Eqs.~(\ref{PLR}) and (\ref{Plong}), which, as argued above, 
are valid in the regime where \mbox{$M,k/a\ll H$}. Thus, the typical
value for the vector field perturbations in the regime in question scales as
%
%\begin{eqnarray}
\begin{equation}
\delta W_{L,R} %& \sim & 
=\sqrt{{\cal P}_+}=\frac{H}{2\pi}={\rm constant}%\\
\qquad{\rm and}\qquad
\delta W_\parallel %& \sim & 
=\sqrt{{\cal P}_\parallel}=
%\frac{3H\sqrt f}{m}
\frac{3H}{M}
\frac{H}{2\pi}\propto a^{-3},
\label{dWs}
\end{equation}
%\end{eqnarray}
where \mbox{$\delta W_\lambda\equiv|\delta\mbox{\boldmath $W$}_\lambda|$} and 
\begin{equation}
{\cal P}_+\equiv\frac12({\cal P}_L+{\cal P}_R)={\cal P}_{L,R}\;,
\label{P+}
\end{equation}
because \mbox{${\cal P}_L={\cal P}_R$} as the theory is parity invariant.

Now, let us find out how the perturbations behave when 
\mbox{$M%=m/\sqrt f
\sim H$}. Eqs.~(\ref{wLRbr}) and (\ref{w+0br}) can be written
as
\begin{eqnarray}
2\sqrt{\frac{H}{\pi}}\left(\frac{k}{H}\right)^{3/2}w_{L,R} & = &
\frac{i}{\sqrt z}J_{1/2}(z)+\frac13 x^3\sqrt z J_{-1/2}(z)\qquad{\rm and}
\label{w+M=H}\\
2\sqrt{\frac{H}{\pi}}\left(\frac{k}{H}\right)^{3/2}w_\parallel & = &
\frac13 x^3\sqrt z J_{1/2}(z)-\frac{1}{\sqrt z}J_{-1/2}(z)\,,
\label{w0M=H}
\end{eqnarray}
where we defined
\begin{equation}
z\equiv\frac{M}{3H}
%\frac{m}{3H\sqrt f}
\propto a^3\qquad{\rm and}\qquad
x\equiv\frac{k}{aH}\ll 1\,.
\label{zx}
\end{equation}
When \mbox{$z\gsim 1$}, for the amplitudes of the oscillating
Bessel functions we have \mbox{$||J_\nu(z)||\approx||J_{-\nu}(z)||$}.
Using this and also that
\mbox{$x\ll 1$} for superhorizon perturbations, we obtain
\begin{equation}
\left.\begin{array}{rl}
2\sqrt{\frac{H}{\pi}}\left(\frac{k}{H}\right)^{3/2}w_{L,R} & 
\simeq\frac{i}{\sqrt z}J_{1/2}(z)\\
 & \\
2\sqrt{\frac{H}{\pi}}\left(\frac{k}{H}\right)^{3/2}w_\parallel & 
\simeq-\frac{1}{\sqrt z}J_{-1/2}(z)
\end{array}\right\}\Rightarrow\;
||w_{L,R}||\approx||w_\parallel||\qquad{\rm for}\;z\gsim 1\,.
\label{equal}
\end{equation}
Indeed, when \mbox{$z\gg 1$} the mode functions approach
\begin{equation}
w_{L,R}=\frac{i}{\sqrt{2H}}\left(\frac{H}{k}\right)^{3/2}\frac{\sin z}{z}
\qquad{\rm and}\qquad
w_\parallel=-\frac{1}{\sqrt{2H}}\left(\frac{H}{k}\right)^{3/2}\frac{\cos z}{z}
\qquad{\rm for}\;z\gg 1\,,
\label{wbigz}
\end{equation}
where we used that
\mbox{$\lim_{_{_{\hspace{-.5cm}z\rightarrow +\infty}}}\hspace{-.2cm}
J_\nu(z)=\sqrt{\frac{2}{\pi z}}\cos[z-\frac{\pi}{2}(\nu-\frac12)]$}. 
Since \mbox{$z\gg 1$}, the frequency of oscillations is very large compared
to the expansion rate $H$. This means that it makes sense to use the average
values of the power spectra over many oscillations. In view of 
Eq.~(\ref{wbigz}), we find
\begin{equation}
\overline{{\cal P}_+}=\overline{{\cal P}_\parallel}=
\frac{1}{2z^2}\left(\frac{H}{2\pi}\right)^2.
\label{Pbars}
\end{equation}
Thus, the typical value for the vector field perturbations in this regime is
\begin{equation}
\delta W_{L,R}=\delta W_\parallel=\frac{1}{\sqrt 2}
%\frac{3H\sqrt f}{m}
\frac{3H}{M}
\frac{H}{2\pi}\propto a^{-3},
\label{dWequal}
\end{equation}
where we used \mbox{$\delta W_\lambda=\sqrt{\overline{{\cal P}_\lambda}}$},
where \mbox{$\lambda=+,\parallel$} and we have 
\mbox{$\delta W_L=\delta W_R\equiv\delta W_+$}.
We have found that, for \mbox{$M%=m/\sqrt f
\gsim H$}, the mode functions 
of the transverse and longitudinal components oscillate rapidly with the same
amplitude, so that the components of the typical perturbations of the vector 
field are equal. 

\subsection{Deviations from scale invariance}

So far we have discussed the case of exact scale invariance for both the 
transverse and longitudinal spectra. A small deviation from scale invariance
is favoured at the moment by the observations. Such a deviation can be achieved
by perturbing 
the scalings of $f$ and $m$. Indeed, introducing the perturbations
\begin{equation}
\alpha=-4(1+\epsilon_f)\qquad{\rm and}\qquad\beta=1+\epsilon_m\;,
\label{eps}
\end{equation}
one can repeat the calculations to find the scale invariance of the spectra.
Parameterising this scale invariance in the usual manner, i.e.
\begin{equation}
{\cal P}_\lambda\propto k^{n_\lambda-1},
\label{nl}
\end{equation}
with \mbox{$\lambda=\parallel,+$}, we have found the following expressions for 
the spectral indexes:
\begin{equation}
n_\parallel-1=-(\epsilon_f+2\epsilon_m)\qquad{\rm and}\qquad
n_+-1=-\frac14(18\epsilon_f+\epsilon_m)\,.
\label{ns}
\end{equation}
We see that, for positive $\epsilon_i$'s, both the spectra are red as required 
by the observations. The spectral indexes are identified if 
\mbox{$\epsilon_m\approx 2\epsilon_f$}. In this case 
\mbox{$n_\lambda\simeq 0.96$} is obtained with 
\mbox{$\epsilon_f\simeq 8\times 10^{-3}$}. Deviation from scale invariance can
be also produced by considering that inflation is not exactly de Sitter, i.e.
\mbox{$\varepsilon\equiv-\dot H/H^2\neq 0$}. According to the usual curvaton 
result, this would introduce a red tilt in the spectral indexes of magnitude 
\mbox{$\delta n_\lambda=-2\varepsilon$}. Thus, \mbox{$\varepsilon\approx 0.02$}
may be enough to satisfy the observations, for \mbox{$(\alpha,\beta)=(-4,1)$}.

\section{\boldmath Case: $f\propto a^{2}$ \& $m\propto a$}\label{fmCASE2}

Here we study the other possibility which allows scale invariant spectra
for all the components of the perturbations of our vector field. Hence, 
in this section we assume the values
\begin{equation}
\alpha=2\quad\&\quad\beta=1\,, 
\label{f2scal}
\end{equation}
thus making \mbox{$f\propto a^{2}$} and \mbox{$M =\,$ constant}. The condition 
that the physical vector field is effectively massless at the time of horizon 
exit in Eq. (\ref{mbound}) suggests that \mbox{$M/H\ll 1$} at all times when 
the scaling above holds. Using this condition and scaling we can calculate 
the power spectra for all the components of the superhorizon vector field 
perturbations generated by the particle production process.

\subsection{The transverse components}

For the choice of scaling above, the equation of motion for the transverse
mode functions in Eq.~(\ref{wLR}) becomes
\begin{equation}
\ddot w_{L,R}+3H\dot w_{L,R}+
\left(\frac{k}{a}\right)^2(1+r^2)w_{L,R}=0\,,
\label{wLR+2}
\end{equation}
which has the same form as the equation of motion for the $\alpha=-4$ case. 
However, because $M$ is constant we may obtain an exact solution to the 
equation of motion above
\begin{equation}
w_{L,R}=-a^{-3/2}\sqrt{\frac{\pi}{4H}}
\left[J_{\tilde\nu}\left(\frac{k}{aH}\right)-
iJ_{-\tilde\nu}\left(\frac{k}{aH}\right)\right],
\label{f2tsol}
\end{equation}
where
\begin{equation}
\tilde\nu\equiv \frac12 \sqrt{9 - 4\left(\frac{M}{H}\right)^2}\simeq \frac32\,.
\end{equation}
As before, we have used the Bunch-Davis vacuum boundary condition in 
Eq.~(\ref{LRvac}) to obtain the constants of integration. Taking the limit of 
this solution at late times (superhorizon scales) we can calculate the power 
spectrum
\begin{equation}
    {\cal P}_+ = \left(\frac{H}{2\pi}\right)^2,
\end{equation}
which is in agreement with Eq.~(\ref{PLR})
and we considered $\tilde\nu\simeq\frac32$. However if we consider the next 
level of perturbation \mbox{$\tilde\nu\simeq\frac32-\frac13 
\left(M/H%\frac{M}{H}
\right)^2$}, then we get a slight deviation from scale invariance 
${\cal P}_+\propto k^{n_+-1}$ where
\begin{equation}
    n_+-1 = \frac23 \left(\frac{M}{H}\right)^2\equiv 2\eta_A\;.
\label{etaA}
\end{equation}
As evident, this deviation from scale invariance gives rise to a blue tilt on
the spectrum. %, which is not favoured by obserations.
Alternatively we may perturb the scaling slightly as
\begin{equation}
\alpha=2(1+\epsilon_f)\qquad{\rm and}\qquad\beta=1+\epsilon_m\;,
\label{eps0}
\end{equation}
%\mbox{$f\propto a^{2(i + \epsilon_f)}$} and $m\propto a^{1 + \epsilon_m}$
in analogy with Eq.~(\ref{eps}). 
If $\epsilon_{f,m}>M/H$ then the scale dependence becomes
\begin{equation}
    n_+-1 = -\frac83\,\epsilon_f\;,
\end{equation}
which allows a red spectrum if $\epsilon_f<0$.

\begin{figure}[t]
\includegraphics[width=170mm,angle=0]{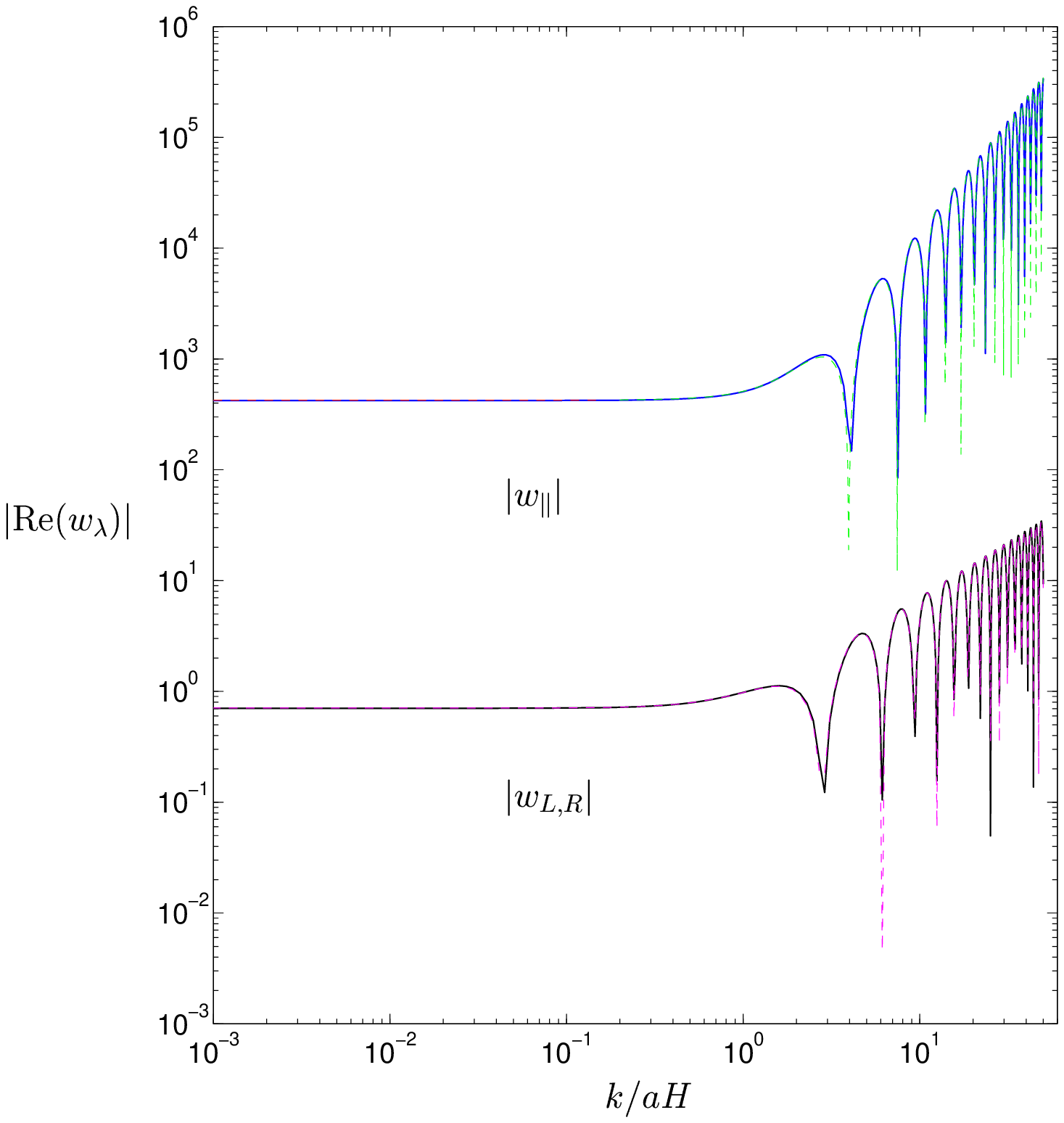}
\vspace{-7cm}
\caption{
Log-log plot of the evolution of the real part of the longitudinal (upper) and 
transverse (lower) mode functions $w_\lambda$ with \mbox{$\lambda=\|,L,R$} of 
the vector field perturbations \underline{\em at a given, fixed $k$}, in terms 
of the physical momentum scale $k/aH$ (weighted by the Hubble scale $H$) in the
case when \mbox{$\alpha=2$} (\mbox{$f\propto a^2$}).
The plots have been normalised with respect to the amplitude of the transverse
mode function at horizon crossing (\mbox{$k/aH=1$}).
The solid lines depict the numerical solutions of the equations of motion in 
Eqs.~(\ref{wLR+2}) and (\ref{wlong+f2}), while the dashed lines depict the 
corresponding analytic solution in Eq.~(\ref{f2tsol}) and the approximation in 
Eqs.~(\ref{w0smallrf2sol}) and (\ref{w0interf2sol}) respectively. 
The precision of the approximation is remarkable as the 
difference can hardly be seen. Notice that there are clearly two regimes for
the evolution of the mode functions: First, when subhorizon, they undergo 
oscillations until horizon crossing (\mbox{$k/aH\sim 1$}), when they enter a
power-law regime. Both mode functions, when superhorizon, 
remain constant with respect to $a$ in agreement, e.g. with 
Eq.~(\ref{w0interf2sol}) for $w_\|$.
Because \mbox{$M\ll H$}, we have \mbox{$w_\|\gg w_{L,R}$}.
}\label{fig2}
\end{figure}

\subsection{The longitudinal component}

Let us concentrate now on the longitudinal mode function.
For the choice of scaling in Eq. (\ref{f2scal}), the equation of motion for the
longitudinal mode function in Eq.~(\ref{wlong}) becomes

\begin{equation}
\ddot w_\parallel+
\left(3+\frac{2}{1+r^2}\right)H\dot w_\parallel+
\left[\left(\frac{k}{a}\right)^2(1+r^2)\right]
w_\parallel = 0\,.
\label{wlong+f2}
\end{equation}
Using the same technique as in the previous case we find exact solutions in 
two regimes

\begin{eqnarray}
w_\parallel = a^{-5/2}\left[
c_5J_{-5/2}\left(\frac{k}{aH}\right)+
c_6J_{5/2}\left(\frac{k}{aH}\right)
\right] & {\rm for} & r\ll 1\qquad{\rm and}
\label{w0smallrf2}\\
w_\parallel = \hat c_5 + \hat c_6 a^{-3} & {\rm for} & r\gg 1
\label{w0interf2}
\end{eqnarray}
where $c_i$, $\hat c_i$ are integration constants. Again we have used 
\mbox{$M/H={\rm constant}\ll 1$}. As in the previous case, using the 
vacuum boundary condition in Eq.~(\ref{longvac}), we can find the values of 
$c_{5,6}$. Then, matching the two solutions and their first derivatives at 
\mbox{$k/aH=M/H$} (i.e. \mbox{$r=1$}) we can find $\hat c_{5,6}$ in the regime 
\mbox{$r\gg 1$}. Thus, the solutions become
\begin{eqnarray}
w_\parallel = -a^{-5/2}\frac{k}{M}\sqrt{\frac{\pi}{4H}}
\left[
J_{-5/2}\left(\frac{k}{aH}\right)+
iJ_{5/2}\left(\frac{k}{aH}\right)
\right] & {\rm for} & %r\ll 1
\frac{k}{aH}\gsim 1\,,
\label{w0smallrf2sol}\\
w_\parallel =\frac{i}{\sqrt{2k}}\frac{H^2}{Mk}
\left\{\left[3i+\frac{2}{45}\left(\frac{M}{H}\right)^5\right]
-\frac19\left(\frac{M}{H}\right)^2\left(\frac{k}{aH}\right)^3\right\}
\simeq-\frac{3}{\sqrt{2k}}\left(\frac{H}{M}\right)\left(\frac{H}{k}\right)
 & {\rm for} & %r\gsim 1
\frac{k}{aH}\ll 1\ll\frac{3H}{M}
\,.
\label{w0interf2sol}
\end{eqnarray}
%Note that Eq.~(\ref{w0interf2sol}) is actually valid after horizon exit, even 
%if \mbox{$r<1$}. 
As in the previous section, the above analytic approximation is remarkably
accurate, as demonstrated in Fig.~\ref{fig2}.

Using Eq.~(\ref{w0interf2sol}), 
we can calculate the power spectrum at arbitrary late times
%, but in fact we find the power spectrum across the two regimes - i.e. at $r\lesssim1$ and $r\gg1$ - to be the same and given by
\begin{equation}
{\cal P}_\parallel = 9\left(\frac{H}{M}\right)^2\left(\frac{H}{2\pi}\right)^2,
\end{equation}
which is in agreement with Eq.~(\ref{Plong}). Since in this case, we have 
\mbox{$M/H={\rm constant}\ll 1$}, we see that the longitudinal power 
spectrum is constant, in contrast to the \mbox{$\alpha=-4$} case. Also, we see
that \mbox{${\cal P}_\parallel\gg {\cal P}_+$}.

Once again if we introduce perturbations to the scaling as in Eq.~(\ref{eps0})
%$f\propto a^{2 + \epsilon_f}$ and $m\propto a^{1 + \epsilon_m}$ 
we can find a slight deviation in the scale dependence
\begin{equation}
    n_\parallel-1 =2(\epsilon_f-\epsilon_m)\,.
\end{equation}

\section{Statistical anisotropy and non-Gaussianity}\label{gfnl}

The theory studied in this paper has two clear advantages. First we can obtain 
a completely isotropic perturbation spectrum for the vector field, which has 
previously never been achieved. As we discuss below, this means that we may 
consider vector fields as dominating the total energy density of the Universe 
when the curvature perturbation is formed. The second advantage is that we can 
also account for a small amount of statistical anisotropy in the curvature 
perturbation spectrum depending on when inflation ends, again by considering 
the vector field alone. We also demonstrate this in what follows. Finally, 
statistical anisotropy can also be present in a correlated manner in the 
bispectrum as well, which characterises the non-Gaussian features of the CMB 
temperature perturbations. In view of the forthcoming observations of the 
recently launched Planck satellite mission this is a particularly promising 
and timely result.

Let us begin with the generic treatment before focusing on cases of particular 
interest. As in Ref.~\cite{GE}, keeping only the leading term in the
anisotropy, the power spectrum may be parameterised in the following fashion

\begin{equation}
{\cal P}_\zeta(\mathbf{k}) =  {\cal P}_\zeta^{\rm iso}(k) 
\left[1+g\left(
%\hat{\mathbf{d}}\cdot \mathbf{\hat{k}}
\mbox{\boldmath $\hat d\cdot\hat k$}
\right)^2\right],
\label{Pzg}
\end{equation}
where ${\cal P}_\zeta^{\rm iso}$ is the isotropic part of the power spectrum,
\mbox{$\mbox{\boldmath $\hat k$}\equiv\mbox{\boldmath $k$}/k$}, 
and \mbox{\boldmath $\hat d$} is a unit vector in some chosen direction. 
In the above the parameter $g$ quantifies the statistical anisotropy in the 
spectrum. In Ref.~\cite{GE}, it has been shown that, currently, the data 
suggest that \mbox{$g\lsim 0.3$} or so.

As shown in Ref.~\cite{stanis}, when the anisotropy in the spectrum is due to
the contribution of a vector field, then at tree level (first order) we have
\begin{equation}
\mbox{\boldmath $\hat d$}=\mbox{\boldmath $\hat N_A$}\,,
\end{equation}
where \mbox{{\boldmath $\hat N_A$}$\,\equiv\,${\boldmath $N_A$}$/N_A$},
\mbox{$N_A\equiv|\mbox{\boldmath $N_A$}|$} and the components of 
\mbox{\boldmath $N_A$} are defined as
\begin{equation}
N_A^i \equiv \frac{\partial N}{\partial W_i}\,,
\end{equation}
with $N$ being the number of remaining e-folds of inflation. The vector 
\mbox{\boldmath $N_A$} quantifies (to first order) the contribution to the
curvature perturbation $\zeta$ from the vector field only; as defined in the 
$\delta N$ formalism \cite{dN}.
In this formalism the magnitude of anisotropy is characterised by \cite{stanis}
\begin{equation}
 g \equiv N^{2}_{A}
\frac{{\cal P}_{\|} - {\cal P}_{+}}{{\cal P}_{\zeta}^{\rm iso}}
\label{g}
\end{equation}
and the isotropic part of the spectrum is
\begin{equation}
 {\cal P}^{\rm iso}_{\zeta}  \equiv   N^{2}_{\phi} {\cal P}_{\phi} + N^2_{A}
 {\cal P}_{+} =  N^{2}_{\phi} {\cal P}_{\phi}
\left(1+%\beta 
\xi\frac{{\cal P}_+}{{\cal P}_\phi}\right),
\label{Piso}
\end{equation}
where we have defined 
\begin{equation}
%\beta 
\xi\equiv \left(\frac{N_{A}}{N_{\phi}}\right)^2
\label{xi}
\end{equation}
and 
\begin{equation}
N_{\phi}\equiv\frac{\partial N}{\partial \phi}\,.
\end{equation}
In the above $\phi$ is a scalar field which also contributes to the curvature 
perturbation. Its contribution is necessary if the anisotropy due to the vector
field is excessive. In this case the vector field can only generate a 
subdominant contribution to $\zeta$, with the dominant contribution due to
some other source, e.g. the scalar field $\phi$. If, however, the anisotropy
due to the vector field is within the observational bounds then one can 
dispense with the scalar field. Thus, if this is the case, we can assume that
the scalar field modulation of $N$ is negligible so that 
\mbox{$N_\phi\rightarrow 0$}. Therefore, dispensing with the scalar field 
contribution is equivalent to \mbox{$\xi%beta
\rightarrow\infty$}.

Now, from Eqs.~(\ref{Piso}) and (\ref{xi})
we can write the anisotropy parameter in Eq.~(\ref{Pzg}) as

\begin{equation}
 g = 
%\beta
\xi\frac{{\cal P}_{\|} - {\cal P}_{+}}{{\cal P}_{\phi}+%\beta 
\xi{\cal P}_{+}}\,.
 \label{e:geq}
\end{equation}
Hence, we can calculate the level of statistical anisotropy in the power 
spectrum.

How do the above translate in the case of the vector curvaton mechanism?
As calculated in Ref.~\cite{stanis}, the tree-level contribution to the 
curvature perturbation spectrum from the vector curvaton field is
\begin{equation}
{\cal P}_{\zeta_A}(\mbox{\boldmath $k$})=\frac49\frac{\hat\Omega_A^2}{W^2}
\left[{\cal P}_++({\cal P}_\parallel-{\cal P}_+)
(\mbox{\boldmath $\hat W\cdot\hat k$})^2\right],
\label{PzA}
\end{equation}
where 
\mbox{$\mbox{\boldmath $\hat W$}\equiv\mbox{\boldmath $W$}/W$} and
\mbox{$W\equiv|\mbox{\boldmath $W$}|$}. In the above $\hat\Omega_A$ is 
defined as
\begin{equation}
\hat\Omega_A\equiv\frac{3\Omega_A}{4-\Omega_A}\sim\Omega_A
\equiv\frac{\rho_A}{\rho}\,,
\label{omega}
\end{equation}
where $\rho_A$ is the density of the vector field. As in the case of the scalar
curvaton paradigm, Eq.~(\ref{PzA}) should be evaluated at the time of decay
of the curvaton field.
%when the scaling of $f$ and $m$ is 
%terminated, regardless whether this moment is during or after inflation
%(see Sec.~??).

As we have shown, in the case when \mbox{$\alpha=-4$} and if
\mbox{$M\gsim H$} by the end of inflation,
\mbox{${\cal P}_+\approx{\cal P}_\parallel$} and statistical anisotropy
in the spectrum can be very small; within the observational bounds.
This means that the vector field can alone give rise to the observed curvature 
perturbation. Thus, if this is the case, we can dispense with the scalar field
$\phi$ and we can write \mbox{${\cal P}_{\zeta}(\mbox{\boldmath $k$})=
{\cal P}_{\zeta_A}(\mbox{\boldmath $k$})$}. Then, from Eqs.~(\ref{Pzg}), 
(\ref{Piso}) and (\ref{PzA}) we find
\begin{equation}
{\cal P}^{\rm iso}_{\zeta}=N_A^2{\cal P}_+\quad{\rm with}\quad
N_A=\frac23\frac{\hat\Omega_A}{W}
\end{equation}
and also \mbox{$g=({\cal P}_\|/{\cal P}_+)-1$}
%\begin{equation}
%g=\frac{{\cal P}_\|-{\cal P}_+}{{\cal P}_+}\,
%\end{equation}
which agrees with Eq.~(\ref{e:geq}) in the limit 
\mbox{$%\beta
\xi\rightarrow\infty$}.

As shown in Ref.~\cite{stanis}, apart from statistical anisotropy in the 
spectrum of the curvature perturbations a vector field can give rise to 
statistical anisotropy in the bispectrum, which characterises the non-Gaussian 
features of the cosmological perturbations.
Non-Gaussianity is an observable of great importance as it can be a powerful
discriminator between mechanisms for the generation of the curvature 
perturbation. It was shown recently in Ref.~\cite{fnlanis} that the statistical
anisotropy in the spectrum and bispectrum are correlated if they are generated 
by a vector field. Such correlation can be a signature prediction of the vector
curvaton scenario. If it is observed it would be direct evidence for a vector 
field contribution to the curvature perturbation.

Non-Gaussianity expresses a non-vanishing 3-point correlator, or equivalently 
a non-zero bispectrum. The bispectrum is a function of three 
\mbox{{\boldmath $k$}$_{1,2,3}$} vectors which may be chosen arbitrarily. 
However we will focus on the two standard configurations; the 
\textit{equilateral} and \textit{squeezed} (or \textit{local}) configurations 
where the magnitudes satisfy $k_1=k_2=k_3$ and $k_1\simeq k_2\gg k_3$ 
respectively. It was shown in Ref.~\cite{fnlanis} that the expressions for 
$f_{\rm NL}$, the non-linearity parameter characterising non-Gaussianity, for a
scale invariant power spectrum in the equilateral and local 
configurations respectively are given by

\begin{equation}
 \frac{6}{5} f_{\rm NL}^{\rm equil} =  
%\beta
\xi^2 {\cal P}^2_+ \frac{3}{2\hat{\Omega}_A}
 \frac{(1+\frac{1}{2}q^2)+[p+\frac{1}{8}(p^2-2q^2)]%A_\perp^2
\hat W_\perp^2}{
({\cal P}_\phi + %\beta
\xi {\cal P}_+)^2}
 \label{e:eq1}
\end{equation}
and
\begin{equation}
 \frac{6}{5} f_{\rm NL}^{\rm local} =  
%\beta
\xi^2 {\cal P}^2_+ \frac{3}{2\hat{\Omega}_A}
 \frac{1+p%A^2_\perp
\hat W_\perp^2}{({\cal P}_\phi + %\beta
\xi {\cal P}_+)^2}\,,
 \label{e:loc1}
\end{equation}
where 
%\mbox{$A_\perp\equiv|\mbox{\boldmath $A$}_\perp|$}, 
\mbox{$\hat W_\perp\equiv|\mbox{\boldmath $\hat W$}_\perp|$}, with 
%{\boldmath $A$}$_\perp$ 
{\boldmath $\hat W$}$_\perp$ 
being the projection of the unit vector 
{\boldmath $\hat W$}
%(or equivalently \mbox{{\boldmath $\hat A$}$\,\equiv\,${\boldmath $A$}$/A$})
to the plane defined by the three \mbox{{\boldmath $k$}$_{1,2,3}$} vectors. 

We have also defined
\begin{equation}
 p \equiv \frac{{\cal P}_{\|} - {\cal P}_{+}}{{\cal P}_{+}} \qquad
 \text{and} \qquad q \equiv \frac{{\cal P}_{-}}{{\cal P}_{+}}\,,
 \label{e:pandq}
\end{equation}
with 
\begin{equation}
{\cal P}_-\equiv\frac12({\cal P}_L-{\cal P}_R)=0\;,
\label{P-}
\end{equation}
i.e. \mbox{$q=0$} since our theory is parity invariant.
%${\cal P}_{-}$ is defined in the same manner as Eq. (\ref{P+}) but as half the difference between the left and right-handed spectrums, hence, for any parity invariant theory we get $q=0$. 

In the following, to calculate $f_{\rm NL}$, we assume that ${\cal P}_\phi$
is due to a light scalar field, in which case
\begin{equation}
 {\cal P}_{\phi} =  \left(\frac{H}{2\pi}\right)^2.
\label{Pphi}
\end{equation}

The curvature perturbation in Eq.~(\ref{Pzg}) and the vector field contribution
to it in Eq.~(\ref{PzA}) should be calculated at the time of the decay of 
the vector field, as in the scalar curvaton scenario. In contrast, the spectra
of the vector field perturbations ${\cal P}_\lambda$ are calculated
either at the onset of the vector field oscillations or at the end
of the scaling of $f(a)$ and $m(a)$; whichever occurs latest, regardless 
of whether this moment is during or after inflation. %(see Sec.~??).
Therefore, at that time, \mbox{$f=1$} and \mbox{$M=\hat m$}, where
\mbox{$\hat m=\,$constant} is the final (vacuum) value of the vector field 
mass. 

As we have shown, the evolution of the components of the vector field 
perturbations goes through the following stages. When subhorizon, the 
perturbations are oscillating (starting of as quantum fluctuations). After 
horizon crossing, they follow a power law evolution of the form
\mbox{$\delta W_\lambda=C_1+C_2 a^{-3}$}, where $C_i$ are constants. In the
\mbox{$\alpha=-4$} case the effective mass $M$ is increasing with time, which
may allow the the vector field and its perturbations to begin oscillations 
(when \mbox{$M\sim H$}) before the end of the scaling behaviour of $f$ and $m$,
i.e. before the end of inflation. 
%If this is so then the perturbations begin oscillating again even 
%before the end of scaling. 
In contrast, in the \mbox{$\alpha=2$} case, since 
\mbox{$M=\,$constant$\,\ll H$}, the perturbations remain in the power-law 
regime until the end of scaling. This suggests that, in this case, 
\mbox{$\hat m<H$}. Thus, when \mbox{$\alpha=2$}, the oscillations commence only
after inflation, when 
$H(t)$ decreases enough such that \mbox{$\hat m\sim H(t)$}. In view of the 
above we can now calculate the statistical anisotropy in the spectrum and 
bispectrum of our model.

\subsection{The power-law regime}

The curvature perturbation is formed in this period of evolution if
\mbox{$\hat m\lesssim H_*$}, where $H_*$ denotes the Hubble scale during 
inflation. Then, from Eqs. (\ref{PLR}) and (\ref{Plong}) we have 
${\cal P}_{\|}=\left(\frac{3H_*}{\hat m}\right)^2{\cal P}_+$, which means that
${\cal P}_{\|}\gtrsim{\cal P}_{+}$. Therefore, our model gives
\begin{equation}
 p=\left(\frac{3H_*}{\hat m}\right)^2-1 \qquad \text{and} \qquad
 g=\left(\frac{%\beta
\xi}{1+%\beta
\xi}\right)
\left[\left(\frac{3H_*}{\hat m}\right)^2-1\right],
 \label{e:pf}
\end{equation}
where %$\beta$ 
$\xi$ is to be calculated at the decay of the vector field%
%(denoted by the subscript `dec')
, which occurs after the onset of the vector field 
oscillations. The latter, in this case, has to occur after the end of 
inflation. 

In this regime we also observe that ${\cal P}_{+}={\cal P}_{\phi}$. 
Substituting Eq. (\ref{e:pf}) into Eqs. (\ref{e:eq1}) and (\ref{e:loc1}) 
(with \mbox{$M\rightarrow\hat m$}) we can now get expressions for the 
non-linearity parameter $f_{\rm NL}$ in terms of
the anisotropy parameter $g$. We find
\begin{eqnarray}
 \frac{6}{5} f_{\rm NL}^{\rm equil}   &=&
 \left(\frac{%\beta
\xi}{1+%\beta
\xi}\right)^2
 \frac{3}{2\hat{\Omega}_A}
 \left(1+\left\{\frac{1}{8}\left[\left(\frac{3H_*}{\hat m}\right)^2+3
\right]^2-2\right\} %A_\perp^2
\hat W_\perp^2\right)\label{res1}\\
 &=& \frac{g^2}{ \left[ \left(\frac{3H_*}{\hat m}\right)^2-1 \right]^2 }
 \frac{3}{2\hat{\Omega}_A}
 \left(1+\left\{\frac{1}{8}\left[\left(\frac{3H_*}{\hat m}\right)^2+3
\right]^2-2\right\}
 %A_\perp^2
\hat W_\perp^2\right) \nonumber
\end{eqnarray}
and
\begin{eqnarray}
 \frac{6}{5} f_{\rm NL}^{\rm local}   &=& 
\left(\frac{%\beta
\xi}{1+%\beta
\xi}\right)^2 \frac{3}{2\hat{\Omega}_A}
\left\{1+\left[\left(\frac{3H_*}{\hat m}\right)^2-1\right]
%A_\perp^2
\hat W_\perp^2\right\}\label{res1a1}\\
 &=& \frac{g^2}{\left[\left(\frac{3H_*}{\hat m}\right)^2-1\right]^2}
 \frac{3}{2\hat{\Omega}_A}
 \left\{1+\left[\left(\frac{3H_*}{\hat m}\right)^2-1\right]%A_\perp^2
\hat W_\perp^2\right\}.
\nonumber
\end{eqnarray}

In analogy with Eq.~(\ref{Pzg}) we can write
\begin{equation}
f_{\rm NL}=f_{\rm NL}^{\rm iso}\left(1+\calg\cdot
%g_{\rm NL}
%{\cal G}_{\rm NL}
\hat W_\perp^2\right),
\label{gNL}
\end{equation}
where $\calg$
quantifies the statistical anisotropy in the bispectrum.
It is evident that the direction of the statistical anisotropy in the 
bispectrum is correlated with the one in the spectrum (cf. Eq.~(\ref{PzA})), 
since they are both
determined by the direction of the unit vector \mbox{\boldmath $\hat W$}
exactly as in Ref.~\cite{fnlanis}. At the moment, observations do not 
provide any information about the value of $\calg$. 

\subsubsection{Highly anisotropic and subdominant limit}

Suppose that, at the end of scaling, \mbox{${\cal P}_{\|} \gg {\cal P}_+$}.
%which corresponds to the curvature perturbation being formed earlier in the 
%Universe's history. 
This is equivalent to \mbox{$(H_*/\hat m)^2\gg 1$}. In this limit we find 
\mbox{$p=(3H_*/\hat m)^2\gg 1$}. Thus, Eq. (\ref{e:geq}) gives
\begin{equation}
 g  = \left(\frac{%\beta
\xi}{1+%\beta
\xi}\right)
 \left(\frac{3H_*}{\hat m}\right)^2.
 \label{g5}
\end{equation}
Were we to dispense with the scalar field, i.e. if the vector field 
contribution to $\zeta$ were dominant, then \mbox{$%\beta
\xi\gg 1$} and the above
would give \mbox{$g=(3H_*/\hat m)^2\gg 1$}, which clearly violates the
observational constraints. Therefore, the contribution of the vector field 
perturbations to the curvature perturbation must be subdominant to that of the 
scalar field. Thus, we must have \mbox{$N_A \ll N_\phi$}, which means 
\mbox{$%\beta
\xi \ll 1$}. Similarly, the vector field contribution to the total 
energy density of the Universe has to be subdominant as well 
\mbox{$\Omega_A \ll 1$}. Hence, Eq.~(\ref{omega}) suggests 
\mbox{$\hat\Omega_A\rightarrow\frac34\Omega_A$}.
Using this, for the case when \mbox{$(H_*/\hat m)^2\gg 1$} we find
\begin{equation}
 \frac{6}{5} f_{\rm NL}^{\rm equil} 
\simeq \frac{2g^2}{\Omega_{A}} 
\left(\frac{\hat m}{3H_*}\right)^4\left[
1+\frac{1}{8}\left(\frac{3H_*}{\hat m}\right)^4
\hat W_\perp^2\right] \label{hadleq1}
\end{equation}
and
\begin{equation}
 \frac{6}{5} f_{\rm NL}^{\rm local}  
\simeq\frac{2g^2}{\Omega_{A}}\left(\frac{\hat m}{3H_*}\right)^4\left[1+ 
\left(\frac{3H_*}{\hat m}\right)^2
\hat W_\perp^2\right].\label{hadlloc1}
\end{equation}

From the above we see that, in this case, \mbox{$f_{\rm NL}\propto g$}, i.e. 
statistical anisotropy in the spectrum of the curvature perturbation would
intensify the non-Gaussianity as is the case in Ref.~\cite{fnlanis}. 
Furthermore, by comparison with Eq.~(\ref{gNL}), we see that, for both 
equilateral and local configurations, for the isotropic part we have
\begin{equation}
\frac{6}{5} f_{\rm NL}^{\rm iso}  
\simeq\frac{2g^2}{\Omega_{A}}\left(\frac{\hat m}{3H_*}\right)^4
=\frac{2\xi^2}{\Omega_{A}}\,,
\label{fnliso}
\end{equation}
which can, in principle, be substantial even if \mbox{$\xi\ll 1$} because 
$\Omega_{A}$ can be very small. Notice however, that,
in both configurations, \mbox{$\calg\gg 1$}, by virtue of 
the condition \mbox{$\hat m<H_*$}. This means that, in this case, the 
statistical anisotropy in the bispectrum is dominant and would be readily 
detected if there is a confirmed detection of a non-zero $f_{\rm NL}$.
Turning this around, if non-Gaussianity is not observed to be predominantly
anisotropic this would rule out the \mbox{$\hat m<H_*$} regime of this model.

From Eqs.~(\ref{hadleq1}) and (\ref{hadlloc1}) we readily obtain that the 
amplitudes of the modulated $f_{\rm NL}$ in both configurations are
\begin{equation}
\frac{6}{5} ||f_{\rm NL}^{\rm local}||=2\frac{g^2}{\Omega_{A}}
\left(\frac{\hat m}{3H_*}\right)^2\quad{\rm and}\quad
\frac{6}{5} ||f_{\rm NL}^{\rm equil}||=\frac14\frac{g^2}{\Omega_{A}}
\label{fNLamps}
\end{equation}
Hence, in general 
\mbox{$||f_{\rm NL}^{\rm local}||<||f_{\rm NL}^{\rm equil}||$}. This means that
the observed upper bound \mbox{$f_{\rm NL}\lsim{\cal O}(10^2)$} \cite{wmap}
should be applied to the equilateral amplitude. Hence we obtain that 
\mbox{$g^2/\Omega_{A}\lsim 10^2$}, which means that 
\mbox{$f_{\rm NL}^{\rm iso}\lsim{\cal O}(1)$}, since \mbox{$\hat m<H_*$} in 
this regime.

Before moving on, it is important to stress that the above results are valid 
for \mbox{$\alpha=-1\pm 3$}, i.e. for both \mbox{$f\propto a^{-4}$} and 
\mbox{$f\propto a^2$}, as the condition \mbox{$\hat m^2\ll H_*^2$} is possible 
in both cases. In fact, this is the only possibility for the case when 
\mbox{$\alpha =2$}.

\subsubsection{Almost isotropic and dominant limit}

In the case when \mbox{$\alpha=-4$}, $M$ is growing with time and the bound in
Eq.~(\ref{mbound}) can be satisfied even with $\hat m$ not smaller than $H_*$.
Therefore, in this case we can investigate the possibility that 
\mbox{$\hat m\sim H_*$}, which corresponds to the edge of the power-law regime 
(which is inaccessible in the \mbox{$\alpha=2$} case). 

Considering this regime we find 
\mbox{${\cal P}_{\|} \sim {\cal P}_+={\cal P}_\phi$}. 
Thus, in this case we find an almost isotropic curvature perturbation, which 
allows us to dispense with the scalar field and assume that the curvature 
perturbation is dominated by the vector field contribution. Hence, we can take
\mbox{$N_A\gg N_\phi$}, i.e. \mbox{$%\beta
\xi \gg 1$}. Then, from Eqs.~(\ref{e:geq}),
(\ref{e:pandq}) and (\ref{e:pf}) we find 
\begin{equation}
\left(\frac{3H_*}{\hat m}\right)^2-1=g=p=\frac{\delta{\cal P}}{{\cal P}_+}\,,
\end{equation}
where \mbox{$\delta{\cal P}\equiv {\cal P}_\|-{\cal P}_+$}. Thus, if the 
fractional difference of the spectra is not excessive, the vector field might 
generate statistical anisotropy in the CMB within the observational bounds.
Using the above, it is straightforward to show that 
\begin{equation}
\frac{6}{5} f_{\rm NL}\simeq\frac{3}{2\hat\Omega_A}\left(1+g%A_\perp^2
\hat W_\perp^2\right),
\end{equation}
in both the local and equilateral configurations. By comparison with 
Eq.~(\ref{gNL}) we see that, in this case, \mbox{$\calg\approx g$}. 
Hence, statistical anisotropy is of the same magnitude in both the spectrum and
bispectrum of the curvature perturbations. Therefore, observational constraints
on $g$ in Ref.~\cite{GE} suggest that $f_{\rm NL}$ may feature an angular 
modulation at a level as large as 30\% or so. 

In the isotropic and dominant limit when \mbox{$\hat m\rightarrow 3H_*$} we 
find that \mbox{$g=0$} and 
\begin{equation}
f_{\rm NL}=\frac{5}{4\hat{\Omega}_A}\,.
\label{fnlcurv}
\end{equation}
As expected, this is equivalent to the scalar curvaton scenario \cite{curv}.
Thus, in this case, substantial non-Gaussianity can be generated if the vector 
field decays before it dominates the Universe, with 
\mbox{${\Omega}_A\ll 1$}.

\subsection{The late-time oscillations regime}

As we have shown in Sec.~\ref{dWosc}, in the case when \mbox{$\alpha=-4$},
it is possible that $M$ grows much larger than $H$ before the end of scaling.
This is indeed so if \mbox{$\hat m\gg H_*$}. Then, after \mbox{$M\sim H$},
the vector field perturbations undergo rapid oscillations, during which, 
\mbox{$\overline{{\cal P}_{\|}}=\overline{{\cal P}_+}$} as shown in 
Eq. (\ref{Pbars}). This means that $g=0$ and $p=0$ 
(cf. Eqs.~(\ref{e:geq}) and (\ref{e:pandq}) respectively), which results in an
isotropic power spectrum. In this regime, from Eqs.~(\ref{Pbars}) and 
(\ref{Pphi}),  we also find that 
\begin{equation}
\overline{{\cal P}}_{\|,+}=
\frac12\left(\frac{3H}{M}\right)^2{\cal P}_{\phi}\,.
\label{PPphi}
\end{equation}
Therefore Eqs. (\ref{e:eq1}) and (\ref{e:loc1}) reduce to
\begin{equation}
 \frac{6}{5} f_{\rm NL} = %\beta
\xi^2\frac{3}{2\hat{\Omega}_A}
 \left[2\left(\frac{\hat m}{3H_*}\right)^2+
%\beta 
\xi\right]^{-2},
 \label{e:eq3}
\end{equation}
which is valid in both the local and equilateral configurations.

Since the vector field perturbations are isotropic we have no need of the
scalar field contribution. As discussed previously, we can dispense with the 
scalar field in the limit \mbox{$%\beta
\xi\rightarrow\infty$}. In this limit
Eq.~(\ref{e:eq3}) reduces to the scalar curvaton expression in 
Eq.~(\ref{fnlcurv}), as expected. However, if there is indeed a contribution to
the curvature perturbation from a scalar field, this may affect the value of 
$f_{\rm NL}$ even if \mbox{$%\beta
\xi\gg 1$}. The reason is
easily understood from Eq.~(\ref{PPphi}), which suggests that, at the end
of scaling \mbox{${\cal P}_{\|,+}\ll{\cal P}_{\phi}$}, since 
\mbox{$M\rightarrow \hat m\gg H_*$}.\footnote{%
This reflects the fact that the amplitude of the oscillating vector field 
perturbations becomes exponentially suppressed during inflation because it 
decreases as $a^{-3}$ before the end of scaling, cf. Eq.~(\ref{dWequal}).} 
Thus, if \mbox{$(\hat m/3H_*)^2>%\beta
\xi/2\gg 1$}, we find
from Eqs. (\ref{e:eq1}) and (\ref{e:loc1})
\begin{equation}
f_{\rm NL}=%\beta
\xi^2\frac{5}{4\hat{\Omega}_A}
\left(\frac{{\cal P}_+}{{\cal P}_{\phi}}\right)^2
=\frac{5}{4\hat{\Omega}_A} 
\left[\frac{%\beta
\xi}{2}\left(\frac{3H_*}{\hat m}\right)^2\right]^2<
\frac{5}{4\hat{\Omega}_A}\,,
\end{equation}
which is smaller than the scalar curvaton result.

\subsection{One-loop corrections}

In the above we considered only the tree level contribution to the spectrum and
bispectrum of the vector field. In principle, one-loop corrections may also 
contribute significantly. Their contribution in the case of vector fields
has been studied in Ref.~\cite{cyd}, where it is shown that the one-loop 
corrections to spectrum and bispectrum dominate only if
\begin{equation}
N_{AA}^2{\cal P}_A
%(2{\cal P}_+-{\cal P}_\|)
>N_A^2,
\label{1loop}
\end{equation}
where \mbox{${\cal P}_A\equiv(2{\cal P}_++{\cal P}_\|)$} and
\mbox{$N_{AA}\equiv|\mbox{\boldmath $N_{AA}$}|$}, with 
\begin{equation}
N_{AA}^{ij} \equiv \frac{\partial^2 N}{\partial W_i\partial W_j}\,.
\end{equation}
As shown in Ref.~\cite{stanis}, for the vector curvaton we have
\begin{equation}
N_A=\frac23\frac{\hat\Omega_A}{W}\quad{\rm and}\quad
N_{AA}=2\frac{\hat\Omega_A}{W^2}\,.
\end{equation}
Using the above we can recast Eq.~(\ref{1loop}) as
\begin{equation}
\delta W\sim\sqrt{{\cal P}_A}>\frac13 W\,,
\end{equation}
which clearly violates our perturbative approach. Hence, we conclude that in 
the vector curvaton case the one-loop correction has to be subdominant.

\section{Evolution of the zero mode}\label{sub:The-zero-mode}

As is evident from Eq.~(\ref{PzA}), in order to calculate the curvature 
perturbation associated with the vector field one needs to study also the 
evolution of the homogeneous zero mode $W$. Combining Eqs.~(\ref{EoMhom}) and 
(\ref{BW}) and using Eq.~(\ref{fa}), we obtain
\begin{equation}
\mbox{\boldmath$\ddot{W}$}+3H\mbox{\boldmath$\dot{W}$}+\left[
\left(1-\frac12\alpha\right)\dot H-\frac14(\alpha+4)(\alpha-2)H^2+M^2\right]
\mbox{\boldmath$W$}=0\,,
\label{EoMalpha}
\end{equation}
where we also considered Eq.~(\ref{M}). 

\subsection{During inflation}

As we have shown, to obtain a scale 
invariant spectrum for the transverse components of the vector field 
perturbations we require $f(a)$ to scale according to Eq.~(\ref{alpha}),
i.e. \mbox{$\alpha=-1\pm 3$}. Using this and considering de Sitter inflation 
(with \mbox{$\dot H\approx 0$}) the above becomes
\footnote{The equation of motion for the zero mode of the canonically 
normalised field is: \mbox{$\mbox{\boldmath$\ddot{B}$}+
H\mbox{\boldmath$\dot{B}$}+(-2H^{2}+M^{2})\mbox{\boldmath$B$}=0$},
%with \mbox{$M=m/\sqrt f$}.
which agrees with the findings in Ref.~\cite{vecurv}.%
}
\begin{equation}
\mbox{\boldmath$\ddot{W}$}+
3H\mbox{\boldmath$\dot{W}$}+M^{2}\mbox{\boldmath$W$}=0\,.
\label{EoM}
\end{equation}
% where we used Eq.~(\ref{BW}). 
We show below that, when \mbox{$M\ll H$} (true at early times when 
\mbox{$\alpha=-4$}; always true when \mbox{$\alpha=2$}),
the solution of the above is well approximated by
\begin{equation}
W\simeq \hat{C}_{1}+\hat{C}_{2}a^{-3},
\label{Wsolu0}
\end{equation}
where %\mbox{$W\equiv|\mbox{\boldmath$W$}|$} and 
$\hat{C}_{i}$ are constants. The dominant term %(growing or decaying mode)
to the solution of Eq.~(\ref{Wsolu0}) is determined by
the initial conditions. A natural choice of initial conditions for the vector 
field zero-mode can be based on energy equipartition grounds. 
As is demonstrated in what follows, if energy equipartition is assumed at the 
onset of inflation, the dominant term turns out to be 
the decaying mode, \mbox{$W\propto a^{-3}$} when \mbox{$\alpha=-4$} and the 
``growing'' mode \mbox{$W=\,$constant}, when \mbox{$\alpha=2$}.

To apply energy equipartition in the initial conditions
we need to consider the energy-momentum tensor for this
theory, which, from Eq.~(\ref{L}), is given by \cite{sugravec}
\begin{eqnarray}
T_{\mu\nu} & = & f\left(\frac{1}{4}g_{\mu\nu}F_{\rho\sigma}F^{\rho\sigma}-
F_{\mu\rho}F_{\nu}^{\;\rho}\right)+m^{2}\left(A_{\mu}A_{\nu}-
\frac{1}{2}g_{\mu\nu}A_{\rho}A^{\rho}\right).
\label{Tmn}
\end{eqnarray}
If we assume that the homogenised vector field lies along the 
$x^{3}$-direction, we can write the above as \cite{sugravec}
\begin{equation}
T_{\mu}^{\,\nu}={\rm diag}(\rho_{A},-p_{\perp},-p_{\perp},+p_{\perp})\,,
\label{Tdiag}
\end{equation}
 where 
 \begin{equation}
\rho_{A}\equiv\rho_{{\rm kin}}+V_{A}\;,
\qquad p_{\perp}\equiv\rho_{{\rm kin}}-V_{A}\;,
\label{rp}
\end{equation}
 with 
 \begin{eqnarray}
\rho_{{\rm kin}} & \equiv & -\frac{1}{4}fF_{\mu\nu}F^{\mu\nu}\;=
\;\frac{1}{2}a^{-2}f\dot{A}^{2}
=\frac{1}{2}\left[\dot{W}+\left(1-\frac12\alpha\right)HW\right]^{2},
\label{rkin}\\
 & & \nonumber \\
V_{A} & \equiv & -\frac{1}{2}m^{2}A_{\mu}A^{\mu}\;=
\;\frac{1}{2}a^{-2}m^{2}A^{2}=\frac{1}{2}M^{2}W^{2},
\label{VA}
\end{eqnarray}
where \mbox{$A\equiv|\mbox{\boldmath$A$}|$}, we used Eqs.~(\ref{BW}) and
(\ref{fa}), and we assumed a negative signature for the metric.

Energy equipartition, therefore, corresponds to
\begin{equation}
(\rho_{\mathrm{kin}})_0\simeq (V_{A})_0\,,
\label{equip}
\end{equation}
 where the subscript `0' indicates the values at some initial time,
e.g. near the onset of inflation.

\subsubsection{Case: $f\propto a^{-4}$} \label{sub:rhoA-f-prop-a-4}

In this case \mbox{$M\propto a^3$} and the solution to Eq.~(\ref{EoM}) is
 \begin{equation}
W=a^{-3}\left[\hat{C}_{3}\sin\left(\frac{M}{3H}\right)+
\hat{C}_{2}\cos\left(\frac{M}{3H}\right)\right].
\label{Wsolu}
\end{equation}
%where %\mbox{$W\equiv|\mbox{\boldmath$W$}|$} and 
%$\hat{C}_{i}$ are constants. 
When \mbox{$M\gsim H$} the above shows that the
amplitude of the oscillating zero mode is decreasing as 
\mbox{$||W||\propto a^{-3}$}. In the opposite regime, when \mbox{$M\ll H$} 
the solution above is well approximated by Eq.~(\ref{Wsolu0}) with 
\mbox{$\hat C_1=\hat C_3a_0^{-3}M_0/3H$}, where we considered that 
\mbox{$a^{-3}M=a_0^{-3}M_0=\,$constant}. Using this, 
the constants $\hat{C}_{2}$ and $\hat{C}_{3}$ in Eq.~(\ref{Wsolu})
can be expressed in terms of initial values of the field amplitude
$W_{0}$ and its velocity $\dot{W}_{0}$ (in field space): 
\begin{equation}
\hat{C}_{2}=-\frac{\dot{W}_{0}}{3H}\;a_{0}^{3}\quad{\rm and}
\quad\hat{C}_{3}=\frac{\left(\dot{W}_{0}+3HW_{0}\right)}{M_{0}}\;a_{0}^{3}\;.
\label{eq:Wconsts}
\end{equation}

Assuming initial equipartition of energy we can relate $W_{0}$ with
$\dot{W}_{0}$. From Eqs.~(\ref{rkin}) and (\ref{VA}), setting 
\mbox{$\alpha=-4$}, we readily obtain
\begin{equation}
\rho_{{\rm kin}}=\frac{1}{2}(\dot{W}+3HW)^{2}\quad{\rm and}\quad
V_{A}=\frac{1}{2}M^{2}W^{2}.
\label{kinV-4}
\end{equation}
Then, using Eq.~(\ref{equip}), we get
\begin{equation}
\dot{W}_{0}\simeq W_{0}
\left(-3H\pm M_{0}\right).
\end{equation}
Substituting this relation into Eq.~(\ref{eq:Wconsts}) we find
that the evolution of the vector field $W$ in Eq.~(\ref{Wsolu})
takes the simple form: 
\begin{equation}
W=W_{0}\left(\frac{a}{a_{0}}\right)^{-3}\sqrt{2}
\cos\left(\frac{M}{3H}\pm\frac{\pi}{4}\right).
\label{eq:W-EoM}
\end{equation}
 Note that this equation is valid for any value of $M$. However, we can see 
that when \mbox{$M\ll H$} the zero mode of the vector field is decreasing as 
\mbox{$W\propto a^{-3}$}, but when \mbox{$M\gg H$} it oscillates
rapidly with a decreasing amplitude proportional to $a^{-3}$. On
this basis we can assume that the typical value of the zero mode during
inflation always scales as 
\begin{equation}
W\propto a^{-3}.
\label{Wa}
\end{equation}

With the assumption of initial equipartition of energy for the vector
field at the onset of inflation we can calculate the kinetic and potential
energy densities.\footnote{By ``potential'' we refer to the energy density 
stored in the mass-term \mbox{$V_A=-\frac12m^2A_\mu A^\mu$}.}
Inserting Eq.~(\ref{eq:W-EoM}) and its derivative
into Eqs.~(\ref{rkin}) and (\ref{VA}) we find 
\begin{equation}
\rho_{\mathrm{kin}}=\left[W_{0}M_{0}
\sin\left(\frac{M}{3H}\pm\frac{\pi}{4}\right)\right]^{2}\quad{\rm and}
\quad V_{\mathrm{A}}=\left[W_{0}M_{0}
\cos\left(\frac{M}{3H}\pm\frac{\pi}{4}\right)\right]^{2}.
\end{equation}
Hence, the total energy density is constant 
\begin{equation}
\rho_{A}=M_{0}^{2}W_{0}^{2}.
\label{eq:rhoA-duringInfl}
\end{equation}
Because this relation is independent of the vector field mass $M$
it is valid in both regimes: when \mbox{$M\ll H$} and $W$ follows a power
law evolution, and when \mbox{$M\gg H$} and $W$ oscillates. This is valid
as long as $f(a)$ and $m(a)$ are varying with time.

\subsubsection{Case: $f\propto a^2$}

In this case, \mbox{$M=\,$constant}, which means that the solution of
Eq.~(\ref{EoM}) is 
\begin{equation}
W=a^{-3/2}\left[\hat C_1a^{\sqrt{\frac94-(\frac{M}{H})^2}}+
\hat C_2a^{-\sqrt{\frac94-(\frac{M}{H})^2}}\right].
%\;\Rightarrow\;
%W\simeq\tilde{C}_{1}+\tilde{C}_{2}a^{-3},
\label{Wsolu2}
\end{equation}
Since in this case \mbox{$M\ll H$}, the above solution is always well 
approximated by Eq.~(\ref{Wsolu0}) and there is no oscillating regime. 

Now, Eqs.~(\ref{rkin}) and (\ref{VA}) take the form
\begin{equation}
\rho_{{\rm kin}}=\frac{1}{2}\dot{W}^{2}
\quad{\rm and}\quad V_{A}=\frac{1}{2}M^{2}W^{2}.
\label{kinV2}
\end{equation}
Combining Eqs.~(\ref{Wsolu0}) and (\ref{kinV2}), we find
\begin{equation}
\rho_{\rm kin}=\frac92H^2\hat C_2^2a^{-6}.
\label{rkin2}
\end{equation}
Thus, assuming energy equipartition at the onset of inflation 
(cf. Eq,~(\ref{equip})) gives
\begin{equation}
\left(1+\frac{\hat C_1}{\hat C_2}a_0^3\right)^2=
\left(\frac{3H}{M_0}\right)^2\gg 1\;\Rightarrow\;
\hat C_1\simeq\pm\frac{3H}{M_0}a_0^{-3}\hat C_2\;,
\end{equation}
where we used that \mbox{$M_0=M\ll H$}. Inserting the above into 
Eq.~(\ref{Wsolu0}) we find
\begin{equation}
W=a_0^{-3}\hat C_2\left[\left(\frac{a_0}{a}\right)^3\pm\frac{3H}{M_0}\right]
\simeq{\rm constant}\simeq W_0\;,
\label{Wa0}
\end{equation}
because, after the onset of inflation, \mbox{$(a_0/a)^3\ll 1\ll 3H/M_0$}.

Therefore, we have found that $W$ remains constant. Since 
\mbox{$M=\,$constant}, this means that 
\mbox{$V_A$} also remains constant. On the other hand, Eq.~(\ref{rkin2})
suggests that \mbox{$\rho_{\rm kin}\propto a^{-6}$}. Thus, since we assumed 
energy equipartition at the onset of inflation, we find that, during inflation,
\mbox{$\rho_{\rm kin}\ll V_A$}. Hence, 
%
%While equation of motion in Eq.~(\ref{EoMhom}) takes the same form
%as Eq.~(\ref{EoM}) but with the constant mass, $M=\mathrm{const}$.
%Because $M\ll H_{*}$ we can rewrite this equation as
%\begin{equation}
%\frac{\mathrm{d}\dot{W}}{\mathrm{d}t}+3H_{*}\dot{W}\approx0.
%\end{equation}
%The solution of this equation is 
%\begin{eqnarray}
%W & = & -\frac{\tilde{C}_{1}}{3H_{*}}\mathrm{e}^{-3H_{*}t}+\tilde{C}_{2},\\
%\dot{W} & = & \tilde{C}_{1}\mathrm{e}^{-3H_{*}t}.
%\end{eqnarray}
%Similary as in sec.~(\ref{sub:rhoA-f-prop-a-4}) we can use equipartition
%energy condition for the vector field at the onset of inflation two
%show that the constant term in these equation is the dominant one.
%From which it becomes clear that the physical vector field is constant,
%$W\approx W_{0}$ and $\dot{W}\rightarrow0$. Plugging this result
%into expressions for kinetic and potential energies in Eqs.~(\ref{eq:rkin-f-prop-a})
%and (\ref{eq:VA-f-prop-a}) we find that the total energy density
%is
\begin{equation}
\rho_{A}\approx V_A\simeq M_{0}^{2}W_{0}^{2},
\label{eq:rhoA-f-prop-a}
\end{equation}
where \mbox{$M=\mathrm{constant}=M_{0}$}. 
This result is the same as in the case 
$f\propto a^{-4}$ in Eq.~(\ref{eq:rhoA-duringInfl}). 

We should stress here that, according to the above, for \mbox{$\alpha=-1\pm 3$}
the typical value of the zero mode scales 
as \mbox{$W\propto a^{(\alpha/2)-1}$}, which means that the zero mode of the 
comoving vector field \mbox{$A=aW/\sqrt f$} remains constant.

\subsection{After the end of scaling}

The simplest choice would be to assume that the scaling of $f(a)$ and $m(a)$ 
during inflation is terminated at the end of inflation. This would imply that
$f$ and $m$ are modulated by a degree of freedom which varies during inflation,
e.g. the inflaton field. We explore this possibility in Sec.~\ref{eg}. Here we 
briefly comment on the possibility of allowing the scaling to end before the 
end of inflation.
%but after the onset of the oscillations of the mode functions and 
%the zero mode. 

At the end of scaling \mbox{$f\rightarrow 1$} and 
\mbox{$m\rightarrow \hat m=\,$constant}. Then, the zero-mode equation of 
motion can be obtained by setting \mbox{$\alpha=0$} in Eq.~(\ref{EoMalpha}).
During de Sitter inflation this gives
\begin{equation}
\mbox{\boldmath $\ddot W$}+3H\mbox{\boldmath $\dot W$}+
(2H^2+\hat m^2)\mbox{\boldmath $W$}=0\,,
\label{EoMend}
\end{equation}
which is solved by 
\begin{equation}
W=a^{-3/2}\left[\hat C_1'a^{\sqrt{\frac14-(\frac{\hat m}{H})^2}}+
\hat C_2'a^{-\sqrt{\frac14-(\frac{\hat m}{H})^2}}\right],
\label{Wsoluend}
\end{equation}
where $\hat C_i'$ are constants. The end of scaling occurs before the onset of
the oscillations if \mbox{$\hat m\ll H$}. In this case the above suggests that
\mbox{$W=\tilde C_1'a^{-1}+\tilde C_2' a^{-2}$}, with $\tilde C_i'$ constants. 
This is not of the same form with Eqs.~(\ref{Wsolu0}) or (\ref{Wsolu2}).
%Thus, one cannot readily match these solutions. 
Therefore the transition at the end of scaling is not smooth but it is indeed
felt by the zero mode if this occurs before the onset of the oscillations.

The end of scaling occurs after the onset of the oscillations if 
\mbox{$\hat m\gsim H$} (which is possible only in the case with 
\mbox{$\alpha=-4$}). In this case Eq.~(\ref{Wsoluend}) suggests that the 
amplitude of the oscillations decreases as \mbox{$||W||\propto a^{-3/2}$},
which agrees with the findings of Ref.~\cite{vecurv}. This is not in agreement
with Eq.~(\ref{Wsolu}), which suggests that, during the oscillations, 
\mbox{$||W||\propto a^{-3}$}. Thus, even if the end of scaling takes place 
after the onset of the oscillations, the transition modifies the 
evolution of the zero-mode.

The vector field perturbations mimic the behaviour of the zero mode if the
end of scaling occurs in the \mbox{$r\gg 1$} regime, i.e. when 
\mbox{$\hat m\gg k/a$}. For such scales, after the end of scaling, 
the equations of motion for the transverse and longitudinal components,
Eqs.~(\ref{EoMtrans}) and (\ref{EoMlong}), become identical and of the same 
form as Eq.~(\ref{EoMend}). Thus, the mode functions satisfy the equation
\begin{equation}
\ddot w_\lambda+3H\dot w_\lambda+(2H^2+\hat m^2)w_\lambda=0\,,
\label{eomend}
\end{equation}
which means that the typical value of the vector field perturbations is
\begin{equation}
\delta W_\lambda=\sqrt{{\cal P}_\lambda}\propto w_\lambda
=a^{-3/2}\left[\tilde c_1a^{\sqrt{\frac14-(\frac{\hat m}{H})^2}}+
\tilde c_2a^{-\sqrt{\frac14-(\frac{\hat m}{H})^2}}\right],
\label{dwsoluend}
\end{equation}
where $\tilde c_i$ are constants. Therefore, similarly to the above, the 
solution is not similar to Eqs.~(\ref{wLR=r}) and (\ref{w+0=r}) for the 
transverse and longitudinal components respectively if \mbox{$\hat m\ll H$}. 
This is also true if \mbox{$\hat m\gsim H$}, i.e. when the end of scaling 
occurs after the onset of the oscillations, for which 
\mbox{$||\delta W_\lambda||\propto a^{-3/2}$}; in contrast to
\mbox{$||\delta W_\lambda||\propto a^{-3}$} before the end of scaling
cf. Eq.~(\ref{dWequal}). Note here that the spectrum for
the scales that exit the horizon after the end of scaling is not scale 
invariant. 

From the above we see that, if the end of the scaling %of $f(a)$ and $m(a)$
occurs before the end of inflation the evolution of the zero-mode and the
vector field perturbations is non-trivially affected, which significantly 
complicates the treatment. Since attributing the scaling of $f(a)$ and $m(a)$
to some degree of freedom which varies during inflation but not afterward 
(e.g. the inflaton field, see Sec.~\ref{eg}) is much more physically motivated
than the alternative considered above, in the following, we assume that this 
is indeed the case and the end of scaling occurs at the end of inflation.

\subsection{After inflation}

At the end of inflation we assume that the scaling of $f$ and $m$ has ended 
and we have
\begin{equation}
f=1\quad\mathrm{and}\quad m=\hat{m}\,.
\label{eq:fm-after-infl}
\end{equation}
Hence, Eqs.~(\ref{eq:rhoA-duringInfl}) and (\ref{eq:rhoA-f-prop-a}) no longer 
apply. The evolution of $\rho_A$ is determined as follows.

As mentioned already, after the end of scaling, \mbox{$\alpha=0$} and 
\mbox{$M=\hat m$}. Then, Eqs.~(\ref{rkin}) and (\ref{VA}) become
\begin{equation}
\rho_{{\rm kin}}=\frac{1}{2}(\dot{W}+HW)^{2}
\quad{\rm and}\quad V_{A}=\frac{1}{2}\hat m^{2}W^{2}.
\label{kinVend}
\end{equation}
%
%\begin{eqnarray}
%\rho_{{\rm \textrm{kin}}} & = & \frac{1}{2}\left(\frac{\dot{A}}{a}\right)^{2}
%=\frac{1}{2}(\dot{W}+HW)^{2},\label{eq:rkin-after-infl}\\
%\nonumber \\V_{A} & = & \frac{1}{2}\hat{m}^{2}\left(\frac{A}{a}\right)^{2}
%=\frac{1}{2}\hat{m}^{2}W^{2}.
%\label{eq:VA-after-infl}
%\end{eqnarray}
The behaviour of $\rho_{\mathrm{kin}}$ and $V_{A}$ depends on whether
the vector field is light or not. To see this let us calculate
the evolution of the field after inflation. With the conditions in 
Eq.~(\ref{eq:fm-after-infl}) the physical vector field of Eq.~(\ref{BW}) 
is \mbox{{\boldmath $W$}={\boldmath $A$}$/a$},
%\begin{equation}
%W=B=\frac{A}{a}.
%\end{equation}
while Eq.~(\ref{EoMalpha}) becomes
\begin{equation}
\mbox{\boldmath$\ddot{W}$}+3H\mbox{\boldmath$\dot{W}$}+
\left(\dot{H}+2H^{2}+\hat{m}^{2}\right)\mbox{\boldmath$W$}=0\,,
\label{eq:EoM-W-after-infl}
\end{equation}
where the Hubble parameter after inflation decreases as 
\mbox{$H(t)=\frac{2}{3\left(1+w\right)t}$}, with
\mbox{$w\equiv p/\rho$} being the barotropic parameter of the Universe.
Solving Eq.~(\ref{eq:EoM-W-after-infl}) we find
\begin{eqnarray}
W & = & t^{\frac{1}{2}(\frac{w-1}{w+1})}
\left[\tilde{C}_{1}J_{d}\left(\hat{m}t\right)+
\tilde{C}_{2}J_{-d}\left(\hat{m}t\right)\right]\qquad{\rm and}
\label{eq:W-afterInfl-gen}\\
\dot{W}+HW & = & \hat{m}\,t^{\frac{1}{2}(\frac{w-1}{w+1})}
\left[\tilde{C}_{1}J_{d-1}\left(\hat{m}t\right)-
\tilde{C}_{2}J_{1-d}\left(\hat{m}t\right)\right],
\label{eq:Wdot-afterInfl-gen}
\end{eqnarray}
where \mbox{$d=\frac{1+3w}{6\left(1+w\right)}$}. One can easily see that
the vector field behaves differently if it is light, \mbox{$\hat{m}t\ll1$},
or heavy, \mbox{$\hat{m}t\gg1$}.

Let us first see what happens if the vector field is light. Then,
Eqs.~(\ref{eq:W-afterInfl-gen})
and (\ref{eq:Wdot-afterInfl-gen}) can be approximated as
\begin{eqnarray}
W & = & t^{\frac{1}{2}(\frac{w-1}{w+1}})
\left[\frac{\tilde{C}_{1}}{\Gamma\left(1+d\right)}
\left(\frac{\hat{m}t}{2}\right)^{d}+
\frac{\tilde{C}_{2}}{\Gamma\left(1-d\right)}
\left(\frac{\hat{m}t}{2}\right)^{-d}\,\right]\qquad{\rm and}\\
\dot{W}+HW & = & \hat m\,t^{\frac{1}{2}(\frac{w-1}{w+1})}
\left[d\,\frac{\tilde{C}_{1}}{\Gamma\left(1+d\right)}
\left(\frac{\hat{m}t}{2}\right)^{d-1}-
\frac{1}{1-d}\frac{\tilde{C}_{2}}{\Gamma\left(1-d\right)}
\left(\frac{\hat{m}t}{2}\right)^{1-d}\,\right].
\end{eqnarray}
Although the solution has one decaying and one growing mode, it might
happen that the decaying mode stays larger than the growing mode.
To check this we calculate constants $\tilde{C}_{1}$ and $\tilde{C}_{2}$
by matching the above equations to the values $W_{\mathrm{end}}$ and
$\dot{W}_{\mathrm{end}}$ at the end of inflation (denoted by `end'). 
Thus, we find that
\begin{eqnarray}
W & = & \frac{2}{3w+1}\left(\frac{a}{a_{\mathrm{end}}}\right)^{\frac{1}{2}
\left(3w-1\right)}\left(W_{\mathrm{end}}+
\frac{\dot{W}_{\mathrm{end}}}{H_{*}}\right)\quad{\rm and}\\
\dot{W}+HW & = & H_{*}\left(\frac{a}{a_{\mathrm{end}}}\right)^{-2}
\left(W_{\mathrm{end}}+\frac{\dot{W}_{\mathrm{end}}}{H_{*}}\right),
\end{eqnarray}
where $H_*$ is the inflationary Hubble scale. Plugging these solutions into 
Eq.~(\ref{kinVend}) (and using that \mbox{$a^{3(1+w)}\propto t^2$}) we obtain
\begin{equation}
\frac{V_A}{\rho_{\rm kin}}=\frac{4}{(3w+1)^2}
\left(\frac{\hat m}{H_*}\right)^2
\left(\frac{t}{t_{\rm end}}\right)^2\simeq(\hat mt)^2\ll1\,,
\end{equation}
which implies that
%(\ref{eq:rkin-after-infl}) and (\ref{eq:VA-after-infl}) 
the total energy density of the light vector field 
%(Eq.~(\ref{rp})) 
is
\begin{equation}
\rho_{A}\simeq\rho_{\rm kin}=\frac{1}{2}\left(\dot{W}_{\mathrm{end}}+
W_{\mathrm{end}}H_{*}\right)^{2}\left(\frac{a}{a_{\mathrm{end}}}\right)^{-4}
\;\Rightarrow\;\rho_A\propto a^{-4}.
\label{eq:rhoA-mllH}
\end{equation}
Therefore, we see that the energy density of the light vector field
scales as that of relativistic particles. %as expected.
This is in striking difference to the scalar field case, in which 
when the field is light its density remains constant even after inflation.
%\begin{equation}
%\rho_{A}\propto a^{-4}.
%\label{eq:rhoA-mllH}
%\end{equation}

On the other hand, if the vector field is heavy, \mbox{$\hat{m}t\gg1$},
the Bessel functions in Eqs.~(\ref{eq:W-afterInfl-gen}) and 
(\ref{eq:Wdot-afterInfl-gen})
are oscillating and the latter can be recast as
\begin{eqnarray}
W & = & \sqrt{\frac{2}{\pi}}\,t^{-\frac{1}{1+w}}
\left[\tilde{C}_{1}\cos\left(\hat{m}t-\frac{1+2d}{4}\,\pi\right)+
\tilde{C}_{2}\cos\left(\hat{m}t-\frac{1-2d}{4}\,\pi\right)\right]
\qquad{\rm and}\\
\dot{W}+HW & = & \sqrt{\frac{2}{\pi}}\,\hat{m}\,t^{-\frac{1}{1+w}}
\left[\tilde{C}_{1}\sin\left(\hat{m}t-\frac{1+2d}{4}\,\pi\right)+
\tilde{C}_{2}\sin\left(\hat{m}t-\frac{1-2d}{4}\,\pi\right)\right].
\end{eqnarray}
As can be seen above, when the vector field is heavy, it oscillates with
a very high frequency and with amplitude decreasing as 
\mbox{$t^{-1/\left(1+w\right)}\propto a^{-3/2}$}.
This was already demonstrated in Ref.~\cite{vecurv}. When we calculate
the energy density of the oscillating vector field from 
Eqs.~(\ref{rp}) 
%(\ref{eq:rkin-after-infl}), (\ref{eq:VA-after-infl})
and (\ref{kinVend}) we find
\begin{equation}
\rho_{A}=\frac{1}{\pi}\hat{m}\,t^{-\frac{2}{1+w}}
\left[\tilde{C}_{1}^{2}+\tilde{C}_{2}^{2}+
2\tilde{C}_{1}\tilde{C}_{2}\cos\left(d\pi\right)\right]
\;\Rightarrow\;\rho_A\propto a^{-3}.
\label{eq:rhoA-mggH}
\end{equation}
%In this case exact values of constants $\tilde{C}_{1}$ and $\tilde{C}_{2}$
%are not important since we can already see that the energy density
%of the massive vector field scales as
%\begin{equation}
%\rho_{A}\propto a^{-3},
%\label{eq:rhoA-mggH}
%\end{equation}
Thus, we see that the energy density of the heavy vector field
scales as that of non-relativistic matter. Furthermore, if calculating
the pressure from Eq.~(\ref{rp}), we find that it oscillates with
a high frequency:
\begin{equation}
p_{\bot}=\frac{1}{\pi}\hat{m}\,t^{-\frac{2}{1+w}}
\left[\tilde{C}_{1}^{2}\sin\left(2\hat{m}t-d\pi\right)+
\tilde{C}_{2}^{2}\sin\left(2\hat{m}t+d\pi\right)+
2\tilde{C}_{1}\tilde{C}_{2}\sin\left(2\hat{m}t\right)\right]
\;\Rightarrow\;\overline{p_\perp}\approx 0\,.
\label{p=0}
\end{equation}
Therefore, we have found that, on average, the oscillating vector field behaves
as pressureless {\em isotropic} matter (see Eq.~(\ref{Tdiag})), which is
in agreement with Ref.~\cite{vecurv}. Hence, the massive vector field 
%acts as the pressureless isotropic matter and it 
can dominate the Universe without generating excessive large scale anisotropy.
This is crucial for the vector curvaton mechanism because, %in order 
to produce the curvature perturbation, the field must dominate 
(or nearly dominate) the Universe without inducing anisotropic expansion. 

\section{Curvaton Physics}

In this section we calculate constraints for our vector curvaton model assuming
that the scaling behaviour of $f(a)$ and $m(a)$ ends when inflation is 
terminated. This implies that the scaling is controlled by some degree of
freedom which varies during inflation, e.g. the inflaton field. A specific
example of this kind is discussed in Sec.~\ref{eg}.

\subsection{Basics}

According to the curvaton scenario \cite{curv} the total curvature
perturbation can be calculated as the sum of individual curvature
perturbations from the constituent components of the universe multiplied
by the appropriate weighting factor. In our scenario this is
written as follows %\cite{Lyth_Ungarelli_Wands(2003)}

\begin{equation}
\zeta=(1-\hat{\Omega}_{A})\zeta_{\mathrm{rad}}+\hat{\Omega}_{A}\zeta_{A},
\label{eq:zeta-general}
\end{equation}
 where $\hat{\Omega}_{A}$ is defined in Eq.~(\ref{omega}). As in the scalar 
curvaton paradigm, the above is to be evaluated at the time of decay of the
curvaton field.

As was discussed in Sec.~\ref{gfnl},
if \mbox{$\hat{m}\gg H_{*}$} at the end of inflation, then the vector field
perturbation spectrum is isotropic and may generate the total curvature
perturbation in the Universe without violating observational bounds
on the statistical anisotropy of the curvature perturbations. 
If this is the case, we can assume that \mbox{$\zeta_{\mathrm{rad}}=0$}.
On the other hand, when \mbox{$\hat{m}\ll H_{*}$}, the amplitude of the
spectrum of the longitudinal component of the vector field perturbations
is substantially larger than the one of the transverse perturbations. Hence,
the curvature perturbation due to the vector field is excessively anisotropic.
To avoid conflict with observational bounds (see Ref.~\cite{GE}), the 
contribution of the vector field to the curvature perturbation
%energy density of the vector field 
has to remain subdominant. Therefore, for this scenario, we have to consider
\mbox{$\zeta_{\mathrm{rad}}\neq0$} and the curvature perturbation already
present in the radiation dominated universe must dominate the one
produced by the vector curvaton field.

For definiteness let us assume that inflation is driven by some inflaton
field, which after inflation oscillates around its VEV until reheating,
when it decays into relativistic particles. The vector field must
be subdominant during this time. But after reheating, the Universe
is radiation dominated with the energy density decreasing as 
\mbox{$\rho_{\mathrm{rad}}\propto a^{-4}$}.
If the vector field at this epoch is heavy and therefore undergoes 
rapid oscillations, its relative energy density increases, 
\mbox{$\rho_{A}/\rho_{\mathrm{rad}}\propto a$},
as can be seen from Eq.~(\ref{eq:rhoA-mggH}). When the field becomes
dominant (or nearly dominant) it can imprint its perturbation spectrum
onto the Universe.

The contribution to the curvature perturbation by the vector field
is calculated as follows. On the spatially flat slicing of spacetime 
we can write for each component of the content of the Universe
\begin{equation}
\zeta_n=-H\frac{\delta\rho_n}{\dot{\rho}_n}\,,
\end{equation}
where $n$ represents different components of the cosmic fluid. Using the
continuity equation %for independent fluids, 
\mbox{$\dot{\rho}_n=-3H\left(\rho_n+p_n\right)$},
one can recast the above equation for the vector field as
\begin{equation}
\zeta_A=
\left.\frac{\delta\rho_{A}}{3\rho_{A}}\right|_{\mathrm{dec}}
\approx\frac{2}{3}\left.\frac{||\delta W||}{||W||}\right|_{\mathrm{dec}}
\simeq\frac23\left.\frac{\delta W}{W}\right|_{\mathrm{end}},
\label{eq:zetaW_general}
\end{equation}
where %the hat means the amplitude of the vector field and 
we considered that the decay of the vector field (labelled by `dec') occurs
after inflation and after the onset of its oscillations so that it is 
pressureless, as shown in Eq.~(\ref{p=0}).
In the last relation we took into account that, after inflation, the equations
of motion are the same for the zero mode and for the superhorizon
perturbations of the field.

In Sec.~\ref{dWosc} it was shown that the typical
value of the field perturbation is \mbox{$\delta W\sim(3H_{*}/M)(H_{*}/2\pi)$}.
If \mbox{$M\ll H_{*}$} this is because the longitudinal component is dominant
over the transverse ones (see Eq.~(\ref{dWs})). 
If \mbox{$M\gg H_{*}$}, then the transverse
and longitudinal components are oscillating with the same amplitudes
(see Eq.~(\ref{dWequal})).
%\footnote{Here we use 
%\mbox{$\delta W=\sqrt{\sum_\lambda(\delta W_\lambda)^2}$}, with 
%\mbox{$\lambda=L,R,\|$}.}

For this reason, at the end of inflation, we can write 
\begin{equation}
\delta W_{\mathrm{end}}\sim\frac{3H_{*}}{\hat{m}}\frac{H_{*}}{2\pi}
\simeq\frac{H_{*}^{2}}{\hat{m}}\,,
\label{eq:dWend}
\end{equation}
where we have taken \mbox{$M=\hat{m}$} and \mbox{$f=1$} at the end of 
inflation. $W_{\mathrm{end}}$ can be found from Eq.~(\ref{eq:rhoA-duringInfl})
by using \mbox{$(\rho_A)_{\rm end}\simeq W_{0}M_{0}\simeq
W_{\mathrm{end}}\hat{m}$} (see Eqs.~(\ref{eq:rhoA-duringInfl}) and
(\ref{eq:rhoA-f-prop-a})). Thus,
\begin{equation}
W_{\mathrm{end}}\sim\frac{\sqrt{(\rho_{A})_{\rm end}}}{\hat{m}}\,.
\end{equation}
 Hence, from Eq.~(\ref{eq:zetaW_general}) we calculate the curvature
perturbation of the vector field 
\begin{equation}
\zeta_A\sim\Omega_{\mathrm{end}}^{-1/2}\frac{H_{*}}{m_{P}}\,,
\label{eq:zetaW}
\end{equation}
where \mbox{$\Omega_{\mathrm{end}}\equiv(\rho_{A}/\rho)_{\mathrm{end}}$}
is the density parameter of the vector field at the end of inflation,
$\rho_{\mathrm{end}}$ is the total energy density dominated by the
inflaton field, and we have used the Friedmann equation:
\mbox{$3m_{P}^{2}H_{*}^{2}=\rho_{\mathrm{end}}$}.
%with the reduced Plank mass $m_{\mathrm{P}}=2.4\times10^{18}\:\mathrm{GeV}$.
Since the vector field must be subdominant during inflation we have
\mbox{$\Omega_{\mathrm{end}}\ll 1$}. Since we also require 
\mbox{$\zeta_A\sim(\delta W/W)_{\rm end}<1$} for our perturbative approach to 
be valid, we obtain the following range for $\Omega_{\mathrm{end}}$:
\begin{equation}
\left(\frac{H_*}{m_P}\right)^2\ll\Omega_{\mathrm{end}}\ll 1\,.
\label{Wrange}
\end{equation}

Eqs.~(\ref{eq:zetaW}) and (\ref{Wrange}) are valid in both 
\mbox{$\alpha=-1\pm3$} cases. 
The only difference is that, in the \mbox{$f\propto a^2$} case, statistically
isotropic curvature perturbations cannot be generated. Hence, only the
considerations for statistically anisotropic perturbations in
Sec.~\ref{sub:Statistically-anisotropic-perturbations} are relevant.%
\footnote{%
The lower bound in Eq.~(\ref{Wrange}) guarantees that \mbox{$\delta W/W\ll 1$}
throughout inflation. The reason is the following. As discussed before 
Eq.~(\ref{eq:dWend}), during inflation \mbox{$\delta W/W\sim H_*^2/MW$}. 
Now, for \mbox{$f\propto a^{-1\pm 3}$} and \mbox{$m\propto a$}
we have \mbox{$M=m/\sqrt f\propto a^{-\frac32(-1\pm 1)}$}. Also, from 
Eqs.~(\ref{Wa}) and (\ref{Wa0}) we obtain 
\mbox{$W\propto a^{\frac32(-1\pm 1)}$}.
Thus, we see that, in all cases considered \mbox{$MW=\,$constant}, which means
that \mbox{$\delta W/W=\,$constant} during inflation. Therefore, 
\mbox{$(\delta W/W)_{\rm end}\ll 1$} is sufficient to guarantee the 
validity of our perturbative approach throughout inflation.}

\subsection{The parameter space}

Here we calculate the parameter space for the model following the method
of Ref.~\cite{sugravec}. First, we note that at the end of inflation
the inflaton field starts oscillating and \mbox{$w\neq -1$}.
%and the gravity becomes attractive by definition, 
Therefore the Hubble parameter decreases as \mbox{$H(t)\sim t^{-1}$}.
In general, the inflaton potential is approximately quadratic around its VEV.
Thus, the coherently oscillating inflaton field corresponds to a collection of
massive particles (inflatons) whose energy density decreases as $a^{-3}$. 
When the Hubble parameter falls bellow the inflaton decay rate $\Gamma$, the 
inflaton particles decay into much lighter relativistic particles reheating
the Universe. After reheating, the Universe becomes radiation dominated with 
energy density scaling as \mbox{$\rho_{\mathrm{rad}}\propto a^{-4}$}.

On the other hand, the evolution of the energy density of the vector field, 
depends on its mass $\hat{m}$. As discussed in 
Sec.~\ref{sub:The-zero-mode}, \mbox{if $\hat{m}\ll H_{*}$} the energy density
scales as \mbox{$\rho_{A}\propto a^{-4}$} until the vector field
becomes heavy and starts 
oscillating. If \mbox{$\hat{m}\gg H_{*}$}, however, the vector field has 
already started oscillating during inflation and \mbox{$\rho_A\propto a^{-3}$}.

To avoid causing an excessive anisotropic expansion period the vector
field must be oscillating before it dominates the Universe and decays. 
This requirement implies that 
\begin{equation}
\Gamma,\:\hat{m}>\Gamma_{A},\: H_{\mathrm{dom}}\;,
\end{equation}
 where $\Gamma_{A}$ is the decay rate of the vector field and 
$H_{\mathrm{dom}}$
is the value of the Hubble parameter when the vector field
dominates the Universe if it has not decayed already. 
Working as in Ref.~\cite{sugravec}, we can estimate $H_{\mathrm{dom}}$ as 
\begin{equation}
H_{\mathrm{dom}}\sim\Omega_{\mathrm{end}}\Gamma^{1/2}
\mathrm{min}\left\{ 1;\frac{\hat{m}}{H_{*}}\right\} ^{2/3}
\mathrm{min}\left\{ 1;\frac{\hat{m}}{\Gamma}\right\} ^{-1/6}.
\end{equation}
Similarly, if the vector field decays before it dominates, the density
parameter just before the decay is given by 
\begin{equation}
\Omega_{\rm dec}\sim\Omega_{\mathrm{end}}
\left(\frac{\Gamma}{\Gamma_{A}}\right)^{1/2}
\mathrm{min}\left\{ 1;\frac{\hat{m}}{H_{*}}\right\} ^{2/3}
\mathrm{min}\left\{ 1;\frac{\hat{m}}{\Gamma}\right\} ^{-1/6}.
\label{eq:Omega-dec}
\end{equation}
where \mbox{$\Omega_{\rm dec}\equiv(\Omega_A)_{\rm dec}$}.
Combining the last two equations and using Eq.~(\ref{eq:zetaW})
we can express the inflationary Hubble scale as 
\begin{equation}
\frac{H_{*}}{m_{P}}\sim\Omega_{\rm dec}^{1/2}\;\zeta_{A}\,
\mathrm{min}\left\{ 1;\frac{\hat{m}}{H_{*}}\right\} ^{-1/3}
\mathrm{min}\left\{ 1;\frac{\hat{m}}{\Gamma}\right\} ^{1/12}
\left(\frac{\max\left\{\Gamma_A;H_{\mathrm{dom}}\right\}}{\Gamma}\right)^{1/4}.
\label{eq:H-expr-gen}
\end{equation}

The bound on the inflationary scale can be obtained by considering that the 
decay rate of the vector field is \mbox{$\Gamma_{A}\sim h^{2}\hat{m}$},
where $h$ is the coupling to the decay products. Then we can write
\mbox{$\max\left\{ \Gamma_{A};H_{\mathrm{dom}}\right\}\gtrsim h^{2}\hat{m}$}.
%where due to the gravitational decay $h\gtrsim\hat{m}/m_{\mathrm{P}}$.
Furthermore, we must consider the possibility of thermal evaporation
of the vector field condensate during the radiation dominated phase. If this
were to occur, all the memory of the superhorizon perturbation spectrum
would be erased. The thermalisation rate is determined by the scattering
rate of the massive boson field with the thermal bath and is given
by \mbox{$\Gamma_{\mathrm{sc}}\sim h^{4}T$}. Requiring that the condensate
does not thermalise before it decays, \mbox{$\Gamma_{\mathrm{sc}}<\Gamma_{A}$},
in Ref.~\cite{nonmin} it was shown that $h$ must satisfy 
\begin{equation}
\frac{\hat{m}}{m_{P}}\lesssim h\lesssim
\left(\frac{\hat{m}}{m_{P}}\right)^{1/4}.
\label{eq:bound-on-g}
\end{equation}
The lower bound in the above is due to decay through gravitational 
interactions, while the upper bound becomes irrelevant if the vector field 
dominates the Universe before it decays (then \mbox{$h\lsim 1$} is sufficient).
This is because, in the latter case, the energy density of the thermal bath is 
exponentially smaller than $\rho_A$ and the vector field condensate does not 
evaporate.

From Eq.~(\ref{eq:H-expr-gen}) one can see that the parameter space
is maximised if the Universe undergoes prompt reheating after inflation,
i.e. if \mbox{$\Gamma\rightarrow H_{*}$}. To find the parameter space
we investigate two separate cases: when \mbox{$\hat{m}\gg H_{*}$} and when 
\mbox{$\hat{m}\ll H_{*}$}.

\subsubsection{Statistically isotropic perturbations}

This possibility can be realised only in the case when \mbox{$\alpha=-4$}.
As mentioned before, if the mass of the vector field at the
end of inflation is larger than the Hubble parameter, \mbox{$\hat{m}>H_*$},
then the field has started oscillating already during inflation. In
this case the amplitudes of the longitudinal and transverse perturbations
are equal and therefore the curvature perturbations induced by the
vector field are statistically isotropic. We can assume, in this case, that
the vector field alone is responsible for the total curvature perturbation
in the Universe without the need to invoke additional perturbations
from other fields. Thus, we can set \mbox{$\zeta_{\mathrm{rad}}=0$} in 
Eq.~(\ref{eq:zeta-general}) and write 
\begin{equation}
\zeta\sim\Omega_{\rm dec}\zeta_{A}\;.
\end{equation}
 Using this and the lower bound on $h$ we find from Eq.~(\ref{eq:bound-on-g})
the lower bound for the inflationary Hubble parameter
\begin{equation}
\frac{H_{*}}{m_{P}}\gsim
\left(\frac{\zeta}{\sqrt{\Omega_{\rm dec}}}\right)^{4/5}
\left(\frac{\hat{m}}{m_{P}}\right)^{3/5},
\label{Hbound0}
\end{equation}
where we have taken into account that the parameter space is maximised
when the universe undergoes prompt reheating, i.e. 
\mbox{$\Gamma\rightarrow H_{*}$}.
From this expression it is clear that the lowest bound is attained
when the vector field dominates the Universe before its decay, 
\mbox{$\Omega_{\rm dec}\rightarrow1$},
and when the oscillations of the vector field 
%starts oscillating just 
commence at the very end of inflation, i.e. \mbox{$\hat{m}\rightarrow H_{*}$}. 
With these values we find the bounds
%minimal expansion rate and inflationary energy scale
\begin{equation}
H_{*}\gsim 10^{9}\:\mathrm{GeV}\quad\Leftrightarrow\quad 
V_{*}^{1/4}\gsim 10^{14}\:\mathrm{GeV}\,,
\label{eq:constraint-H-heavy-vFd}
\end{equation}
where $V_{*}^{1/4}$ denotes the inflationary energy scale and we used
that $\zeta\approx5\times10^{-5}$ from the observations of the Cosmic 
Background Explorer.

In view of the above, we can obtain a lower bound for the decay rate of the
vector field. Indeed, using Eqs.~(\ref{eq:bound-on-g}) and 
(\ref{eq:constraint-H-heavy-vFd}) we find
\begin{equation}
\Gamma_{A}\gsim\frac{\hat{m}^{3}}{m_{P}^{2}}
\gsim\frac{H_*^{3}}{m_{P}^{2}}\gsim10^{-9}\:\mathrm{GeV}\,.
\label{GAbound}
\end{equation}
From the above we find that the temperature of the Universe after the decay
of the vector field is 
\mbox{$T_{\rm dec}\sim\sqrt{m_P\Gamma_A}\gsim 10^4\,$GeV}, which is comfortably
higher than the temperature at Big Bang Nucleosynthesis (BBN) %with 
\mbox{$T_{\rm BBN}\sim 1\,$MeV} (i.e. the decay occurs much earlier than BBN),
and also higher than the electroweak phase transition, 
i.e. the decay precedes possible electroweak baryogenesis processes.
%Were the vector curvaton to decay too late it might threaten the successful
%predictions of the Big Bang Nucleosynthesis (BBN). For this reason, the vector
%field must decay and restore the radiation dominated Universe in advance of 
%BBN. Thus, we require \mbox{$\Gamma_A\gg H_{\rm BBN}$}, where the 
%the Hubble parameter at the BBN is of order 
%\begin{equation}
%H_{\mathrm{BBN}}\sim\frac{T_{\mathrm{BBN}}^{2}}{m_{P}}\sim10^{-23}
%\:\mathrm{GeV}\,,
%\end{equation}
%with the temperature at BBN being \mbox{$T_{\rm BBN}\sim 1\,$MeV}. 
%By comparing the above with Eq.~(\ref{GAbound}) we conclude that,
%in this scenario, the vector filed decays well before BBN takes place.

Since \mbox{$\hat m>H_*$}, Eq.~(\ref{eq:constraint-H-heavy-vFd}) corresponds 
to a lower bound on $\hat m$. An upper bound on $\hat m$ can be obtained as 
follows. Because, \mbox{$\hat m>H_*\gsim\Gamma$}, 
%and \mbox{$\Gamma_A\sim h^2\hat m\lsim\hat m$}, 
Eq.~(\ref{eq:Omega-dec}) becomes
\begin{equation}
\Omega_{\rm dec}\sim\Omega_{\mathrm{end}}
\sqrt{\frac{\Gamma}{\Gamma_{A}}}\;.
\label{Odec}
\end{equation}
From Eq.~(\ref{eq:bound-on-g}) we have \mbox{$\Gamma_A\gsim\hat m^3/m_P^2$}.
Combining this with the above we obtain
\begin{equation}
\hat m^3\lsim\left(\frac{\Omega_{\rm end}}{\Omega_{\rm dec}}\right)
\Gamma m_P^2.
\label{mG}
\end{equation}
Now, when \mbox{$\alpha=-4$} we have \mbox{$M\propto a^3$} during inflation.
Since the end of scaling occurs when inflation is terminated, for 
\mbox{$a<a_{\rm end}$} we can write
\begin{equation}
\hat m=\left(\frac{a_{\rm end}}{a}\right)^3M\simeq e^{3N_{\rm osc}}H_*\;,
\label{mH}
\end{equation}
where we considered that the field begins oscillating when 
\mbox{$M\simeq H_*$}
and $N_{\rm osc}$ is the number of remaining e-folds of inflation when the 
oscillations begin. Inserting the above into Eq.~(\ref{mG}) we find
\begin{equation}
N_{\mathrm{osc}}\lsim N_{\rm osc}^{\rm max}\equiv\frac29\left[
\ln\left(\frac{\Omega_{\rm end}}{\Omega_{\rm dec}}\right)+
\ln\sqrt{\frac{\Gamma}{H_{*}}}
+\ln\left(\frac{m_P}{H_*}\right)\right]<
\frac29\ln\left(\frac{m_P}{\Omega_{\rm dec}H_*}\right)\,,
\label{Nosc}
\end{equation}
where, in the last inequality, we used that \mbox{$\Omega_{\rm end}<1$} and
\mbox{$\Gamma\lsim H_*$}. Now, considering that \mbox{$\hat m\gsim H_*$},
Eq.~(\ref{Hbound0}) gives
\begin{equation}
\frac{\Omega_{\rm dec}H_*}{m_P}\gsim\zeta^2.
\label{Hbound1}
\end{equation}
Hence, combining Eqs.~(\ref{Nosc}) and (\ref{Hbound1}) we obtain
\begin{equation}
N_{\rm osc}^{\rm max}<-\frac49\ln\zeta=4.4\;.
\label{Noscbound}
\end{equation}
Thus, in view of Eq.~(\ref{mH}), we obtain the bound 
\mbox{$\hat m\lsim e^{3N_{\rm osc}^{\rm max}}H_*$}, 
which results in the following parameter space for $\hat m$:
\begin{equation}
1\lsim \hat m/H_*< 10^6,
\end{equation}
where we used Eq.~(\ref{Noscbound}). The above range is reduced if the
decay of the curvaton occurs more efficiently than through gravitational 
couplings, i.e. if \mbox{$h>\hat m/m_P$}. Nevertheless, we see that the 
parameter
space in which the vector field undergoes isotropic particle production and
can alone account for the curvature perturbation, is not small but may well be 
exponentially large. Indeed, repeating the above calculation with 
\mbox{$\Gamma_A\sim\hat m$} (i.e. \mbox{$h\sim 1$}) it is easy to find that
\begin{equation}
N_{\mathrm{osc}}=\frac23\left[
\ln\left(\frac{\Omega_{\rm end}}{\Omega_{\rm dec}}\right)+
\ln\sqrt{\frac{\Gamma}{H_{*}}}
%+\ln\left(\frac{m_P}{H_*}\right)
\right].
\label{Nosc1}
\end{equation}
Hence, using that \mbox{$\Omega_{\rm end}<1$} and \mbox{$\Gamma\lsim H_*$}
we obtain
\begin{equation}
N_{\rm osc}^{\rm max}=-\frac23\ln\Omega_{\rm dec}\lsim 3.1
\quad\Rightarrow\quad 1\lsim \hat m/H_*< 10^4,
\end{equation}
where we used \mbox{$\Omega_{\rm dec}\gsim 10^{-2}$}. This is because, in the
case considered, $f_{\rm NL}$ is given by Eq.~(\ref{fnlcurv}), so a smaller
$\Omega_{\rm dec}$ would violate the current observational bounds on the
non-Gaussianity in the CMB temperature perturbations \cite{wmap}. 

Still, it seems that, to obtain an exponentially large parameter space for 
$\hat m$, we need $\rho_A$ not to be too much smaller than $V_*$ during 
inflation and also inflationary reheating to be efficient. 
%(\mbox{$\Gamma\sim H_*$}).
In the case of gravitational decay (\mbox{$\Gamma_A\sim \hat m^3/m_P^2$})
Eq.~(\ref{Nosc}) has a weak dependence on both $\Omega_{\rm end}$ and $\Gamma$:
\mbox{$\hat m\propto (\Omega_{\rm end}^2\Gamma)^{1/3}$}, which means that 
the allowed range of values for $\hat m$ remains large even when 
$\Omega_{\rm end}$ and $\Gamma$ are substantially reduced. 
This is not necessarily so when \mbox{$\Gamma_A\sim h^2\hat m$}, with 
\mbox{$h\gg\hat m/m_P$}. Indeed, in this case it can be easily shown that
\mbox{$\hat m\propto h^{-2}\Omega_{\rm end}^2\Gamma$}. Therefore, if $\Gamma$
is very small it may eliminate the available range for $\hat m$. Fortunately,
the decay coupling $h$ can counteract this effect without being too small.%
%For example, demanding that the reheating temperature 
%\mbox{$T_{\rm reh}\sim\sqrt{\Gamma m_P}$} is small enough to avoid gravitino 
%overproduction requires \mbox{$\Gamma\lsim 1\,$GeV}. 

\subsubsection{Statistically anisotropic perturbations}
\label{sub:Statistically-anisotropic-perturbations}

If the vector field is not responsible for the total curvature perturbation in 
the Universe, the parameter space is more relaxed. In this case, the vector 
field may start oscillating after inflation and hence its mass is 
\mbox{$\hat{m}\ll H_{*}$}. However, this means that the curvature perturbation 
due to the vector field is strongly statistically anisotropic. For this reason
we can no longer set $\zeta_{\mathrm{rad}}$ to zero because the curvature
perturbation present in the radiation dominated Universe must be dominant.
In other words, the parameter %$\beta$ 
$\xi$ defined in Eq.~(\ref{xi}) needs to be
very small, \mbox{$%\beta
\xi\ll1$}. 

From the Eq.~(\ref{dWs}) it is clear that if inflation ends when
\mbox{$M=\hat{m}\ll H_{*}$}, the longitudinal power spectrum is much greater
than the transverse ones, \mbox{$\mathcal{P}_{\parallel}\gg\mathcal{P}_{+}$}.
Thus, from Eq.~(\ref{g}) we find 
\begin{equation}
g\simeq\frac{N_A^2{\cal P}_\|}{{\cal P}_\zeta^{\rm iso}}
\simeq\xi\,\frac{{\cal P}_\|}{{\cal P}_\phi}\,,
\end{equation}
where we also considered Eq.~(\ref{Piso}) with \mbox{$\xi\ll 1$}. 
Since the anisotropic contribution to the spectrum is subdominant \cite{GE}, 
to first order we can write 
\begin{equation}
%\left(\frac{\zeta}{\delta W_\|}\right)^2
\frac{\zeta^2}{\delta W_\|^2}
\simeq\frac{{\cal P}_\zeta^{\rm iso}}{{\cal P}_\|}\,.
\end{equation}
Hence, 
\begin{equation}
N_A^2\delta W_\|^2\simeq g\zeta^2.
\label{NAP}
\end{equation}
Now, from the $\delta N$ formalism at tree level we have \cite{stanis}%
\footnote{Technically, here we should have 
\mbox{$N_A\delta W\rightarrow\,${\boldmath $N_A\cdot$}$\delta${\boldmath $W$}}
but the dot-product is dominated by one term, that of the longitudinal 
component, since \mbox{$\delta W\simeq\delta W_\|$} as shown in 
Eq.~(\ref{dWdom}).}
\begin{equation}
\zeta=N_\phi\delta\phi+N_A\delta W.
\end{equation}
Comparing the above with Eq.~(\ref{eq:zeta-general}) we can equate the 
contributions to $\zeta$ from the vector field. Thus, we find
\begin{equation}
N_A\delta W=\hat\Omega_A\zeta_A\;.
\label{NAdW}
\end{equation}
Using that 
\begin{equation}
\frac{\delta W^2}{\delta W_\|^2}=
\frac{\sum_\lambda\delta W_\lambda^2}{\delta W_\|^2}\simeq
\frac{2{\cal P}_++{\cal P}_\|}{{\cal P}_\|}\approx 1
\label{dWdom}
\end{equation}
and combining Eqs.~(\ref{NAP}) and (\ref{NAdW}) we obtain
\begin{equation}
\zeta\sim g^{-1/2}\Omega_{A}\zeta_{A}.
\label{zzA}
\end{equation}
Inserting this expression into Eq.~(\ref{eq:H-expr-gen}) and considering again 
that the lowest decay rate of the vector field is through gravitational decay,
\mbox{$\mathrm{max}\left\{\Gamma_{A};H_{\mathrm{dom}}\right\} 
\geq\hat{m}^{3}/m_P^{2}$} we find
\begin{equation}
\frac{H_{*}}{m_{P}}>
\left(\frac{g\,\zeta^2}{\Omega_{\rm dec}}\right)^{3/4}
\left(\frac{\hat{m}}{m_{P}}\right)^{5/8}
\left(\frac{\Gamma}{m_{P}}\right)^{-3/8}
\mathrm{min}\left\{ 1;\frac{\hat{m}}{\Gamma}\right\} ^{1/8}.
\label{eq:Hbound-anisotropic}
\end{equation}
The above suggests that the lower bound on $H_*$ is minimised for prompt 
reheating with \mbox{$\Gamma\rightarrow H_*$}. Also, from observations
we know that the statistically anisotropic contribution to the curvature
perturbation must be subdominant. Thus,
%with \mbox{$g\lsim 0.3$} \cite{GE}. Thus, 
the vector field should not dominate the Universe before its decay. Hence, 
using \mbox{$\Gamma\rightarrow H_*$} and
%the maximum allowed anisotropy and 
\mbox{$\Omega_{A}<1$}
we obtain
\begin{equation}
\frac{H_{*}}{m_{P}}>\sqrt g\,\zeta\,\sqrt{\frac{\hat{m}}{m_{P}}}\,.
\label{Hanisbound}
\end{equation}
From this expression it is clear that the parameter space for $H_*$ is 
maximised for the lowest mass value. The minimum mass of the vector field can
be estimated from the requirement that the field decays before BBN.
Because the lowest decay rate is the gravitational decay, this 
condition reads \mbox{$\hat{m}^{3}/m_{P}^{2}\gsim T_{\mathrm{BBN}}^{2}/m_{P}$},
with \mbox{$T_{\rm BBN}\sim 1\,$MeV}, which corresponds to 
\mbox{$\hat{m}\gsim 10^4\;\mathrm{GeV}$}. Using this, we find
that the parameter space for the vector curvaton model with the statistically 
anisotropic curvature perturbations is 
\begin{equation}
H_{*}>\sqrt g\;10^7\:\mathrm{GeV}\quad\Leftrightarrow\quad 
V_{*}^{1/4}>g^{1/4}10^{13}\:\mathrm{GeV}\,,
\end{equation}
i.e. it is somewhat relaxed compared to the statistically isotropic case
(cf. Eq.~(\ref{eq:constraint-H-heavy-vFd}))
depending on the magnitude of the statistical anisotropy in the spectrum,
for which \mbox{$g\lsim 0.3$} \cite{GE}.
This result is valid for both \mbox{$\alpha=-1\pm 3$} cases.
From the above it is evident that there is ample parameter space for 
the mass of the vector field
\begin{equation}
10\;{\rm TeV}\lsim\hat m\ll H_*\;.
\end{equation}

We can readily use the above to briefly discuss an indicative example. 
Suppose that \mbox{$\hat m\sim 10^{-2}H_*$}. Then, from Eq.~(\ref{Hanisbound})
we obtain \mbox{$H_*>10^{-2}g\zeta^2m_P$}. Assume now that statistical 
anisotropy in the power spectrum is observed with \mbox{$g\sim 0.01$}. Thus,
we obtain \mbox{$H_*>10^{-13}m_P\sim 100\;$TeV}. Let us assume also that
non-Gaussianity is observed with predominant angular modulation that peaks in
the equilateral configuration with amplitude 
\mbox{$||f_{\rm NL}^{\rm equil}||\sim 10$}. Then, according to 
Eq.~(\ref{fNLamps}), we have \mbox{$g^2/\Omega_{\rm dec}\sim 10$}, i.e.
\mbox{$\Omega_{\rm dec}\sim 10^{-5}$}. Also, it is easy to check that
\mbox{$\Omega_{\rm end}\sim\Omega_{\rm dec}\sqrt{\Gamma_A/\hat m}$}
comfortably satisfies the bound in Eq.~(\ref{Wrange}) if 
\mbox{$\Gamma_A>T_{\rm BBN}^2/m_P$}. Finally, from the above we can
estimate the amplitude of the non-Gaussianity in the
local configuration \mbox{$||f_{\rm NL}^{\rm local}||\sim 10^{-3}$} 
[cf. Eq.~(\ref{fNLamps})] and also %the isotropic part 
\mbox{$f_{\rm NL}^{\rm iso}\sim 10^{-7}$} [cf. Eq.~(\ref{fnliso})].

\subsection{Constraints from isocurvature perturbations}

As discussed, when \mbox{$\hat m\ll H_*$}, particle production is strongly 
anisotropic and the curvaton has to be subdominant when it decays.
The fact that the curvaton decays while subdominant allows the possibility to
generate a sizable isocurvature perturbation, which needs to comply with
observational constraints. Of course, if the curvaton decays early enough into
relativistic particles which join the preexisting thermal bath then there is
a possibility that no isocurvature perturbation is generated. For this to be so
we require all the components of the late time Universe to be relativistic and
in thermal equilibrium so that they can be produced by the decay products of 
the vector curvaton field. Hence, dark matter needs to be thermal. For WIMP
dark matter we require curvaton decay to occur before the breaking of 
electroweak unification (i.e. the temperature at curvaton decay should be
\mbox{$T_{\rm dec}>1\;$TeV}) so that the decay products can produce WIMPs.
In this case baryons are also generated by the curvaton decay products and so
are neutrinos. However, if the vector curvaton decay occurs later or if the 
dark matter is not thermal (e.g. axions) then an isocurvature perturbation can 
be generated. An estimate of the maximum isocurvature perturbation is then 
given by
\begin{equation}
{\cal S}_{\rm max}=\left.\frac{\delta\rho}{\rho}\right|_{\rm after}
-\left.\frac{\delta\rho}{\rho}\right|_{\rm before}\approx
\frac25[(\zeta_{\rm bef}+\Omega_{\rm dec}\zeta_A)-\zeta_{\rm bef}]=
\frac25\Omega_{\rm dec}\zeta_A\,,
\end{equation}
where \mbox{$\Omega_{\rm dec}\ll 1$},
\mbox{$(\delta\rho/\rho)_{\rm before}\approx\frac25\zeta_{\rm bef}$} 
corresponds to the perturbation which preexists curvaton decay, while 
 \mbox{$(\delta\rho/\rho)_{\rm after}\approx
\frac25(\zeta_{\rm bef}+\Omega_{\rm dec}\zeta_A)$} 
[c.f Eq.~(\ref{eq:zeta-general})]
corresponds to the perturbation after curvaton decay which includes the
contribution to the perturbation due to the curvaton. The above assumes that
one of the constituents of the Universe content (e.g. dark matter) does not 
receive any contribution from the curvaton decay products while another 
such constituent receives 100\% contribution, hence this is the maximum 
isocurvature perturbation (Realistically, $\cal S$ is model dependent and it is
determined by the branching ratio of the vector curvaton decay to the various 
constituents of the Universe content). 

Observational constraints require that an uncorrelated isocurvature 
perturbation cannot exceed 11\% of the adiabatic mode \cite{isoc}. 
This means that
\begin{equation}
\Omega_{\rm dec}\zeta_A\lsim 0.1\zeta
\;\Rightarrow\;g\lsim 0.01\,,
\label{gisoc}
\end{equation}
where we considered also Eq.~(\ref{zzA}). Thus, even the maximum possible 
isocurvature perturbation produced by our model allows the generation of
statistical anisotropy at the level of a few percent. 

It is important to note here that the isocurvature perturbation is 
uncorrelated to the adiabatic one in contrast to the scalar curvaton model. 
This is because, in our model, when \mbox{$\hat m\ll H_*$}, $\zeta$
{\em is not generated by the vector curvaton field}
but it preexists curvaton decay. The vector curvaton contribution to
$\zeta$ is negligible (i.e. \mbox{$\zeta\approx\zeta_{\rm bef}$})
and its effect amounts to generating 
statistical anisotropy only. In the scalar curvaton model, however, it is the 
curvaton field that generates $\zeta$ (i.e. $\zeta_{\rm bef}$ is negligible), 
which means that, when curvaton decay occurs before domination, the 
isocurvature perturbation is fully (anti)correlated with the one of the 
curvaton. 

\section{Scalar fields as modulators}\label{eg}

Throughout this paper we have taken the modulation of the kinetic function $f$
and the mass of the vector field for granted and assumed that it is due to some
degree of freedom which varies during inflation. The most natural choice for 
such a degree of freedom is, of course, the inflaton field itself but other 
choices are also possible. In this section we briefly explore a couple of such
possibilities, inspired by beyond the standard model theories such as 
supergravity or superstrings. 

\subsection{The inflaton as modulator}

In string theory the modulation of parameters such as masses or kinetic 
functions is due to so-called moduli fields. The moduli 
are scalar fields which parametrise the size and shape of the extra dimensions.
In that sense they are not fundamental scalar fields, 
but appear so from the view-point of the 4-dimensional observer. Typically, 
the dependence of masses and couplings on canonically normalised (i.e. with 
canonical kinetic terms) moduli fields is exponential. 

Consider the following kinetic function
\begin{equation}
f(\phi)\propto e^{-\frac{\alpha}{\mu}\phi/m_P},
\label{fphi}
\end{equation}
where \mbox{$\mu>0$} is a real constant. Comparing this with Eq.~(\ref{fa})
we see that
\begin{equation}
m\propto a \propto e^{-\frac{1}{\mu}\phi/m_P},
\label{maphi}
\end{equation}
where we considered also Eq.~(\ref{m}). From the above we readily obtain
\begin{equation}
\dot\phi=-\mu H m_P\;,
\label{dotphi}
\end{equation}
which suggests that
\begin{equation}
(\rho_\phi)_{\rm kin}\equiv\frac12\dot\phi^2=\frac16\mu^2\rho\,
\label{rphikin}
\end{equation}
and we have used the flat Friedmann equation \mbox{$\rho=3(Hm_P)^2$}.

Now, suppose that $\phi$ is also driving quasi-de Sitter inflation, with 
\mbox{$H\simeq\,$constant}. This requires
\mbox{$\rho=\rho_\phi\simeq V(\phi)\gg(\rho_\phi)_{\rm kin}$}, which demands
\begin{equation}
\mu<\sqrt 6\;.
\label{mubound1}
\end{equation}
To drive quasi-de Sitter inflation a scalar field needs to follow a slow-roll
attractor solution, in which the acceleration term $\ddot\phi$ in its 
Klein-Gordon equation of motion is negligible. Thus, the field equation for
the inflaton is the well known slow-roll equation
\begin{equation}
3H\dot\phi\simeq-V'(\phi)\,,
\label{SR}
\end{equation}
where the prime denotes derivative with respect to $\phi$.
\footnote{Because \mbox{$\rho_A\ll V_*$} we can ignore the backreaction
of the vector field to the scalar field dynamics. Indeed, if the modulation
of $f$ and $m$ are due to a single scalar field $\phi$ then the Klein-Gordon 
equation of motion of the latter obtains a source term of the form
$$
\frac{\partial{\cal L}}{\partial\phi}=
-\frac14f'F_{\mu\nu}F^{\mu\nu}+mm'A_\mu A^\mu=
\frac{H}{\dot\phi}(\alpha\rho_{\rm kin}-2V_A)\sim
-\frac{H}{\dot\phi}\rho_A\;,
$$
where $\cal L$ is given in Eq.~(\ref{L}) and we considered 
\mbox{$f\propto m^\alpha$} and \mbox{$m\propto a$}. If $\rho_A$ is small enough
then the above can be negligible compared with the $-V'$ source term in the
equation of motion.}
Combining the above with Eq.~(\ref{dotphi}) and taking 
\mbox{$V(\phi)\simeq 3(Hm_P)^2$} we obtain
\begin{equation}
V(\phi)\simeq V_0e^{\mu\phi/m_P}\propto a^{-\mu^2},
\label{V}
\end{equation}
where $V_0$ is a density scale and we used Eq.~(\ref{maphi}). Thus, we find 
that the inflaton is characterised by an exponential potential, which is 
reasonable for a modulus field. The above also suggests that, for quasi-de 
Sitter inflation, when \mbox{$V(\phi)\simeq\rho_\phi=\rho\simeq\,$constant}, 
we need to have
\begin{equation}
\mu<1\,,
\label{mubound2}
\end{equation}
which is somewhat stronger than the bound in Eq.~(\ref{mubound1}). Combining 
Eqs.~(\ref{SR}) and (\ref{V}) we get
\begin{equation}
\Delta\phi=\frac{2}{\mu}m_P\ln\left(1-\frac12\mu^2\Delta N\right)\,,
\label{Dphi}
\end{equation}
where \mbox{$\Delta N=H\Delta t$} is the number of the elapsing e-folds
during which the inflaton varies by $\Delta\phi$. Notice that the above is 
consistent with Eq.~(\ref{dotphi}) but it is valid only when 
\mbox{$\mu^2\ll 2/\Delta N$}.\footnote{It turns out that 
Eq.~(\ref{dotphi}) remains always valid even if Eq.~(\ref{Dphi}) is not.}
The violation of this condition is equivalent with the violation of the 
``robustness'' assumption for $H$ (i.e \mbox{$H\simeq\,$constant}).
That is, it corresponds to a significant variation in $H$, i.e. 
\mbox{$-\Delta H\sim H^2\Delta t$}. Technically, this can occur even under
slow-roll conditions if $\Delta t$ is large enough. 
Since a ``robust'' $H$ is necessary for the generation of a
scale invariant spectrum of perturbations, it should last at least as much
as it takes for the entire range of the observable cosmological scales to exit 
the horizon. This range corresponds to about \mbox{$\Delta N\simeq 9$} e-folds,
which induces the bound
\begin{equation}
\mu<\frac{\sqrt 2}{3}\,,
\label{mubound3}
\end{equation}
which is tighter than the bounds in Eqs.~(\ref{mubound1}) and (\ref{mubound2}).

Let us attempt to obtain an estimate of $\mu$. The slow roll parameters 
for the model in Eq.~(\ref{V}) are
\begin{eqnarray}
\varepsilon & \equiv & 
-\dot H/H^2\simeq\frac12 m_P^2\left(\frac{V'}{V}\right)^2=\frac12\mu^2,
\label{vareps}\\
\eta_\phi & \equiv & m_P^2\frac{V''}{V}=\mu^2.
\label{eta}
\end{eqnarray}

If the inflaton is the source of the dominant contribution to the curvature 
perturbation of the Universe then, the spectral index of ${\cal P}_\zeta$ is
\begin{equation}
n_s-1=2\eta_\phi-6\varepsilon=-\mu^2=-2\varepsilon\,.
\label{nsphi}
\end{equation}
If, on the other hand, the dominant contribution to the curvature 
perturbation of the Universe is due to the vector curvaton (only possible
in the \mbox{$\alpha=-4$} case with \mbox{$\hat m\gsim H_*$}) then we have
\begin{equation}
n_s-1=2\eta_A-2\varepsilon\approx-2\varepsilon\,,
\label{nsphi2}
\end{equation}
where we considered that \mbox{$\eta_A$} is negligible. Indeed, in almost all 
cases studied in the previous sections we had \mbox{$\eta_A=0$} when 
\mbox{$\epsilon_f,\epsilon_m=0$}, the reason being that the power spectra for
the vector field perturbations are exactly scale invariant, when inflation is
considered to be de Sitter (i.e. \mbox{$H=\,$constant}), which means that any
deviations from scale invariance would be due to \mbox{$\varepsilon\neq 1$}
only. The only exception was the transverse component in the case when 
\mbox{$\alpha=2$}, where the $\eta_A$ parameter (defined in Eq.~(\ref{etaA})) 
may not be negligible. However, in the following, we assume that 
\mbox{$M\ll H$} is sufficiently small to ignore $\eta_A$ even in this case.

From Eqs.~(\ref{nsphi}) and (\ref{nsphi2}) we see that, regardless of whether
the dominant contribution to the curvature perturbation is from the inflaton
or from the vector curvaton field, the spectral index is 
\begin{equation}
n_s=1-2\varepsilon\,.
\label{nseps}
\end{equation}
Comparing the above with the observed value \cite{wmap}
\mbox{$n_s=0.960\pm 0.014$}
(for negligible tensors) we find
\begin{equation}
\mu=0.200\pm 0.036\,,
\label{muvalue}
\end{equation}
which satisfies the bound in Eq.~(\ref{mubound3}). In view of Eq.~(\ref{Dphi}),
the above value guarantees that $H$ remains ``robust'' for about 
\mbox{$\Delta N\simeq 50$}, which comfortably encompasses the cosmological 
scales.

Note that, for the model in Eq.~(\ref{V}), the slow-roll parameters in 
Eqs.~(\ref{vareps}) and (\ref{eta}) remain constant and smaller than unity.
Hence, inflation never ends. One can remedy this by either invoking some kind
of hybrid mechanism to remove the inflaton from the slow-roll trajectory and 
send it to the true vacuum or
by modifying the model such that slow-roll naturally breaks down eventually.%
\footnote{%
For example, in the toy-model with inflaton scalar potential
\mbox{$V(\phi)=V_0[\cosh(\mu\phi/m_P)-1]$} the slow-roll parameters
are \mbox{$\eta=\frac12\mu^2+\varepsilon$} and 
\mbox{$\varepsilon=\frac12\mu^2\coth^2\left(\mu\phi/2m_P\right)$}. The
model approximates Eq.~(\ref{V}) when \mbox{$\phi\gg m_P/\mu$} but 
allows inflation to end at \mbox{$\phi_{\rm end}=
\frac{1}{\mu}m_P\ln\left(\frac{\sqrt 2+\mu}{\sqrt 2-\mu}\right)
\simeq\sqrt 2\,m_P$}, where \mbox{$\varepsilon(\phi_{\rm end})\equiv 1$} and
we considered \mbox{$\mu\ll 1$} in the last equality.}
%We consider such a modification in the toy-model below.

\subsection{Higgsed vector curvaton}

Another possibility is that the vector field is Higgsed, which means that
\mbox{$m\propto\varphi$}, where $\varphi$ is a Higgs field. Hence, from
Eq.~(\ref{m}) we require \mbox{$\varphi\propto a$} during inflation. Suppose
that $\varphi$ is rolling down a hilltop potential of the form
\begin{equation}
V(\varphi)=V_{\rm top}-\frac12 m_\varphi^2\varphi^2+\cdots,
\label{Vvphi}
\end{equation}
where the ellipsis corresponds to terms of higher order which stabilise the 
potential but are negligible during inflation and $V_{\rm top}$ is a constant
density scale. The equation of motion of $\varphi$ is
\footnote{As in the previous subsection we ignore the backreaction from the 
vector field.}
\begin{equation}
\ddot\varphi+3H\dot\varphi-m_\varphi^2\varphi\simeq 0\,,
\end{equation}
whose solution has the following growing mode 
\begin{equation}
\varphi\propto a^{-\frac32[1-\sqrt{1+\frac49(m_\varphi/H)^2}]}.
\end{equation}
Therefore, the requirement that \mbox{$\varphi\propto a$} is achieved if the
effective mass of the Higgs field during inflation is
\begin{equation}
m_\varphi=2H\;.
\end{equation}

It turns out that such a value for the effective mass of the scalar field 
during inflation is quite reasonable in the context of supergravity theories. 
Indeed, in Ref.~\cite{randall} it has been demonstrated that K\"{a}hler
corrections to the scalar potential are expected to give a contribution of 
order $H$ to the masses of scalar fields. Hence, these fields would be 
fast-rolling during inflation down the potential slopes. 

Moreover, in this case, $f$ is the gauge kinetic function, which, in 
supergravity theories, is a holomorphic function of the scalar fields of the
theory. Hence, it is natural to expect that \mbox{$f=f(\varphi)$} and 
the rolling $\varphi$ would modulate the kinetic function as well as the mass.
Indeed, if \mbox{$m_\varphi=2H$}, then to satisfy Eq.~(\ref{fa}) one simply 
needs \mbox{$f(\varphi)\propto\varphi^\alpha$} with \mbox{$\alpha=-1\pm 3$}.

Of course, $f$ may also depend on other fast-rolling scalar fields
\mbox{$f=f(\phi_1, \phi_2,\ldots\phi_n)$}. If we assume that
\mbox{$f\propto\prod_i^n\phi_i^{\alpha_i}$} with \mbox{$\alpha_i={\cal O}(1)$} 
then
\begin{equation}
\frac{\dot f}{f}=\sum_i^n\alpha_i\frac{\dot\phi_i}{\phi_i}=
\frac32 H\sum_i^n\alpha_i
%\left[\sqrt{1+\frac49\left(\frac{m_i}{H}\right)}-1\right],
\left(\sqrt{1+\frac49\frac{m_i^2}{H^2}}-1\right),
\qquad{\rm where}\qquad m_i^2\equiv\frac{\partial^2V}{\partial\phi_i^2}\,.
\end{equation}
%where \mbox{$m_i^2\equiv\frac{\partial^2V}{\partial\phi_i^2}$}.
Since K\"{a}hler corrections result in \mbox{$m_i^2\sim H^2$}, we find that
\mbox{$\dot f/f\sim H$} as required. We would still need to tune $\alpha_i$
and $m_i^2$ such that \mbox{$\dot f/f=(-1\pm 3)H$}, but it is evident that
the required values can be naturally attained in the context of supergravity, 
as also discussed in Ref.~\cite{sugravec}. 

Having said that, since $f$ is the gauge 
kinetic function in this case, we have \mbox{$f\sim 1/e^2$}, where 
\mbox{$e\propto a^{-\alpha/2}$} is the gauge coupling. This means that, because
\mbox{$f\rightarrow 1$} at the end of inflation, the vector field becomes
strongly coupled during inflation if \mbox{$\alpha>0$}. Thus, the physically
motivated case is only \mbox{$\alpha=-4$}, where \mbox{$e\propto a^2$} and
the vector field is always weakly coupled. Fortunately, it is this 
case which has the richest phenomenology, as we have shown.

\section{Summary of results}

The model studied in this paper, albeit simple, has been shown to have a 
rich phenomenology, without suffering from any instabilities. 
%as is for example the non-minimally coupled vector curvaton in 
%Refs.~\cite{rcurv,stanis,fnlanis}.
By studying particle production in this model we have found that a scale 
invariant spectrum of vector perturbations for the transverse components can
be obtained if \mbox{$\alpha=-1\pm 3$} and the field is light
%effectively massless
at horizon exit. We also found that, to get a scale invariant spectrum of 
vector field perturbations for the longitudinal component we additionally
require that \mbox{$m\propto a$}. We have assumed that this scaling continues 
throughout inflation as would be the case if both $f$ and $m$ are modulated by 
some degree of freedom which varies during inflation, e.g. the inflaton field.
We also assumed that inflation is quasi-de Sitter.

In all cases we have solved the equations of motion for the mode functions of 
the components of the vector field perturbations both numerically and through 
analytic approximations, normalising them appropriately at the vacuum. A
comparison between the two has shown that our analytic expressions approximate 
the numerical solutions with very high precision. We have found that, after 
horizon exit, the mode functions $w_\lambda$ of the components of the vector 
field perturbations cease oscillating and follow a power-law evolution of the
form
\begin{equation}
w_\lambda=C_A+C_Ba^{-3},
\label{powerlaw}
\end{equation}
where $C_A,C_B$ are constants determined by the vacuum initial conditions.
Depending on the case considered, the dominant term in the above is either
the constant or the decaying mode.

The effective mass of the vector field during inflation is [cf. Eq.~(\ref{M})]
\begin{equation}
M=\frac{m}{\sqrt f}\propto a^{1-(\alpha/2)}.
%\frac{\alpha}{2}}.
%\;\Rightarrow\;M=e^{(\frac{\alpha}{2}-1)N}\hat m\,,
\end{equation}
%where $N$ is the remaining e-folds of inflation and we used that 
%\mbox{$f=e^{\alpha N}$}, since \mbox{$f\rightarrow 1$} at the end of 
%inflation, when \mbox{$M\rightarrow\hat m$}. 
As mentioned, scale invariance for the 
transverse spectra requires \mbox{$\alpha=-1\pm 3$} plus the vector field 
needs to be light when cosmological scales exit the horizon 
\mbox{$M_*\ll H_*$}, where \mbox{$H_*\simeq\,$constant} is the 
inflationary Hubble scale. In the case when \mbox{$\alpha=2$} (i.e. 
\mbox{$f\propto a^2$}) we see that \mbox{$M=\,$constant} so that
\mbox{$\hat m=M_*\ll H_*$}, i.e the vector 
field remains always light until the end of inflation. In this case, the 
power-law regime for the evolution of the mode functions of the vector field
perturbations is valid until inflation terminates, while the dominant term in
Eq.~(\ref{powerlaw}) is the constant one for all $w_\lambda$ 
(see Fig.~\ref{fig2}). 
%Therefore, 
%when \mbox{$\alpha=2$} our model produces the results outlined above for the
%case \mbox{$\hat m<H_*$}.

In contrast, when \mbox{$\alpha=-4$} (i.e. \mbox{$f\propto a^{-4}$}) we have 
\mbox{$M\propto a^3$}. This allows the possibility to have
\mbox{$M_*\ll H_*\ll\hat m$}, i.e. while the field is light when the 
cosmological scales exit the horizon it can become heavy before the end of 
inflation. If this is the case then the power-law regime is terminated before
the end of inflation when the vector field becomes heavy and begins coherent 
oscillations. Hence, when \mbox{$\alpha=-4$} we could have either 
\mbox{$\hat m<H_*$} or \mbox{$\hat m\gsim H_*$}. Studying the power-law regime
we have shown that the dominant term in 
Eq.~(\ref{powerlaw}) for the mode functions is the constant term for the 
transverse components but not for the longitudinal one, for which the dominant
term is the decaying mode \mbox{$w_\|\propto a^{-3}$} (see Fig.~\ref{fig1}). 
If \mbox{$\hat m<H_*$} then inflation ends before the power-law regime is 
concluded. In this case, the results are the same as in the 
\mbox{$\alpha=2$} case. However, if \mbox{$\hat m>H_*$} the power-law regime
is terminated before the end of inflation. 

Consider first that \mbox{$\hat m<H_*$}, which is possible for 
\mbox{$\alpha=-1\pm 3$}. As mentioned, in this case the power-law regime 
continues until the end of inflation.
If this is so we have shown that the power spectra for the transverse and 
longitudinal components of the vector field superhorizon perturbations are 
given by
\begin{equation}
{\cal P}_+=\left(\frac{H_*}{2\pi}\right)^2\qquad{\rm and}\qquad
{\cal P}_\|=\left(\frac{3H_*}{\hat m}\right)^2\left(\frac{H_*}{2\pi}\right)^2,
\label{spectra}
\end{equation}
where ${\cal P}_\|$ is the spectrum of the longitudinal component,
\mbox{${\cal P}_+\equiv\frac12({\cal P}_L+{\cal P}_R)$}, with 
${\cal P}_L,{\cal P}_R$ being the spectra of the left and right polarisations
of the transverse components and \mbox{${\cal P}_L={\cal P}_R$} since the model
is parity conserving. Because of the condition \mbox{$\hat m<H_*$}, the
particle production process is found to be more efficient in the longitudinal
direction than the transverse ones. Hence, in this case, the contribution of 
the vector field to the curvature perturbation is strongly anisotropic. Indeed,
the anisotropy parameter in Eq.~(\ref{Pzg}) in the spectrum is found to be
\begin{equation}
g=\xi\left(\frac{3H_*}{\hat m}\right)^2,
\end{equation}
where $\xi$ is defined in Eq.~(\ref{xi}) and quantifies the level of the 
contribution of the vector field to the overall curvature perturbation $\zeta$ 
of the Universe. Ref.~\cite{GE} suggests that \mbox{$g\lsim 0.3$}, which 
implies that \mbox{$\xi\ll 1$} in this case. Hence, the vector field 
contribution to $\zeta$ has to be subdominant, with the dominant component due
to some other source, e.g. the inflaton field.

Non-Gaussianity in the curvature perturbation is found to be also 
statistically anisotropic with its magnitude and direction correlated 
with statistical anisotropy in the spectrum as in Ref.~\cite{fnlanis}.
Indeed, for the isotropic part we found [cf. Eq.~(\ref{fnliso})]
\begin{equation}
%\frac{6}{5} 
f_{\rm NL}^{\rm iso}  
\simeq\frac{5g^2}{3\Omega_{A}}\left(\frac{\hat m}{3H_*}\right)^4=
\frac{5\xi^2}{3\Omega_{\rm dec}}
\end{equation}
which can be substantial even if \mbox{$\xi\ll 1$}, where $\Omega_{\rm dec}$
is the density parameter at the decay of the vector field.
%, c.f. Eq.~(\ref{omega}). 
The anisotropic part of 
$f_{\rm NL}$ depends on the configuration but we have shown that 
\mbox{$\calg\gg 1$} in both the equilateral and local cases, where
$\calg$ is defined in Eq.~(\ref{gNL}). Thus, we see that the bispectrum
is predominantly anisotropic, which suggests that, if non-Gaussianity is 
observed without significant angular modulation, then the case 
\mbox{$\hat m<H_*$} in our model will be ruled out.

Consider now that \mbox{$\hat m\gg H_*$}, which is possible only for 
\mbox{$\alpha=-4$}. If this is the case, then before 
the end of inflation the mode 
functions begin oscillating with frequency increasingly larger than $H_*$, 
while for the average power spectra we have [cf. Eq.~(\ref{Pbars})]
\begin{equation}
\overline{{\cal P}_+}=\overline{{\cal P}_\parallel}=
\frac12\left(\frac{3H_*}{\hat m}\right)^2\left(\frac{H_*}{2\pi}\right)^2.
\label{avrgP}
\end{equation}
Since particle production in this case is isotropic there is no statistical
anisotropy in the spectrum and the bispectrum, i.e. \mbox{$g=\calg=0$}.
Therefore, the vector field can alone generate the curvature perturbation 
of the Universe without any contribution from other sources. The model can 
indeed generate non-Gaussianity in the curvature perturbation with
\begin{equation}
f_{\rm NL}=\frac{5}{4\hat{\Omega}_{\rm dec}}\;,
\label{fNLfin}
\end{equation}
where $\hat{\Omega}_A$ is defined in Eq.~(\ref{omega}). The above result is 
identical to the scalar curvaton mechanism. If there are other significant
contributions to $\zeta$ beyond the vector field then $f_{\rm NL}$ is
smaller than the above.

It is interesting to consider the case when \mbox{$\hat m\sim H_*$}, again
possible only when \mbox{$\alpha=-4$} (i.e. \mbox{$f\propto a^{-4}$}). In this 
case the
mode functions are about to begin oscillating at the end of inflation, while 
their value is comparable but not necessarily identical. This case 
allows the possibility that \mbox{$|\delta{\cal P}|<{\cal P}_+$}, where 
\mbox{$\delta{\cal P}={\cal P}_\|-{\cal P}_+$}. Then, the vector field can 
alone generate the total curvature perturbation, but it can also produce 
statistical anisotropy in the spectrum and bispectrum, which would be at equal 
level
\begin{equation}
\calg
%{\cal G}_{\rm NL}
\simeq g=\frac{\delta{\cal P}}{{\cal P}_+}\lsim 0.3\,.
\end{equation}
In this case we see that the anisotropy in the bispectrum is subdominant, with 
the dominant component $f_{\rm NL}^{\rm iso}$ given by Eq.~(\ref{fNLfin}).

To investigate the parameter space for this model we have studied
the evolution of the zero-mode of the vector field assuming energy 
equipartition at the onset of inflation. When \mbox{$\alpha=2$} we 
found that \mbox{$W\simeq\,$constant} throughout inflation. In contrast, when
\mbox{$\alpha=-4$} we found that the zero mode during inflation scales as 
\mbox{$W\propto a^{-3}$} during the power-law regime, while its amplitude 
scales as \mbox{$||W||\propto a^{-3}$} during the oscillations, when the vector
field becomes heavy. In all cases though, the density
of the vector field $\rho_A$ remains constant during inflation, regardless 
whether the field is oscillating or not. After inflation, the density
scales as radiation \mbox{$\rho_A\propto a^{-4}$} or matter
\mbox{$\rho_A\propto a^{-3}$} when the field is light \mbox{$\hat m<H(t)$}
or heavy \mbox{$\hat m>H(t)$} respectively. This is different from scalar 
fields, whose density remains constant when they are light. We have also 
verified that, when the field is heavy and undergoes oscillations, it acts as
a pressureless isotropic fluid, which can dominate the Universe without 
generating a large-scale anisotropy, in accordance to the findings in 
Ref.~\cite{vecurv}. 

First, we considered the case when \mbox{$\hat m\gg H_*$}, possible only if
\mbox{$\alpha=-4$}. As there is no anisotropy in this case we assumed that the
vector curvaton alone generates $\zeta$. By taking into account all relevant
bounds on the decay rate of the inflaton and vector curvaton fields we have 
found that the scenario works when \mbox{$H_*\gsim 10^9\,$GeV}. The vector 
curvaton begins its oscillations before the end of inflation but no earlier 
than in the last few e-folds, as \mbox{$N_{\rm osc}\lsim 4$}. Still, 
this allows an exponentially large parameter space for the value of $\hat m$,
which may be as large as $10^6H_*$. The parameter space is reduced if the
decay of the inflaton is late or if the contribution of the vector field to
the energy budget is very small during inflation. This, however, can be 
counteracted if the vector field decay rate is also small. 
If \mbox{$\hat m\sim H_*$} then the vector field can alone generate $\zeta$
but may also produce statistical anisotropy within the observational bounds.
In this case it is easy to show that \mbox{$H_*\gsim 10^9\,$GeV} as well.
Finally, the case when \mbox{$\hat m<H_*$} allows a slightly lower 
inflationary scale since the lower bound to $H_*$ is relaxed by a factor 
$<10^{-2}\sqrt g$.
%to \mbox{$H_*\gsim 10^5\,$GeV}. 
In this case the vector curvaton contribution to the curvature perturbation is 
strongly anisotropic, which means that \mbox{$\Omega_A\ll 1$} at decay. 

\section{Conclusions}

In this paper we studied a particularly promising vector curvaton model 
consisting of a massive Abelian vector field, with a Maxwell type kinetic term
and with varying kinetic function $f$ and mass $m$ during inflation. The model
is rather generic, it does not suffer from instabilities such as ghosts 
and may be naturally realised in the context of theories
beyond the standard model such as supergravity and superstrings. 

We have parametrised the time dependence of the kinetic function as 
\mbox{$f\propto a^\alpha$}, where \mbox{$a=a(t)$} is the scale factor. 
Our model offers two distinct possibilities. If \mbox{$\hat m<H_*$} 
(possible for \mbox{$\alpha=-1\pm 3$}) the vector 
field can only produce a subdominant contribution to the curvature perturbation
$\zeta$, but it can be the source of statistical anisotropy in the spectrum
and bispectrum. In fact, non-Gaussianity in this case is predominantly 
anisotropic, which means that, if a non-zero $f_{\rm NL}$ is observed without
angular modulation, then our model is falsified in the \mbox{$\hat m<H_*$} 
case. The second possibility (possible for \mbox{$\alpha=-4$} only)
corresponds to \mbox{$\hat m\gsim H_*$}. In this case the vector field can
alone generate the curvature perturbation $\zeta$ without any contribution from
other sources such as scalar fields. If \mbox{$\hat m\gg H_*$} particle 
production is isotropic and the model does not generate any statistical 
anisotropy. The vector field begins oscillating a few e-folds before the end 
of inflation but its density remains constant until inflation ends. The 
parameter space for this case can be exponentially large, i.e. 
\mbox{$1\ll \hat m/H_*<10^6$}. Significant non-Gaussianity 
can be generated, provided the vector field decays before it dominates the 
Universe, in which case $f_{\rm NL}$ is found to be identical to the scalar 
curvaton scenario. In other words, if \mbox{$\hat m\gg H_*$}, our vector 
curvaton can reproduce the results of the scalar curvaton paradigm.
Finally, if \mbox{$\hat m\sim H_*$} the vector field can alone generate the 
curvature perturbation $\zeta$ but it can also generate statistical anisotropy
in the spectrum and bispectrum. In this case, the anisotropy in $f_{\rm NL}$ is
subdominant and equal to the statistical anisotropy in the spectrum,
which is a characteristic signature of this possibility. 

We have also found that inflation has to occur at relatively high energies, 
with \mbox{$H_*\gsim 10^9\,$GeV} in the (almost) isotropic and 
\mbox{$H_*>\sqrt g\,10^7\,$GeV} in the anisotropic case 
(with \mbox{$\hat m\gsim 10\;$TeV}). These 
bounds correspond to prompt reheating, with \mbox{$T_{\rm reh}\sim V_*^{1/4}$}
which could result in gravitino overproduction. However, if the vector curvaton
dominates the Universe, its decay could release enough entropy to efficiently 
dilute the density of the gravitinos. Furthermore, as in Ref.~\cite{sugravec},
one could substantially reduce the inflationary scale through introducing
an increment to the mass of the vector field, say at a phase transition, after 
the end of inflation, following the mechanism first suggested for the scalar 
curvaton scenario in Ref.~\cite{low}.

For our model to work $f$ and $m$ should vary in a specific manner, which 
requires tuning. We have outlined two possibilities for achieving the desired
modulation for these quantities. First, we considered that the quantities in 
question are modulated by a string modulus field, which could also play the 
role of the inflaton. In this case we found that the potential of the modulus 
has to be approximately exponential, which is reasonable. The second 
possibility which we discussed was that of a Higgsed vector curvaton in the
context of supergravity theories, where scalar fields during inflation 
obtain an effective mass of order the Hubble scale. We showed that a Higgs 
field with tachyonic mass $2H_*$ suffices to account for the desired modulation
for $m$, while it is natural to expect that the gauge kinetic function is 
modulated by the fast-rolling scalar fields of the theory such that 
\mbox{$\dot f/f\sim H_*$} as required. These examples demonstrate that the
tuning of the modulation of $f$ and $m$ can be attained in a realistic manner.
This should be contrasted with the traditional case of generating the 
curvature perturbation using scalar fields, where their effective mass needs
to be fine-tuned at least by ${\cal O}(10^{-2})$ against K\"{a}hler corrections
to produce an approximately scale-invariant spectrum 
(the famous $\eta$-problem).
In our model, the vector field also needs to be effectively massless 
(\mbox{$M_*<H_*$}) when the cosmological scales exit the horizon but, to our
knowledge, there is no compelling reason why this should not be so. Note 
also, that the vector field can become heavy by the end of inflation in
the \mbox{$\alpha=-4$} case. We considered de~Sitter inflation in our treatment
so that the obtained spectra were exactly scale invariant if 
\mbox{$f\propto a^{-1\pm 3}$} and \mbox{$m\propto a$}. Deviations from the
Harrison-Zel'dovich spectrum can be attained either by perturbing the 
modulation of $f$ and $m$ or by considering quasi-de Sitter inflation, with
\mbox{$\varepsilon\equiv-\dot H/H^2\sim 10^{-2}$}.

Even though in our specific examples we employed scalar fields to modulate
$f$ and $m$, in principle their variation can be controlled by any kind
of degree of freedom which varies during inflation. In that sense, this model
can generate the curvature perturbation in the Universe without direct 
involvement of fundamental scalar fields. Of course, theories beyond the 
standard model are abundant with scalar fields and vector fields alike. 
The next step, therefore, is to realise this vector curvaton model 
in the context of realistic extensions of the standard model. 

%\begin{acknowledgements}
\acknowledgements

This work was supported (in part) by the European Union through the Marie 
Curie Research and Training Network "UniverseNet" (MRTN-CT-2006-035863) 
and by STFC Grant ST/G000549/1.
M.K. and J.M.W. are also supported by the Lancaster University Physics 
Department.

%I would like to thank D.~H.~Lyth and J.~McDonald for discussions and the 
%referee for insightful comments. 
%\end{acknowledgements}

\end{widetext}

\begin{thebiblio}{03}

\bibitem{vecurv}
K.~Dimopoulos,
%``Can a vector field be responsible for the curvature perturbation in the
%universe?,''
Phys.\ Rev.\  D {\bf 74} (2006) 083502.
%[arXiv:hep-ph/0607229].
%%CITATION = PHRVA,D74,083502;%%

\bibitem{curv}
D.~H.~Lyth and D.~Wands,
%``Generating the curvature perturbation without an inflaton,''
Phys.\ Lett.\ B {\bf 524} (2002) 5;
K.~Enqvist and M.~S.~Sloth,
%``Adiabatic CMB perturbations in pre big bang string cosmology,''
Nucl.\ Phys.\ B {\bf 626} (2002) 395;
%[arXiv:hep-ph/0109214].
%%CITATION = HEP-PH 0109214;%%
T.~Moroi and T.~Takahashi,
%``Effects of cosmological moduli fields on cosmic microwave background,''
Phys.\ Lett.\ B {\bf 522} (2001) 215
[Erratum-ibid.\ B {\bf 539} (2002) 303].
%[arXiv:hep-ph/0110096].
%%CITATION = HEP-PH 0110096;%%

\bibitem{early}
S.~Mollerach,
%``Isocurvature Baryon Perturbations And Inflation,''
Phys.\ Rev.\ D {\bf 42}, 313 (1990);
%%CITATION = PHRVA,D42,313;%%
A.~D.~Linde and V.~Mukhanov,
%``Nongaussian isocurvature perturbations from inflation,''
Phys.\ Rev.\ D {\bf 56}, R535 (1997).
%[arXiv:astro-ph/9610219]
%%CITATION = ASTRO-PH 9610219;%%

\bibitem{liber}
K.~Dimopoulos and D.~H.~Lyth,
%``Models of inflation liberated by the curvaton hypothesis,''
Phys.\ Rev.\  D {\bf 69} (2004) 123509;
%[arXiv:hep-ph/0209180].
%%CITATION = PHRVA,D69,123509;%%
%\bibitem{liber2}
T.~Moroi, T.~Takahashi and Y.~Toyoda,
%``Relaxing constraints on inflation models 
%with curvaton,''
Phys.\ Rev.\ D {\bf 72} (2005) 023502;
%[arXiv:hep-ph/0501007].
%%CITATION = HEP-PH 0501007;%%
T.~Moroi and T.~Takahashi,
%``Implications of the curvaton on inflationary 
%cosmology,''
Phys.\ Rev.\ D {\bf 72} (2005) 023505.
%[arXiv:astro-ph/0505339].
%%CITATION = ASTRO-PH 0505339;%%

\bibitem{pmfrev}
D.~Grasso and H.R.~Rubinstein,
  %``Magnetic fields in the early universe,''
  Phys.\ Rept.\  {\bf 348} (2001) 163;
%  [arXiv:astro-ph/0009061].
  %%CITATION = PRPLC,348,163;%%
M.~Giovannini,
  %``The magnetized universe,''
  Int.\ J.\ Mod.\ Phys.\  D {\bf 13} (2004) 391.
%  [arXiv:astro-ph/0312614].
  %%CITATION = IMPAE,D13,391;%%

\bibitem{TW}
M.~S.~Turner and L.~M.~Widrow,
  %``Inflation Produced, Large Scale Magnetic Fields,''
  Phys.\ Rev.\  D {\bf 37} (1988) 2743.
  %%CITATION = PHRVA,D37,2743;%%

\bibitem{mine}
A.~C.~Davis, K.~Dimopoulos, T.~Prokopec and O.~Tornkvist,
  %``Primordial spectrum of gauge fields from inflation,''
  Phys.\ Lett.\  B {\bf 501} (2001) 165;
%  [Phys.\ Rev.\ Focus {\bf 10} (2002) STORY9]
%  [arXiv:astro-ph/0007214].
  %%CITATION = 00627,10,STORY9;%%
K.~Dimopoulos, T.~Prokopec, O.~Tornkvist and A.~C.~Davis,
  %``Natural magnetogenesis from inflation,''
  Phys.\ Rev.\  D {\bf 65} (2002) 063505.
%  [arXiv:astro-ph/0108093].
  %%CITATION = PHRVA,D65,063505;%%

\bibitem{pmfinf}
W.~D.~Garretson, G.~B.~Field and S.~M.~Carroll,
  %``Primordial magnetic fields from pseudoGoldstone bosons,''
  Phys.\ Rev.\  D {\bf 46} (1992) 5346;
%  [arXiv:hep-ph/9209238].
  %%CITATION = PHRVA,D46,5346;%%
F.~D.~Mazzitelli and F.~M.~Spedalieri,
  %``Scalar electrodynamics and primordial magnetic fields,''
  Phys.\ Rev.\  D {\bf 52} (1995) 6694;
%  [arXiv:astro-ph/9505140].
  %%CITATION = PHRVA,D52,6694;%%
M.~Novello, L.~A.~R.~Oliveira and J.~M.~Salim,
  %``Direct Electrogravitational Couplings And The Behavior Of Primordial Large
  %Scale Magnetic Fields,''
  Class.\ Quant.\ Grav.\  {\bf 13} (1996) 1089;
  %%CITATION = CQGRD,13,1089;%%
A.~Dolgov,
  %``Breaking Of Conformal Invariance And Electromagnetic Field Generation In
  %The Universe,''
  Phys.\ Rev.\  D {\bf 48} (1993) 2499;
%  [arXiv:hep-ph/9301280].
  %%CITATION = PHRVA,D48,2499;%%
B.~Ratra,
  %``Cosmological 'seed' magnetic field from inflation,''
  Astrophys.\ J.\  {\bf 391} (1992) L1;
  %%CITATION = ASJOA,391,L1;%%
E.~A.~Calzetta, A.~Kandus and F.~D.~Mazzitelli,
  %``Primordial magnetic fields induced by cosmological particle creation,''
  Phys.\ Rev.\  D {\bf 57} (1998) 7139;
%  [arXiv:astro-ph/9707220].
  %%CITATION = PHRVA,D57,7139;%%
O.~Bertolami and D.~F.~Mota,
  %``Primordial magnetic fields via spontaneous breaking of Lorentz
  %invariance,''
  Phys.\ Lett.\  B {\bf 455} (1999) 96;
%  [arXiv:gr-qc/9811087].
  %%CITATION = PHLTA,B455,96;%%
M.~Giovannini,
  %``Magnetogenesis and the dynamics of internal dimensions,''
  Phys.\ Rev.\  D {\bf 62} (2000) 123505;
%  [arXiv:hep-ph/0007163].
  %%CITATION = PHRVA,D62,123505;%%
T.~Prokopec and E.~Puchwein,
  %``Photon mass generation during inflation: de Sitter invariant case,''
  JCAP {\bf 0404} (2004) 007;
%  [arXiv:astro-ph/0312274].
  %%CITATION = JCAPA,0404,007;%%
%T.~Prokopec and E.~Puchwein,
  %``Nearly minimal magnetogenesis,''
  Phys.\ Rev.\  D {\bf 70} (2004) 043004;
%  [arXiv:astro-ph/0403335].
  %%CITATION = PHRVA,D70,043004;%%
K.~Enqvist, A.~Jokinen and A.~Mazumdar,
  %``Seed perturbations for primordial magnetic fields from MSSM flat
  %directions,''
  JCAP {\bf 0411} (2004) 001;
%  [arXiv:hep-ph/0404269].
  %%CITATION = JCAPA,0411,001;%%
M.~R.~Garousi, M.~Sami and S.~Tsujikawa,
  %``Generation of electromagnetic fields in string cosmology with a massive
  %scalar field on the anti D-brane,''
  Phys.\ Lett.\  B {\bf 606} (2005) 1;
%  [arXiv:hep-th/0405012].
  %%CITATION = PHLTA,B606,1;%%
A.~Ashoorioon and R.~B.~Mann,
  %``Generation of cosmological seed magnetic fields from inflation with
  %cutoff,''
  Phys.\ Rev.\  D {\bf 71} (2005) 103509;
%  [arXiv:gr-qc/0410053].
  %%CITATION = PHRVA,D71,103509;%%
J.~E.~Madriz Aguilar and M.~Bellini,
  %``Stochastic gravitoelectromagnetic inflation,''
  Phys.\ Lett.\  B {\bf 642} (2006) 302;
%  [arXiv:gr-qc/0605043].
  %%CITATION = PHLTA,B642,302;%%
F.~Agustin Membiela and M.~Bellini,
  %``Power spectrum of large-scale magnetic fields from Gravitoelectromagnetic
  %inflation with a decaying cosmological parameter,''
%  arXiv:
0712.3032 [hep-th];
  %%CITATION = ARXIV:0712.3032;%%
L.~Campanelli, P.~Cea, G.~L.~Fogli and L.~Tedesco,
  %``Inflation-Produced Magnetic Fields in Nonlinear Electrodynamics,''
  Phys.\ Rev.\  D {\bf 77} (2008) 043001;
%L.~Campanelli, P.~Cea, G.~L.~Fogli and L.~Tedesco,
  %``Inflation-Produced Magnetic Fields in R^n F^2 and I F^2 models,''
  Phys.\ Rev.\  D {\bf 77} (2008) 123002.
%  [arXiv:0802.2630 [astro-ph]].
  %%CITATION = PHRVA,D77,123002;%%

\bibitem{gaugekin}
M.~Giovannini,
  %``On the variation of the gauge couplings during inflation,''
  Phys.\ Rev.\  D {\bf 64} (2001) 061301;
%  [arXiv:astro-ph/0104290].
  %%CITATION = PHRVA,D64,061301;%%
K.~Bamba and J.~Yokoyama,
  %``Large-scale magnetic fields from inflation in dilaton electromagnetism,''
  Phys.\ Rev.\  D {\bf 69} (2004) 043507;
%  [arXiv:astro-ph/0310824].
  %%CITATION = PHRVA,D69,043507;%%
%K.~Bamba and J.~Yokoyama,
  %``Large-scale magnetic fields from dilaton inflation in noncommutative
  %spacetime,''
  Phys.\ Rev.\  D {\bf 70} (2004) 083508;
%  [arXiv:hep-ph/0409237].
  %%CITATION = PHRVA,D70,083508;%%
O.~Bertolami and R.~Monteiro,
  %``Varying electromagnetic coupling and primordial magnetic fields,''
  Phys.\ Rev.\  D {\bf 71} (2005) 123525;
%  [arXiv:astro-ph/0504211].
  %%CITATION = PHRVA,D71,123525;%%
J.~M.~Salim, N.~Souza, S.~E.~Perez Bergliaffa and T.~Prokopec,
  %``Creation of cosmological magnetic fields in a bouncing cosmology,''
  JCAP {\bf 0704} (2007) 011;
%  [arXiv:astro-ph/0612281].
  %%CITATION = JCAPA,0704,011;%%
K.~Bamba and M.~Sasaki,
  %``Large-scale magnetic fields in the inflationary universe,''
  JCAP {\bf 0702} (2007) 030;
%  [arXiv:astro-ph/0611701].
  %%CITATION = JCAPA,0702,030;%%
J.~Martin and J.~Yokoyama,
  %``Generation of Large-Scale Magnetic Fields in Single-Field Inflation,''
  JCAP {\bf 0801} (2008) 025;
%  [arXiv:0711.4307 [astro-ph]].
  %%CITATION = JCAPA,0801,025;%%
K.~Bamba and S.~D.~Odintsov,
  %``Inflation and late-time cosmic acceleration in non-minimal Maxwell-$F(R)$
  %gravity and the generation of large-scale magnetic fields,''
  JCAP {\bf 0804} (2008) 024;
%  [arXiv:0801.0954 [astro-ph]].
  %%CITATION = JCAPA,0804,024;%%
K.~Bamba, C.~Q.~Geng and S.~H.~Ho,
  %``Large-scale magnetic fields from inflation due to Chern-Simons-like
  %effective interaction,''
  JCAP {\bf 0811} (2008) 013;
%  [arXiv:0806.1856 [astro-ph]].
  %%CITATION = JCAPA,0811,013;%%
V.~Demozzi, V.~Mukhanov and H.~Rubinstein,
  %``Magnetic fields from inflation?,''
  JCAP {\bf 0908} (2009) 025.
%  [arXiv:0907.1030 [astro-ph.CO]].
  %%CITATION = JCAPA,0908,025;%%

\bibitem{nonmin}
K.~Dimopoulos and M.~Kar\v{c}iauskas,
  %``Non-minimally coupled vector curvaton,''
  JHEP {\bf 0807} (2008) 119.
%  [arXiv:0803.3041 [hep-th]].
  %%CITATION = JHEPA,0807,119;%%

\bibitem{sugravec}
K.~Dimopoulos,
  %``Supergravity inspired Vector Curvaton,''
  Phys.\ Rev.\  D {\bf 76} (2007) 063506.
%  [arXiv:0705.3334 [hep-ph]].
  %%CITATION = PHRVA,D76,063506;%%

\bibitem{GE}
N.~E.~Groeneboom and H.~K.~Eriksen,
  %``Bayesian analysis of sparse anisotropic universe models and application to
  %the 5-yr WMAP data,''
  Astrophys.\ J.\  {\bf 690} (2009) 1807;
%  [arXiv:0807.2242 [astro-ph]].
  %%CITATION = ASJOA,690,1807;%%
N.~E.~Groeneboom, L.~Ackerman, I.~K.~Wehus and H.~K.~Eriksen,
  %``Bayesian analysis of an anisotropic universe model: systematics and
  %polarization,''
  arXiv:0911.0150 [astro-ph.CO].
  %%CITATION = ARXIV:0911.0150;%%

\bibitem{stanisplanck}
A.~R.~Pullen and M.~Kamionkowski,
  %``Cosmic Microwave Background Statistics for a 
  %Direction-Dependent Primordial Power Spectrum,''
  Phys.\ Rev.\  D {\bf 76} (2007) 103529.
%  [arXiv:0709.1144 [astro-ph]].
  %%CITATION = PHRVA,D76,103529;%%

\bibitem{AoE}
K.~Land and J.~Magueijo,
%``The axis of evil,''
  Phys.\ Rev.\ Lett.\  {\bf 95} (2005) 071301.
% [arXiv:astro-ph/0502237].
  %%CITATION = PRLTA,95,071301;%%

\bibitem{spins}
A.~Slosar {\it et al.},
  %``Galaxy Zoo: Chiral correlation function of galaxy spins,''
%  arXiv:
0809.0717 [astro-ph].
  %%CITATION = ARXIV:0809.0717;%%

\bibitem{align}
M.~J.~Longo,
  %``Does the Universe Have a Handedness,''
%  arXiv:
astro-ph/0703325.
  %%CITATION = ASTRO-PH/0703325;%%

\bibitem{foreground}
A.~de Oliveira-Costa and M.~Tegmark,
%``CMB multipole measurements in the presence of foregrounds,''
  Phys.\ Rev.\  D {\bf 74} (2006) 023005.
%  [arXiv:astro-ph/0603369].
  %%CITATION = PHRVA,D74,023005;%%

\bibitem{hansen}
H.~K.~Eriksen, F.~K.~Hansen, A.~J.~Banday, K.~M.~Gorski and P.~B.~Lilje,
  %``Asymmetries in the CMB anisotropy field,''
  Astrophys.\ J.\  {\bf 605} (2004) 14
  [Erratum-ibid.\  {\bf 609} (2004) 1198];
%  [arXiv:astro-ph/0307507].
  %%CITATION = ASJOA,605,14;%%
F.~K.~Hansen, A.~J.~Banday and K.~M.~Gorski,
  %``Testing the cosmological principle of isotropy: local power spectrum
  %estimates of the WMAP data,''
  Mon.\ Not.\ Roy.\ Astron.\ Soc.\  {\bf 354} (2004) 641.
%  [arXiv:astro-ph/0404206].
  %%CITATION = MNRAA,354,641;%%

\bibitem{soda}
  S.~Yokoyama and J.~Soda,
  %``Primordial statistical anisotropy generated at the end of inflation,''
  JCAP {\bf 0808} (2008) 005.
%  [arXiv:0805.4265 [astro-ph]].
  %%CITATION = JCAPA,0808,005;%%

\bibitem{stanis0}
L.~Ackerman, S.~M.~Carroll and M.~B.~Wise,
  %``Imprints of a Primordial Preferred Direction on the Microwave Background,''
  Phys.\ Rev.\  D {\bf 75} (2007) 083502.
%  [arXiv:astro-ph/0701357].
  %%CITATION = PHRVA,D75,083502;%%

\bibitem{stanis+}
%C.~Pitrou, T.~S.~Pereira and J.~P.~Uzan,
%  %``Predictions from an anisotropic inflationary era,''
%  JCAP {\bf 0804} (2008) 004;
%%  [arXiv:0801.3596 [astro-ph]];
%  %%CITATION = JCAPA,0804,004;%%
S.~Kanno, M.~Kimura, J.~Soda and S.~Yokoyama,
  %``Anisotropic Inflation from Vector Impurity,''
  JCAP {\bf 0808} (2008) 034;
%  [arXiv:0806.2422 [hep-ph]];
  %%CITATION = JCAPA,0808,034;%%
M.~A.~Watanabe, S.~Kanno and J.~Soda,
  %``Inflationary Universe with Anisotropic Hair,''
  Phys.\ Rev.\ Lett.\  {\bf 102} (2009) 191302.
%  [arXiv:0902.2833 [hep-th]].
  %%CITATION = PRLTA,102,191302;%%

\bibitem{stanis}
K.~Dimopoulos, M.~Kar\v{c}iauskas, D.~H.~Lyth and Y.~Rodriguez,
  %``Statistical anisotropy of the curvature perturbation from vector field
  %perturbations,''
  JCAP {\bf 0905} (2009) 013.
%  [arXiv:0809.1055 [astro-ph]].
  %%CITATION = JCAPA,0905,013;%%

\bibitem{dN}
A.~A.~Starobinsky,
  %``Multicomponent de Sitter (Inflationary) Stages and the Generation of
  %Perturbations,''
  JETP Lett.\  {\bf 42} (1985) 152
  [Pisma Zh.\ Eksp.\ Teor.\ Fiz.\  {\bf 42} (1985) 124];
  %%CITATION = ZFPRA,42,124;%%
M.~Sasaki and E.~D.~Stewart,
  %``A General Analytic Formula For The Spectral Index Of The Density
  %Perturbations Produced During Inflation,''
  Prog.\ Theor.\ Phys.\  {\bf 95} (1996) 71;
%  [arXiv:astro-ph/9507001];
  %%CITATION = PTPKA,95,71;%%
D.~H.~Lyth, K.~A.~Malik and M.~Sasaki,
  %``A general proof of the conservation of the curvature perturbation,''
  JCAP {\bf 0505} (2005) 004.
%  [arXiv:astro-ph/0411220].
  %%CITATION = JCAPA,0505,004;%%

\bibitem{triad}
M.~C.~Bento, O.~Bertolami, P.~V.~Moniz, J.~M.~Mourao and P.~M.~Sa,
%``On the cosmology of massive vector fields with SO(3) global symmetry,''
Class.\ Quant.\ Grav.\  {\bf 10} (1993) 285.
%[arXiv:gr-qc/9302034].
%%CITATION = GR-QC 9302034;%%

\bibitem{VI}
L.~H.~Ford,
%``Inflation Driven By A Vector Field,''
Phys.\ Rev.\ D {\bf 40} (1989) 967;
%%CITATION = PHRVA,D40,967;%%
C.~M.~Lewis,
%``Vector inflation and vortices,''
Phys.\ Rev.\ D {\bf 44} (1991) 1661.
%%CITATION = PHRVA,D44,1661;%%

\bibitem{vinf}
A.~Golovnev, V.~Mukhanov and V.~Vanchurin,
  %``Vector Inflation,''
  JCAP {\bf 0806} (2008) 009;
%  [arXiv:0802.2068 [astro-ph]];
  %%CITATION = JCAPA,0806,009;%%
%A.~Golovnev, V.~Mukhanov and V.~Vanchurin,
  %``Gravitational waves in vector inflation,''
  JCAP {\bf 0811} (2008) 018;
%  [arXiv:0810.4304 [astro-ph]];
  %%CITATION = JCAPA,0811,018;%%
T.~Chiba,
  %``Initial Conditions for Vector Inflation,''
  JCAP {\bf 0808} (2008) 004;
%  [arXiv:0805.4660 [gr-qc]];
  %%CITATION = JCAPA,0808,004;%%
A.~Golovnev and V.~Vanchurin,
  %``Cosmological perturbations from vector inflation,''
  Phys.\ Rev.\  D {\bf 79} (2009) 103524;
%  [arXiv:0903.2977 [astro-ph.CO]].
  %%CITATION = PHRVA,D79,103524;%%
A.~Golovnev,
  %``Linear perturbations in vector inflation and stability issues,''
%  arXiv:
0910.0173 [astro-ph.CO].
  %%CITATION = ARXIV:0910.0173;%%

\bibitem{vinf+}
Y.~Zhang,
  %``The Slow-Roll and Rapid-Roll Conditions in the Space-like Vector Field
  %Scenario,''
  Phys.\ Rev.\  D {\bf 80} (2009) 043519.
%  [arXiv:0903.3269 [astro-ph.CO]].
  %%CITATION = PHRVA,D80,043519;%%

\bibitem{mota}
T.~S.~Koivisto and D.~F.~Mota,
  %``Vector Field Models of Inflation and Dark Energy,''
  JCAP {\bf 0808} (2008) 021.
%  [arXiv:0805.4229 [astro-ph]].
  %%CITATION = JCAPA,0808,021;%%

\bibitem{vinf0}
A.~Tartaglia and N.~Radicella,
  %``Vector field theories in cosmology,''
  Phys.\ Rev.\  D {\bf 76} (2007) 083501.
%  [arXiv:0708.0675 [gr-qc]];
  %%CITATION = PHRVA,D76,083501;%%
S.~Koh and B.~Hu,
  %``Timelike Vector Field Dynamics in the Early Universe,''
%  arXiv:
0901.0429 [hep-th];
  %%CITATION = ARXIV:0901.0429;%%
S.~Koh,
  %``Vector Field and Inflation,''
%  arXiv:
0902.3904 [hep-th].
  %%CITATION = ARXIV:0902.3904;%%

\bibitem{vde}
C.~Armendariz-Picon,
%``Could dark energy be vector-like?,''
JCAP {\bf 0407} (2004) 007;
%[arXiv:astro-ph/0405267].
%%CITATION = ASTRO-PH 0405267;%%
H.~Wei and R.~G.~Cai,
  %``Interacting vector-like dark energy, the first and second cosmological
  %coincidence problems,''
  Phys.\ Rev.\  D {\bf 73} (2006) 083002;
%  [arXiv:astro-ph/0603052];
  %%CITATION = PHRVA,D73,083002;%%
C.~G.~Boehmer and T.~Harko,
  %``Dark energy as a massive vector field,''
  Eur.\ Phys.\ J.\  C {\bf 50} (2007) 423;
%  [arXiv:gr-qc/0701029];
  %%CITATION = EPHJA,C50,423;%%
T.~Koivisto and D.~F.~Mota,
  %``Accelerating Cosmologies with an Anisotropic Equation of State,''
  Astrophys.\ J.\  {\bf 679} (2008) 1;
%  [arXiv:0707.0279 [astro-ph]];
  %%CITATION = ASJOA,679,1;%%
H.~Wei and R.~G.~Cai,
  %``Cheng-Weyl Vector Field and its Cosmological Application,''
  JCAP {\bf 0709} (2007) 015;
%  [arXiv:astro-ph/0607064];
  %%CITATION = JCAPA,0709,015;%%
J.~B.~Jimenez and A.~L.~Maroto,
  %``A cosmic vector for dark energy,''
  Phys.\ Rev.\  D {\bf 78} (2008) 063005;
%  [arXiv:0801.1486 [astro-ph]];
  %%CITATION = PHRVA,D78,063005;%%
%J.~B.~Jimenez and A.~L.~Maroto,
  %``Vector models for dark energy,''
%  arXiv:
0807.2528 [astro-ph];
  %%CITATION = ARXIV:0807.2528;%%
%J.~B.~Jimenez and A.~L.~Maroto,
  %``Avoiding the dark energy coincidence problem with a cosmic vector,''
  AIP Conf.\ Proc.\  {\bf 1122} (2009) 107;
%  [arXiv:0812.1970 [astro-ph]];
  %%CITATION = APCPC,1122,107;%%
%J.~B.~Jimenez and A.~L.~Maroto,
  %``Cosmological evolution in vector-tensor theories of gravity,''
  Phys.\ Rev.\  D {\bf 80} (2009) 063512;
%  [arXiv:0905.1245 [astro-ph.CO]].
  %%CITATION = PHRVA,D80,063512;%%
J.~B.~Jimenez, R.~Lazkoz and A.~L.~Maroto,
  %``Cosmic vector for dark energy: constraints from SN, CMB and BAO,''
%  arXiv:
0904.0433 [astro-ph.CO].
  %%CITATION = ARXIV:0904.0433;%%

\bibitem{YMinf}
K.~Bamba, S.~Nojiri and S.~D.~Odintsov,
  %``Inflationary cosmology and the late-time accelerated expansion of the
  %universe in non-minimal Yang-Mills-$F(R)$ gravity and non-minimal
  %vector-$F(R)$ gravity,''
  Phys.\ Rev.\  D {\bf 77} (2008) 123532;
%  [arXiv:0803.3384 [hep-th]];
  %%CITATION = PHRVA,D77,123532;%%
K.~Bamba and S.~Nojiri,
  %``Cosmology in non-minimal Yang-Mills/Maxwell theory,''
%  arXiv:
0811.0150 [hep-th].
  %%CITATION = ARXIV:0811.0150;%%

\bibitem{YMde}
W.~Zhao and Y.~Zhang,
  %``The state equation of the Yang-Mills field dark energy models,''
  Class.\ Quant.\ Grav.\  {\bf 23} (2006) 3405;
%  [arXiv:astro-ph/0510356];
  %%CITATION = CQGRD,23,3405;%%
 W.~Zhao,
  %``Evolution of magnetic component in Yang-Mills condensate dark energy
  %models,''
  Int.\ J.\ Mod.\ Phys.\  D {\bf 16} (2007) 1735;
%  [arXiv:gr-qc/0701136];
  %%CITATION = IMPAE,D16,1735;%%
%W.~Zhao,
  %``Attractor Solution and Coincidence Problems in Coupled Yang-Mills field
  %Dark Energy Models,''
%  arXiv:
0810.5506 [gr-qc].
  %%CITATION = ARXIV:0810.5506;%%

\bibitem{forms}
C.~Germani and A.~Kehagias,
  %``P-nflation: generating cosmic Inflation with p-forms,''
  JCAP {\bf 0903} (2009) 028;
%  [arXiv:0902.3667 [astro-ph.CO]];
  %%CITATION = JCAPA,0903,028;%%
%C.~Germani and A.~Kehagias,
  %``Scalar perturbations in p-nflation: the 3-form case,''
  JCAP {\bf 0911} (2009) 005;
%  [arXiv:0908.0001 [astro-ph.CO]].
  %%CITATION = JCAPA,0911,005;%%
T.~Kobayashi and S.~Yokoyama,
  %``Gravitational waves from p-form inflation,''
  JCAP {\bf 0905} (2009) 004;
%  [arXiv:0903.2769 [astro-ph.CO]];
  %%CITATION = JCAPA,0905,004;%%
T.~S.~Koivisto, D.~F.~Mota and C.~Pitrou,
  %``Inflation from N-Forms and its stability,''
  JHEP {\bf 0909} (2009) 092;
%  [arXiv:0903.4158 [astro-ph.CO]].
  %%CITATION = JHEPA,0909,092;%%
T.~S.~Koivisto and N.~J.~Nunes,
  %``Inflation and dark energy from three-forms,''
  Phys.\ Rev.\  D {\bf 80} (2009) 103509;
%  [arXiv:0908.0920 [astro-ph.CO]].
  %%CITATION = PHRVA,D80,103509;%%
%T.~S.~Koivisto and N.~J.~Nunes,
  %``Three-form cosmology,''
%  arXiv:
0907.3883 [astro-ph.CO].
  %%CITATION = ARXIV:0907.3883;%%

\bibitem{tri}
C.~A.~Valenzuela-Toledo and Y.~Rodriguez,
  %``Non-gaussianity from the trispectrum and vector field perturbations,''
%  arXiv:
0910.4208 [astro-ph.CO].
  %%CITATION = ARXIV:0910.4208;%%

\bibitem{fnlanis}
M.~Kar\v{c}iauskas, K.~Dimopoulos and D.~H.~Lyth,
  %``Anisotropic non-Gaussianity from vector field perturbations,''
  Phys.\ Rev.\  D {\bf 80} (2009) 023509.
%  [arXiv:0812.0264 [astro-ph]].
  %%CITATION = PHRVA,D80,023509;%%

\bibitem{nonAbel}
N.~Bartolo, E.~Dimastrogiovanni, S.~Matarrese and A.~Riotto,
  %``Anisotropic bispectrum of curvature perturbations from primordial
  %non-Abelian vector fields,''
  JCAP {\bf 0910} (2009) 015;
%  [arXiv:0906.4944 [astro-ph.CO]].
  %%CITATION = JCAPA,0910,015;%%
%N.~Bartolo, E.~Dimastrogiovanni, S.~Matarrese and A.~Riotto,
  %``Anisotropic Trispectrum of Curvature Perturbations Induced by Primordial
  %Non-Abelian Vector Fields,''
%  arXiv:
0909.5621 [astro-ph.CO].
  %%CITATION = ARXIV:0909.5621;%%

\bibitem{peloso}
B.~Himmetoglu, C.~R.~Contaldi and M.~Peloso,
  %``Instability of the ACW model, and problems with massive vectors during
  %inflation,''
  Phys.\ Rev.\  D {\bf 79} (2009) 063517;
%  [arXiv:0812.1231 [astro-ph]];
  %%CITATION = PHRVA,D79,063517;%%
%B.~Himmetoglu, C.~R.~Contaldi and M.~Peloso,
  %``Instability of anisotropic cosmological solutions supported by vector
  %fields,''
  Phys.\ Rev.\ Lett.\  {\bf 102} (2009) 111301.
%  [arXiv:0809.2779 [astro-ph]].
  %%CITATION = PRLTA,102,111301;%%
%B.~Himmetoglu, C.~R.~Contaldi and M.~Peloso,
  %``Ghost instabilities of cosmological models with vector fields nonminimally
  %coupled to the curvature,''
%  arXiv:
0909.3524 [astro-ph.CO].
  %%CITATION = ARXIV:0909.3524;%%

\bibitem{carroll}
S.~M.~Carroll, T.~R.~Dulaney, M.~I.~Gresham and H.~Tam,
  %``Instabilities in the Aether,''
  Phys.\ Rev.\  D {\bf 79} (2009) 065011;
%  [arXiv:0812.1049 [hep-th]].
  %%CITATION = PHRVA,D79,065011;%%
T.~R.~Dulaney, M.~I.~Gresham and M.~B.~Wise,
  %``Classical stability of a homogeneous, anisotropic inflating space-time,''
  Phys.\ Rev.\  D {\bf 77} (2008) 083510
  [Erratum-ibid.\  D {\bf 79} (2009) 029903];
%  [arXiv:0801.2950 [astro-ph]];
  %%CITATION = PHRVA,D77,083510;%%

\bibitem{inst}
K.~Dimopoulos, M.~Karciauskas and J.~M.~Wagstaff,
  %``Vector Curvaton without Instabilities,''
%  arXiv:
0909.0475 [hep-ph].
  %%CITATION = ARXIV:0909.0475;%%

\bibitem{tikto}
  J.~M.~Cornwall, D.~N.~Levin and G.~Tiktopoulos,
  %``Derivation Of Gauge Invariance From High-Energy Unitarity Bounds On The S
  %Matrix,''
  Phys.\ Rev.\  D {\bf 10} (1974) 1145
  [Erratum-ibid.\  D {\bf 11} (1975) 972].
  %%CITATION = PHRVA,D10,1145;%%

\bibitem{wmap}
E.~Komatsu {\it et al.}  [WMAP Collaboration],
  %``Five-Year Wilkinson Microwave Anisotropy Probe (WMAP\altaffilmark 1 )
  %Observations:Cosmological Interpretation,''
  Astrophys.\ J.\ Suppl.\  {\bf 180} (2009) 330.
%  [arXiv:0803.0547 [astro-ph]].
  %%CITATION = APJSA,180,330;%%

\bibitem{cyd}
C.~A.~Valenzuela-Toledo, Y.~Rodriguez and D.~H.~Lyth,
  %``Non-gaussianity at tree- and one-loop levels from vector field
  %perturbations,''
%  arXiv:
0909.4064 [astro-ph.CO].
  %%CITATION = ARXIV:0909.4064;%%

\bibitem{isoc}
I.~Sollom, A.~Challinor and M.~P.~Hobson,
  %``Cold Dark Matter Isocurvature Perturbations: Constraints and Model
  %Selection,''
  Phys.\ Rev.\  D {\bf 79} (2009) 123521.
%  [arXiv:0903.5257 [astro-ph.CO]].
  %%CITATION = PHRVA,D79,123521;%%

\bibitem{randall}
%\cite{Dine:1995kz}
%\bibitem{Dine:1995kz}
M.~Dine, L.~Randall and S.~Thomas,
%``Baryogenesis from flat directions of the 
%supersymmetric standard model,''
Nucl.\ Phys.\ B {\bf 458} (1996) 291;
%[arXiv:hep-ph/9507453]
%%CITATION = HEP-PH 9507453;%%
%\cite{Dine:1995uk}
%\bibitem{Dine:1995uk}
%M.~Dine, L.~Randall and S.~Thomas,
%``Supersymmetry breaking in the 
%early universe,''
Phys.\ Rev.\ Lett.\  {\bf 75} (1995) 398;
%[arXiv:hep-ph/9503303].
%%CITATION = HEP-PH 9503303;%%
D.~H.~Lyth and T.~Moroi,
  %``The masses of weakly-coupled scalar fields in the early universe,''
  JHEP {\bf 0405} (2004) 004.
%  [arXiv:hep-ph/0402174].
  %%CITATION = JHEPA,0405,004;%%

\bibitem{low}
K.~Dimopoulos, D.~H.~Lyth and Y.~Rodriguez,
%``Low scale inflation and the curvaton mechanism,''
JHEP {\bf 0502} (2005) 055.
%[arXiv:hep-ph/0411119].
%%CITATION = JHEPA,0502,055;%%

\end{thebiblio}
\end{document}